\documentclass[aps,prb,amsmath,amssymb,preprintz,twocolumn]{revtex4-2}

\usepackage{graphicx}
\usepackage{amsmath}
\usepackage{braket}
\usepackage{amssymb}
\usepackage{multirow}
\usepackage{bbm}
\usepackage{bm}
\usepackage{mathtools}
\usepackage{hyperref}
\usepackage{natbib}
\usepackage{mathtools}
\usepackage{enumerate}
\usepackage[normalem]{ulem}
\usepackage[x11names]{xcolor}

\begin{document}

 \title{Correlation functions of the Kitaev model with a\\ spatially modulated phase in the superconducting order parameter}
	\author{Fabian G. Medina Cuy}
	\author{Fabrizio Dolcini}
	\affiliation{Dipartimento di Scienza Applicata e Tecnologia, Politecnico di Torino, corso Duca degli Abruzzi 24, 10129 Torino (Italy)}
	\date{\today}
	
	\begin{abstract}
		The  Kitaev chain model with a  spatially  modulated phase in the superconducting order parameter exhibits two types of topological transitions, namely a band topology transition between trivial and topological gapped phases, and a Fermi surface Lifshitz transition   from a gapped to a gapless superconducting state. We investigate the correlation functions of the model  for arbitrary values of   superconducting coupling~$\Delta_0$,   chemical potential $\mu$, and phase modulation wavevector $Q$, characterizing the current flowing through the system. In the cases   $\mu=0$  or  $Q=\pm \pi/2$ the model turns out to exhibit special symmetries, which are proven to induce an even/odd effect in the correlations as a function of  the distance $l$ between two lattice sites, as they are non-vanishing or strictly vanishing   depending on the  parity of $l$, measured in the lattice spacing unit.  
  We identify a clear difference between the band topology and the Lifshitz transition through the $Q$-dependence of the short distance correlation functions, which, in particular,   exhibit   pronounced  cusps with discontinuous derivatives across the Lifshitz transition.  
   We also determine the long distance behavior of correlations, finding that in the gapped phase there can be various types of exponential decays and that in the gapless phase the algebraic decay is characterized by two different spatial periods, depending on the model parameters. 
Furthermore, we establish a connection between  the gapless superconducting phase of the Kitaev chain  and the  chiral phase of spin models with  Dzyaloshinskii-Moriya interaction.

	\end{abstract}

	\maketitle

	\section{Introduction}
Correlation functions are one of the most  powerful tools to characterize the properties of quantum systems. In
topological phase transitions, which cannot be directly signaled by the onset of a spontaneously broken local order parameter~\cite{wen-book,wen_PRB_2010,ortiz-review}, their role becomes particularly important.
In topological insulators~\cite{hasan2010,zhang-review,chiu2016}, for instance, topological indices can be extracted from the ground-state  correlation functions, given on any system portion of the order of the  correlation length~\cite{kraus_PRB_2011}. In inversion-symmetric Dirac models, correlation functions are closely connected to the Fourier component of the Berry connection (in 1D) and  of the Berry curvature (in 2D) ~\cite{sigriest_PRB_2017,schnyder_2019}. Also, quantities like entanglement entropy, fidelity and discord, borrowed from  quantum information theory and harnessed for detecting topological quantum phase transitions~\cite{chamon_PRB_2008,zanardi_PRA_2008,johannesson_PRA_2009,li_PRA_2010}, ultimately require the evaluation of correlations. Moreover,  single-particle correlation functions of non-interacting systems can be used as a training set in machine learning techniques to predict topological phases of interacting systems~\cite{ercolessi_scipost_2023}.

One of the most interesting and widely studied topological quantum systems is the Kitaev chain model~\cite{kitaev2001}, which effectively captures the essential properties of a $p$-wave topological superconductor. The ground state of the model is characterized by two topologically distinct gapped phases and, in the topological non-trivial phase, it exhibits two  Majorana quasi-particle  (MQP) edge states, whose braiding properties could offer the opportunity to realize topologically protected quantum information~\cite{kitaev2003,alicea2012,aguado2017,zhang2018,beenakker2020,dassarma2023,aghaee2023}. For this reason, various implementations of the 1D Kitaev chain have been proposed, 
based on quantum spin Hall edge states contacted to   ferromagnets~\cite{fu-kane2008,fu-kane2009}, proximitized spin-orbit nanowires~\cite{lutchyn2010,oreg2010}, ferromagnetic atom  chains~\cite{choy2011,nadjperge2013,simon2013,glazman2013,franz2013,kotetes2014}, and cold atoms in optical lattices~\cite{kraus-zoller_2012,buchler_2014,lewenstein_2021},  receiving promising, although not
yet conclusive, support from experiments~\cite{kouwenhoven2012,furdyna2012,heiblum2012,kouwenhoven2018,yacoby2014,yu2021non,yazdani2014,tang2016,loss-meyer_2016}

So far, most studies of the correlation functions in the Kitaev chain have focused on two aspects. Firstly, the analysis of edge correlations in the case of  1D and 2D  lattices with open boundary conditions, with the purpose of finding a hallmark of the topological transition between the two gapped phases~\cite{zhou_PRL_2017,chen_PRB_2017,zhou_scirep_2018,murakami_PRB_2021,ma-song-ArXiv_2023}. Second, the effect of long-ranged hopping and superconducting terms, which do not allow for  the customary topological classification. These terms can lead to an algebraic decay of correlation functions even in  
gapped phases \cite{ercolessi_PRL_2014,ercolessi_NJP_2016,dellanna_PRB_2020,dellanna_PRB_2022} and their realization in ultracold atom setups has  been proposed~\cite{lewenstein_2021}.

However, in the experiments conducted on superconductor/semiconducting nanowire setups, where  signatures of MQPs are often searched for in the zero bias conductance peaks, an electrical current flow is driven across the system. This has recently motivated  research groups to investigate the effects of a spatial modulation in the phase of the superconducting order parameter of the Kitaev chain~\cite{takasan2022,kotetes2022,maiellaro2023,ma2023,FFF2024}, where the wavevector $Q$ is related to the net momentum of a Cooper pair, and is non vanishing in the presence of a current flow. In particular, in the regime $\Delta_0>w$, where the magnitude $\Delta_0$ of the superconducting order parameter is larger than the bandwidth parameter $w$, it has been shown that the spatial modulation wavevector $Q$ reduces the boundaries of the topological phase. Additionally, correlation functions have been found to exhibit off-diagonal long-range order at specific parameter points~\cite{ma2023}. A recent study has shown that, in the more realistic regime $\Delta_0<w$, an even richer scenario arises: Two types of transitions can occur as a function of $Q$~\cite{FFF2024}. The first one is related to the band topology and is the customary topological transition between the two gapped phases, while the second one is related to the Fermi surface topology, and is a Lifshitz transition~\cite{volovik2017,Volovik2018} between a gapped and a gapless superconducting phase, which further reduces the parameter range of observability of MQPs~\cite{FFF2024}. Furthermore, by treating  $Q$ as the the wavevector of an extra synthetic dimension, such a Lifshitz transition in the 1D Kitaev chain can also be seen as a transition between a type-I to type-II 2D  Weyl semimetal.

Motivated by such insightful results, in this paper we investigate the bulk correlation functions of the 1D Kitaev chain in the presence of the $Q$-modulation in the superconducting parameter phase, considering arbitrary values of the model parameters and of the distance  between the lattice sites of the Kitaev chain. We   shall  identify the behavior  of the correlation functions across the Lifshitz transition and  highlight the differences from the more conventional band topology transition between the two gapped phases, at both short and long distances $l$, measured in units of the lattice spacing.

The article is structured as   follows. In Sec.~\ref{sec2} we describe the model and summarize   those aspects of Ref.\cite{FFF2024} that are needed for the present analysis. Then, in Section~\ref{sec3} we introduce the normal and anomalous correlation functions that we investigate, $\mathcal{C}(l)$ and $\mathcal{A}(l)$  respectively, illustrating the way $Q$ affects their behavior. Moreover, we present our first result: For some noteworthy cases, namely for $\mu=0$ or for $Q=\pm \pi/2$, the model acquires some additional symmetries, resulting into an even/odd effect for the     correlation functions  $\mathcal{C}(l)$ and $\mathcal{A}(l)$, which are non vanishing or vanishing depending on the parity of the site distance~$l$. In   Sec.\ref{sec4} we relax the above parameter constraints, and analyze the behavior of $\mathcal{C}(l)$ and $\mathcal{A}(l)$ for {\it short} distances ($l=1,2$) as a function of the system parameters. We show that their  behavior  as a function of $Q$ when crossing the Lifshitz transition line from gapped to gapless superconductor phase is quite different from the case of the customary transition from topological to trivial gapped phases. Section~\ref{sec5} focusses instead on the {\it long} distance behavior ($l \gg 1$), and we determine the asymptotic behavior  both in the gapped and in the gapless phases.
Furthermore, in Sec.\ref{sec6} we discuss the relation between the Kitaev chain with superconducting modulation $Q$ and the XY spin model characterized by Dzyaloshinskii-Moriya interaction, highlighting a link in terms of correlation functions between a gapless superconductor and a spin chiral phase. 
Finally,  we summarize our results and  draw our conclusions   in  Sec.~\ref{sec-conclusions}.  \\

	\section{\label{sec2} Model, Excitation spectrum and current carrying state.}

	In order to model a 1D $p$-wave TS  crossed by a current flow, we include  a spatial modulation in the phase of the superconducting order parameter of the Kitaev chain, and consider the following Hamiltonian on a 1D lattice
	\begin{align}
		\begin{split}
			\mathcal{H} =& \sum_{j}\left\{ w \left(c^{\dagger}_{j}c^{}_{j + 1} + c^{\dagger}_{j+1}c_{j} \right) -\mu\left(c^{\dagger}_{j}c^{}_{j} - \frac{1}{2}\right) +\right. \\
			+& \left. \Delta_{0} \left(e^{-iQ\left(2j + 1\right)} c^{\dagger}_{j}c^{\dagger}_{j + 1} + e^{iQ\left(2j + 1\right)} c_{j+1}c_{j}    \right) \right\}\quad.
		\end{split}
		\label{KTV-Ham}
	\end{align}
	Here, $c^{}_{j} \, (c^{\dagger}_{j})$ corresponds to the annihilation (creation) operator at the lattice site $j$, while $w > 0$  denotes the tunneling amplitude of the hopping term determining the bare bandwidth $4w$, and $\mu$ is the chemical potential. Moreover,   the second line of Eq.(\ref{KTV-Ham}) represents the superconducting term, with $ \Delta_{0} > 0$ denoting the   superconducting coupling, and $Q$ the spatial modulation of its phase, related to the finite momentum $-2Q$ of a Cooper pair in the presence of a current flow. Assuming   an infinitely long chain, where the number of sites is $N_{s} \gg 1$, we can adopt periodic boundary conditions (PBCs), and treat  $Q$  as a continuum  variable. 

While details about the model Eq.(\ref{KTV-Ham})
can be found in Ref.\cite{FFF2024}, here we shall briefly recall the essential aspects that are needed   to discuss how $Q$ affects the  normal and the anomalous correlation functions. By  applying the   Fourier transform  and by introducing the Nambu spinors $\Psi^{\dagger}_{k;Q} = (c^{\dagger}_{k-Q} \,\,, c_{-k-Q})$,  the Hamiltonian~(\ref{KTV-Ham}) can be rewritten as 
\begin{equation}\label{KTV-Ham-k}
    \mathcal{H} =\frac{1}{2} \sum_{k}\Psi^\dagger_{k;Q} H(k;Q) \Psi^{}_{k;Q}\quad,
\end{equation}
where 
	\begin{equation}
		H\left(k; Q\right) = h_{0}(k;Q) \sigma_{0} + \bm{h}\left(k; Q\right)\cdot \bm{\sigma}
		\label{BdG}
	\end{equation}
	is the Bogolubov-de Gennes (BdG) Hamiltonian,  ${\bm{\sigma}} = \left(\sigma_{x}, \sigma_{y}, \sigma_{z}\right)$ denote the Pauli matrices, $\sigma_{0}$ the $2 \times 2$ identity,  
	\begin{eqnarray}
		h_{0}(k;Q) &=& 2w\sin{k}\, \sin{Q} \label{h0-def}\\
		{\bm{h}} (k;Q ) &=& (0, \,  -\text{Im}\left\{\Delta(k)\right\}, \,  \xi (k;Q )) \quad, \label{h-def}
	\end{eqnarray}
with
\begin{eqnarray} 
		\xi (k; Q ) &=& 2w \cos{Q} \, \cos {k}\,- \mu  \label{xi-def}  \\
  \Delta (k) &=& 2\Delta_{0}i \sin{k}
  \label{Delta(k)-def} \quad.
	\end{eqnarray}
Denoting $ h(k;Q) = |{\bm{h}}(k;Q) | =  [\xi^{2}(k;Q) + |\Delta(k)|^{2}]^{1/2}$, the spectrum	of the single-particle eigenvalues of the BdG Hamiltonian Eq.(\ref{BdG}) reads  
	\begin{equation}
		E_{\pm}(k;Q) = h_{0}(k;Q) \pm h(k;Q)
		\label{EigenV} \quad,
	\end{equation}
where the two bands  fulfill the mutual relation
\begin{equation}\label{Epm-rel}
E_{-}(k;Q)=-E_{+}(-k;Q)\quad,
\end{equation}
stemming from the redundancy of degrees of freedom in the Nambu formalism. Moreover, the eigenvectors $(u_Q(k),v_Q(k))^T$ and $(-v^*_Q(k),u_Q(k))^T$ of Eq.(\ref{BdG}), where   
	\begin{align}
		\begin{split}
		u_Q(k) =& \sqrt{\frac{1}{2}\left(1 + \frac{\xi(k;Q)}{h(k;Q)}\right)},\\
		v_Q(k) =& -i\,\text{sgn}\left(\sin(k)\right)\sqrt{\frac{1}{2}\left(1 - \frac{\xi(k;Q)}{h(k;Q)}\right)} 
		\end{split}\quad,
		\label{u_v}
	\end{align}
enable one to rewrite the Hamiltonian Eq.(\ref{KTV-Ham}) into its diagonal form
	\begin{equation}
		\mathcal{H} = \sum_{k}E_{+}(k;Q)\left(\gamma^{\dagger}_{k-Q}\gamma^{}_{k-Q}- \frac{1}{2}\right)
		\label{Diagonal_Hamiltonian}
	\end{equation}
in terms of   the Bogolubov quasiparticles, 
	\begin{align}
		\begin{split}
			\gamma_{k-Q} =& u_Q(k)c_{k-Q} + v_Q^{*}(k)c^{\dagger}_{-k-Q} \\
			\gamma^{\dagger}_{-k-Q} =& - v_Q(k)c_{k-Q} + u_Q(k)c^{\dagger}_{-k-Q}
		\end{split}.
		\label{Bogoliubov-quasi}
	\end{align}
The two terms appearing in the spectrum Eq.(\ref{EigenV})  allow us to highlight the twofold effect of the superconducting modulation wavevector $Q$. On the one hand, $Q$ renormalizes the bare tunneling amplitude $w\rightarrow w \cos{Q}$ [see Eq.(\ref{xi-def})] that appears in the term  $h(k;Q)$ of the spectrum Eq.(\ref{EigenV}). On the other hand, $Q$ introduces   in Eq.(\ref{EigenV}) the $h_0$-term Eq.(\ref{h0-def}), which is not present in the standard Kitaev chain ($Q=0$). The first effect   alters the boundaries between the topological and trivial gapped phases~\cite{ma2023}, whereas the second effect can have even more dramatic consequences. Indeed $h_0(k;Q)$ breaks the symmetry for $k \rightarrow -k$ of the spectrum and, when its magnitude overcomes $h(k;Q)$  for some $k$'s, it leads to $E_{+}(k;Q)<0$. Importantly, this alters  the occupancy of the $E_{+}$ band and the nature of the many-particle ground state~\cite{kotetes2022,FFF2024}. Indeed,   Eq.(\ref{Diagonal_Hamiltonian}) implies that, depending on whether $E_{+}(k;Q)>0$ or $E_{+}(k;Q)<0$, it is energetically more favorable for the system to have the $k$-state empty or occupied. For these reasons, differently from the customary Kitaev chain ($Q=0$), where the lower band $E_{-}$ is completely filled, or equivalently the upper band $E_{+}$ is completely empty,  the presence of the superconducting modulation 
can induce an {\it indirect} closing of the gap and  lead the system to a  {\it gapless}   ground state $\ket{G(Q)}$ of the model (\ref{KTV-Ham}). 
As   discussed in details in Ref.[\onlinecite{FFF2024}],    such ground state  can be expressed in general as  
	\begin{equation}
		\ket{G(Q)} =\!\! \prod_{\substack{0< k <\pi \\ k \in S_{p} }}\!\!\left(u_Q(k) + v^{*}_Q (k)c^{\dagger}_{-k-Q}c^{\dagger}_{k-Q}\right)\prod_{k \in S_e} c^{\dagger}_{k-Q}\ket{0},
		\label{G-gen}
	\end{equation}
where the Brillouin zone ({\rm BZ}) gets decomposed  in three sectors, ${\rm BZ}=S_{p}  \cup S_{e} \cup S_{h}$, identified through the three possible values
	\begin{equation}
		\eta(k;Q) = \left\{
		\begin{array}{cc}
			1 & k \in S_{h}\\
			0 & k\in S_{p}\\
			-1 & k\in S_{e}
		\end{array}\right. .
		\label{values_eta}
	\end{equation}
of the  spectral asymmetry function
	\begin{equation}
		\eta(k;Q) = \frac{1}{2}\left\{\text{sgn}E_{+}(k;Q) + \text{sgn}E_{-}(k;Q)\right\} \quad.
		\label{spectral-asymmetry}
	\end{equation} 
The sector $S_p$  represents the pair sector, as it involves in the ground state Eq.(\ref{G-gen}) Cooper pairs with a total momentum $-2Q$, while $S_e$ represents the unpaired electron sector, since only single electrons appear in Eq.(\ref{G-gen}) for $k \in S_e$,   and $S_h$  is the unpaired hole sector, since no electron state appears in Eq.(\ref{G-gen}) for $k \in S_h$. 
Note that, as a consequence of Eq.(\ref{Epm-rel}), the unpaired hole sector $S_h$  can be seen as the  mirror of the unpaired electron sector $S_e$ under $k\rightarrow -k$, while the pair sector $S_p$ is self-mirrored under such transformation.\\

\begin{figure*}
		\centering
        \includegraphics[scale = 0.95]{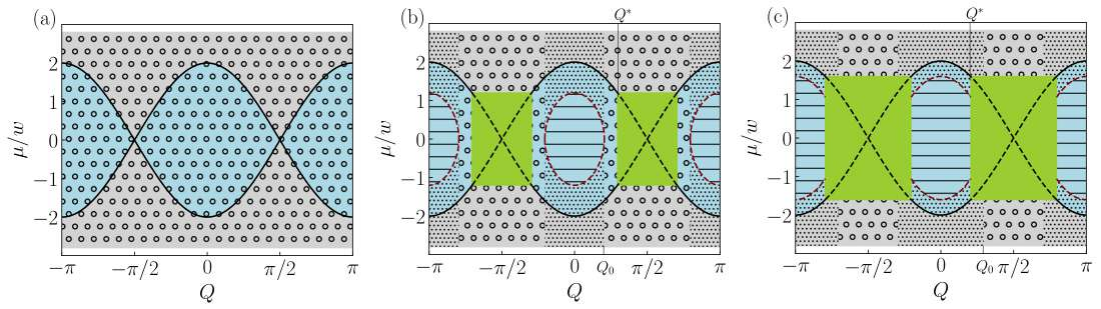}
		\caption{ \label{Fig1}     Phase diagram of the Kitaev chain as a function of the wavevector $Q$ (spatial modulation of the superconducting order parameter) and chemical potential $\mu$, for three different values of the superconducting coupling: (a) $\Delta_0=1.3w$; (b) $\Delta_0=0.8w$; (c) $\Delta_0=0.6w$. Grey and cyan areas denote the trivial and topological gapped phases, respectively, while the green region denotes the gapless superconducting phase. Different symbols in the gapped phases identify three different types of  exponential decay in the correlation functions, at long distance $l\gg 1$. 
  In regions marked with ``$\circ$", and ``$\cdot$",   correlations  behave as a linear combination of two exponential decays,   with or without a relative sign $(-1)^l$ alternating with the distance $l$ [see Eqs.(\ref{Cp-asym-(a)-1})-(\ref{A-asym-(a)-1}) and Eqs.(\ref{Cp-asym-(a)-2})-(\ref{A-asym-(a)-2}), respectively], while inside the elliptic  region marked by ``$-$" they decay as one single exponential decay with additional spatial oscillations [Eqs.(\ref{Cp-asym-(b)})-(\ref{A-asym-(b)})].  Inside the gapless phase   correlations decay algebraically with  additional two-period oscillations [see Eqs.(\ref{Cp-asym-gapless})-(\ref{A-asym-gapless})].
The crossing between situations (b) and (c), where the elliptic gapped region and the gapless region touch, occurs at the value $\Delta_0^*=w/\sqrt{2}$.
 The wavevector $Q^*$, given in Eq.(\ref{Q-star-def}), identifies the gapped/gapless boundary, while   $Q_0$, given in Eq.(\ref{Q0-def}), determines the boundary between the ``$\circ$" and ``$\cdot$" regions.}
	\end{figure*}
 
The ground state of the system can be in two possible type of phases.\\

\textit{Gapped phase.} 
	The gapped phase is characterized by  the condition $E_{+}(k;Q)>0 \,\, \forall k \in {\rm BZ}$, which in turn implies $E_{-}(k;Q)<0 \,\, \forall k \in {\rm BZ}$, on account of Eq.(\ref{Epm-rel}). Then, Eq.(\ref{spectral-asymmetry}) implies $\eta(k;Q) \equiv 0 \,\,\, \forall k\in {\rm BZ}$, and from Eq.(\ref{values_eta}) one deduces that the pair sector $S_p$ covers the entire Brillouin zone, leaving the $S_e$ and $S_h$ sectors empty 
\begin{equation}
\begin{array}{l}
S_p  \equiv {\rm BZ} \leftrightarrow k \in [-\pi,\pi]\\ \\
S_e = S_h = \emptyset
\end{array} \quad. \label{sectors-gapped}
\end{equation}
In this case, the general expression  (\ref{G-gen}) of the ground state reduces to the standard form consisting of Cooper pairs only.
The gapped phase occurs if and only if one of the following three parameter conditions is fulfilled\cite{FFF2024}
	\begin{equation}
		\begin{array}{lcl}
			i) &  |\mu|>2w  \mbox{ \& } & \forall \Delta_0>0 \mbox{ \& } \forall Q\\ 
			ii) & |\mu|<2w  \mbox{ \&} & \sqrt{w^2-\mu^2/4}<\Delta_0 \,\mbox{ \& }   \left|\cos{Q}\right| \neq |\mu|/2w\\ 
			iii) & |\mu|<2w \mbox{ \& } & w|\sin{Q}|<\Delta_0<\sqrt{w^2-\mu^2/4} 
		\end{array}.
		\label{gapped-regime}
	\end{equation}
	As is well known, there exist two topological distinct gapped phases, separated by the curves $\left|\mu\right|= 2w\left|\cos{Q}\right|$ in the parameter space, where the gap in the excitation spectrum closes {\it directly} at either $k^* = 0$ or $k^* = \pi$, i.e.   when $E_{+}(k^*;Q)=E_{-}(k^*;Q)=0$.\\

\textit{Gapless phase.} 
When, however,   $E_{+}(k;Q)<0$ for some~$k$, the gap between the two bands closes {\it indirectly} since  $E_{-}(-k;Q)>0$.  In this case Eq.(\ref{values_eta})
implies that the unpaired electron sector $S_e$ is not an empty set, just like its $k$-mirror set $S_h$.  Cooper pairs are present only in a portion $S_p$ of the Brillouin zone and the ground state  is strictly mixed, as given by  Eq.(\ref{G-gen}). Such type of state  arises also in other contexts, such as $s$-wave paired superfluids with rotationally symmetric confinement potentials  \cite{prem2017,tada2018} and Fermi gases with two species of Fermions \cite{liu2003}.
The ground state of  model~(\ref{KTV-Ham}) is in the gapless phase when both the following conditions are fulfilled\cite{FFF2024}
\begin{equation}\label{gapless-regime}
\sqrt{\Delta_0^2+\frac{\mu^2}{4}} < w  \,\,\, \& \,\,\,
 \Delta_0   < w |\sin{Q}|  
\quad.
\end{equation}
In particular, one can show that the unpaired fermions (electron or holes) sector $S_u=S_e+S_h$ of the Brillouin zone is given by
\begin{equation}
\begin{array}{c}
 S_u \,\,=\left\{ \,\, k \,\, | \,\,
   \left|k^{*}_{-}\right| <\left|k\right| < \pi - \left|k^{*}_{+}\right| \right\}
\label{gapless_lim}
\end{array}
\end{equation}
while the pair sector is
\begin{eqnarray}
S_p =  \left\{ k\,\,| \,\, 
  0 <  |k| <  |k^{*}_{-}|  \,\, \mbox{or}\, \,\pi - \left|k^{*}_{+}\right| < \left|k\right| < \pi
 \right\} \quad. \label{Sp-gapless}
\end{eqnarray}
where 
\begin{equation}
k^{*}_{\pm} = \arcsin\left(\frac{\cos{Q}}{\sqrt{1- \frac{\Delta^{2}_{0}}{w^{2}}}}\right) \pm \arcsin\left(\frac{\mu}{2w\sqrt{1- \frac{\Delta^{2}_{0}}{w^{2}}}}\right). \label{k-star-pm}
\end{equation}

\noindent We conclude this section by recalling the  phase diagram of the Kitaev chain (\ref{KTV-Ham}), obtained in Ref.[\onlinecite{FFF2024}]. Here it is shown   in Fig.\ref{Fig1} as a function of the superconducting modulation wavevector $Q$ and the chemical potential $\mu$, for three different values of $\Delta_0$. Here cyan and gray areas denote the topological and trivial gapped phases, respectively, while the green area denotes the gapless phase.  
While for $\Delta_0>w$ only the gapped phase exists [Fig.\ref{Fig1}(a)], and the ground state only consists of Cooper pairs, for the physically realistic regime $\Delta_0<w$ also the gapless phase appears. It exists for the parameter values (\ref{gapless-regime}), and is the more extended the lower the values of $\Delta_0$ [Figs.\ref{Fig1}(b) and (c)]. The additional symbols appearing in   Fig.\ref{Fig1} identify different long distance behavior of the correlation functions, as we shall explain in detail in Sec.\ref{sec5}.

 	%%%%%%%%%%%%%%%%%%%%%%%%%%%%%%%%%%%%%%%%%%%%%%%
   	%%%%%%%%%%%%%%%%%%%%%%%%%%%%%%%%%%%%%%%%%%%%%%%
    %%%%%%%%%%%%%%%%%%%%%%%%%%%%%%%%%%%%%%%%%%%%%%%
    %%%%%%%%%%%%%%%%%%%%%%%%%%%%%%%%%%%%%%%%%%%%%%%
	\section{Correlation functions: expressions and even/odd effect}
	\label{sec3}  
	We now determine the real-space behavior of the correlation functions of the model (\ref{KTV-Ham}). Specifically, we shall consider the normal and the anomalous correlation, defined as     
	\begin{eqnarray}
		\mathcal{C}(l)&=&e^{i Q l }\langle  c^{\dagger}_{j}c^{}_{j+l} \rangle   \label{C-def} \\
		\mathcal{A}(l)&=& e^{-i Q (2j+l) } \langle  c^{\dagger}_{j }c^{\dagger}_{j+l} \rangle  \quad, 
		\label{A-def}
	\end{eqnarray}
 respectively. Here the expectation values $\langle \cdots \rangle$ are computed with respect to the ground state $\ket{G(Q)}$, whose general expression is given by Eq.(\ref{G-gen}). 
 Due the translational invariance of the model, $\mathcal{C}$ and $\mathcal{A}$ are independent of the site location $j$  and only depend  on the site distance~$l$,   which is  assumed to be $l \neq 0$. All distances are expressed in terms of the lattice spacing. Straightforward algebra, whose details are given in Appendix \ref{appendixA}, enables one to re-express the  correlations Eqs.(\ref{C-def})-(\ref{A-def}) as integrals in momentum space, and to identify the different contributions related to the  sector $S_p$,  $S_e$, $S_h$ characterizing the ground state  (\ref{G-gen}). 
Specifically, one finds  
	\begin{equation}
		\mathcal{C}(l) =  \mathcal{C}_{p}(l) + i \,\mathcal{C}_{u}(l)\,,
		\label{C-Re-Im}
	\end{equation} 
	where
	\begin{equation}
		\mathcal{C}_{p}(l) = - \frac{1}{4\pi}\intop_{S_p} dk \frac{\cos(k l)\xi(k;Q)}{h(k;Q)}  
		\label{Cp-k}
	\end{equation}
represents the Cooper pair contribution, and  is the real part of $\mathcal{C}(l)$, whereas 
	\begin{eqnarray}
		\mathcal{C}_{u}(l) &=&  \frac{1}{2\pi}\intop_{S_e}dk \sin (k l )  
				\label{Cu-k} 
	\end{eqnarray}
represents the unpaired   electron   contribution, and is the imaginary part of $\mathcal{C}(l)$. Similarly,  the anomalous correlation function can be re-expressed as 
\begin{eqnarray}
		\mathcal{A}	(l) =    - \frac{\Delta_{0}}{2\pi}\intop_{S_p} dk \frac{\sin(k l)\sin{k}}{h(k;Q)}  ,\,\,	\label{A-k}
	\end{eqnarray}
 and only consists of contributions from Cooper pairs, as expected. 

A comment is in order about how $Q$ enters the expressions Eqs.(\ref{Cp-k})-(\ref{Cu-k}) and (\ref{A-k}) of the correlation functions. On the one hand, 
the integrand functions in these equations depend on $Q$ only through 
the effect of renormalization $w\rightarrow w \cos{Q}$  of the tunneling amplitude, encoded in the functions $h(k;Q)$ and $\xi(k;Q)$. On the other hand, the integration domains $S_p$ and $S_e$, which are determined by the spectral asymmetry values Eq.(\ref{values_eta}), also 
depend on the $h_0(k;Q)$ term of the spectrum  Eq.(\ref{EigenV}), which is responsible for the possible change induced by $Q$ in the band occupancy. 

Let us now turn to the evaluation of the correlation functions Eqs.(\ref{Cp-k})-(\ref{Cu-k}) and (\ref{A-k}). Firstly, we note that the unpaired fermion contribution $\mathcal{C}_u$  can be given an analytical  exact expression for arbitrary parameter values. Indeed,  in the gapped phase, the spectral asymmetry   always vanishes, $\eta \equiv 0\,\, \forall k \in {\rm BZ}$ [see Eq.(\ref{values_eta})], and one has $\mathcal{C}_u=0$. In contrast, in the gapless phase, one can rewrite Eq.(\ref{Cu-k}) as 
\begin{eqnarray}
\mathcal{C}_{u}(l) &=&  - \frac{1}{4\pi}\intop_{S_u}dk \sin (kl )\eta(k;Q) = \nonumber \\
&=&
-  \frac{\text{sgn}(Q)}{2\pi}\frac{1}{l}\left( \cos(k^{*}_{-}l) - (-1)^{l}\cos(k^{*}_{+}l)    \right)\quad,
\label{Cu-exact}
\end{eqnarray}
where $k^*_\pm$ are given in Eq.(\ref{k-star-pm}) and Eqs.(\ref{values_eta}) and (\ref{gapless_lim}) have been used.
For the Cooper pair contributions $\mathcal{C}_p(l)$ to the normal correlator, and for the anomalous correlator $\mathcal{A}(l)$, analytical results are not available in general. Nevertheless, we
can obtain such  correlations by numerically exact  integration of Eqs.(\ref{Cp-k})  and (\ref{A-k}) and, in some limits, we can provide analytical expressions.
Here and in the next sections we shall  present these results, pointing out the   effect of the $Q$-wavevector related to the current flow.  

%%%%%%%%%%%%%%%%%%%%%%%%%

\subsection{Even/odd effect for the special cases \\$\mu=0$ or $Q = \pm \pi/2$}
\label{sec3-A}
We start by discussing two noteworthy cases, namely   $\mu = 0$  and  $Q =  \pm \pi/2 $, where   the correlation functions  can be rigorously shown to exhibit an even/odd effect. Indeed $\mathcal{C}(l)$ and $\mathcal{A}(l)$ are non-vanishing or vanishing depending on the  even/odd parity of the site distance $l$, measured in units of the lattice spacing, as summarized in  Tables \ref{table1} and \ref{table2}. 

Specifically, Table \ref{table1} refers to the case $\mu=0$ and shows that both the normal and the anomalous correlation functions vanish at any {\it even} site distance $l$, for any value of $Q$ and $\Delta_0$, both in the gapped and in the gapless phase.
Table \ref{table2} illustrates the case $Q=\pm \pi/2$.  In this case, while  the anomalous correlation $\mathcal{A}$ still vanishes at any {\it even} site distance $l$,  the real part $\mathcal{C}_p$ of the normal correlation~$\mathcal{C}$, which originates from the Cooper pairs [see Eq.(\ref{C-Re-Im})], vanishes for any {\it odd} site distance $l$. Again, this holds both in the gapped and in the gapless phase. Note that  in the gapless phase where unpaired fermions are present, $\mathcal{C}$ is   purely imaginary for odd $l$, $\mathcal{C}(l)=i\mathcal{C}_u(l)$,  as it   takes contribution from the unpaired fermions, while for even $l$ it is purely real as it takes contribution from  Cooper pairs, $\mathcal{C}(l)=\mathcal{C}_p(l)$.  

\begin{figure}
\centering
\includegraphics[scale = 0.45]{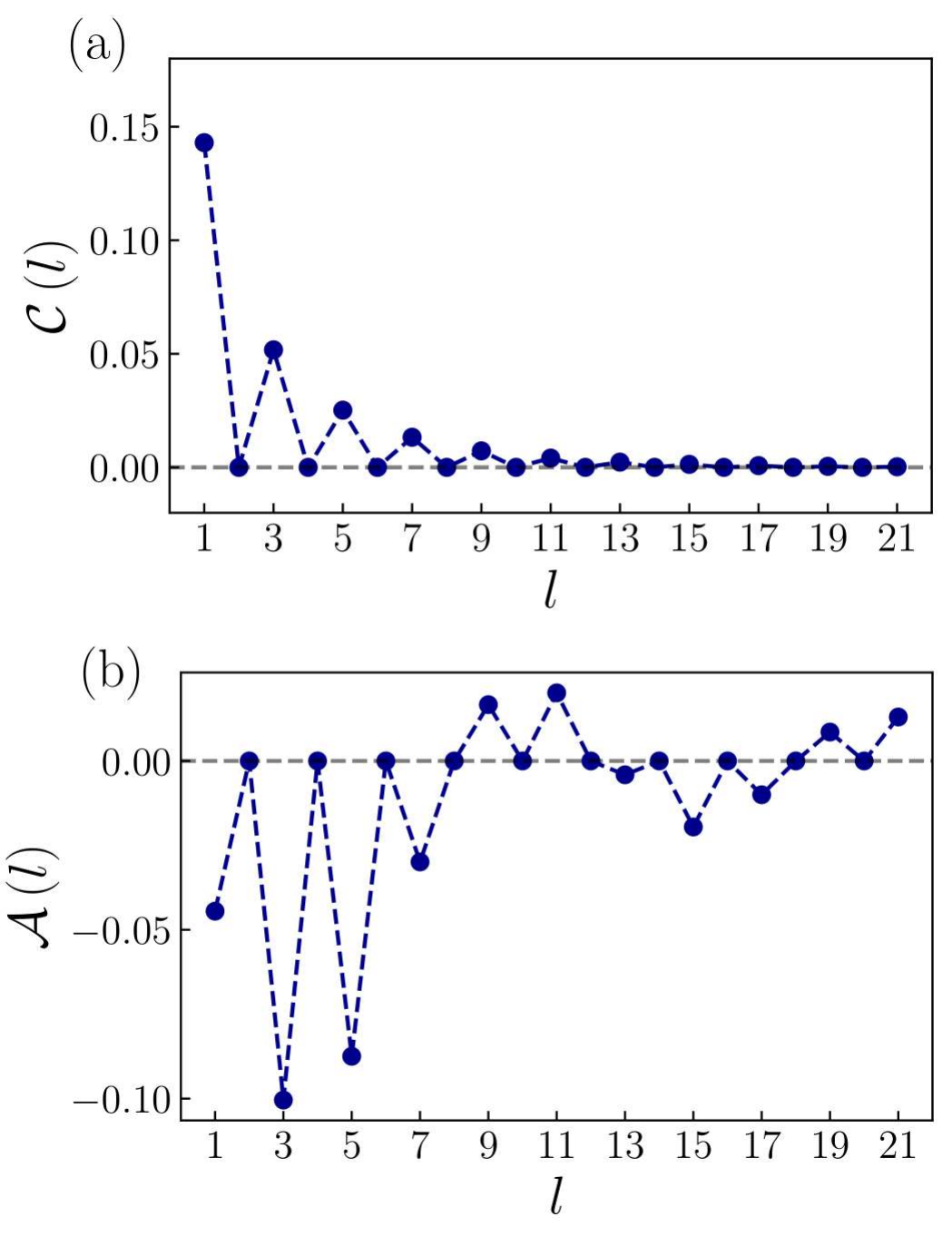}
\caption{ \label{Fig2} The even/odd effect of the correlation functions. Panel (a) is an example of Table \ref{table1} and shows the normal correlation function (\ref{C-def}) as a function of $l>0$, for $\mu = 0$, $Q = 0.6 \pi$ and $\Delta_{0} = 1.3$ (ground state   in the gapped phase).  Panel (b) is an example of Table \ref{table2} and shows the anomalous correlation  (\ref{A-def}) as a function of $l>0$, for the values $\mu = 0.5$, $Q = \pi/2$ and $\Delta_{0} = 0.8$ (ground state  in the gapless phase). In both cases the correlation functions vanish exactly at $l$ even.} 
\end{figure}

Two examples of even/odd effect   are shown in   Fig.\ref{Fig2}. In particular panel (a) displays $\mathcal{C}$ at $\mu=0$  as a function of~$l$ for $\Delta_0=1.3w$, $Q=0.6 \pi$ (gapped phase), while in panel (b) the nomalous correlation $\mathcal{A}$ is plotted as function of $l$ for $\Delta_0=0.8w$ and $\mu=0.5 w$ (gapless phase). Note that, in the two panels, the non-vanishing values of the correlations decay differently at long distance $l \gg 1$. In particular, the decay exhibited by $\mathcal{C}$ in panel (a) is one of the three possible exponential decays characterizing the gapped phase, namely the one highlighted  as circles   ``$\circ$" in  Fig.\ref{Fig1}. Specifically, it is a decay without oscillations, where the decay  length depends of the values of $\Delta_0/w$ and $Q$. By contrast,  $\mathcal{A}$ in panel (b) exhibits oscillations with a slower algebraic decay. As we shall see, this is typical of the gapless phase  and, for the specific parameters of panel (b), the oscillation period is controlled by the chemical potential.  
The  two special cases shown in Fig.\ref{Fig2} are just  examples of   asymptotic behavior of correlation functions, whose thorough analysis  will be presented in details in Sec.\ref{sec5}, for arbitrary values of $\mu$ and $Q$.

\begin{table}[h]
\centering
\begin{tabular}{| c | c | c | c |}
\hline
 &  gapped phase   &  gapless phase   &  \\
 \hline
 \multirow{2}{0.6cm}{$\mathcal{C}(l)$}  & $0$ & $0$ & $l\,\,\text{even}$ \\
&  \checkmark   &  \checkmark   &  $l \,\, \text{odd}$ \\
  \hline                                      
  \multirow{2}{0.6cm}{$\mathcal{A}(l)$} &      $0$     &        $0$      &   $l \,\, \text{even}$ \\
& \checkmark  &     \checkmark     & $l \,\, \text{odd}$\\
    \hline
\end{tabular} \hspace{0.5cm} $(\mu=0)$
\caption{\label{table1} The case $\mu = 0$: Correlation functions for $l \neq 0$. Both the normal and the anomalous correlation function (\ref{C-def}) and (\ref{A-def}) are strictly vanishing for any even value of the site distance $l  \neq 0$, while they are non-vanishing for $l$ odd. This holds both in the gapped and in the gapless phase. }
\end{table}

\begin{table}[h]
\centering
\begin{tabular}{| c | c | c | c |}
\hline
&  gapped phase   &  gapless phase   &  \\
 \hline
 \multirow{2}{0.6cm}{$\mathcal{C}(l)$}  & \checkmark & $\equiv \mathcal{C}_p(l)$  & $l\,\,\text{even}$ \\
&  $0$   &  $\equiv i\mathcal{C}_u(l)$   &  $l \,\, \text{odd}$ \\
  \hline                                      
  \multirow{2}{0.6cm}{$\mathcal{A}(l)$} &      $0$     &        $0$      &   $l \,\, \text{even}$ \\
& \checkmark  &     \checkmark     & $l \,\, \text{odd}$\\
    \hline
\end{tabular} \hspace{0.5cm} $(Q=\pm \frac{\pi}{2})$
\caption{\label{table2} The case $Q=\pm \pi/2$: Correlation functions for $l \neq 0$. While the anomalous correlation function  (\ref{A-def}) vanishes for any even values of $l$, the normal correlation function exhibits a different behavior depending on whether the system ground state is in its gapped or in the gapless phase. In the former case it strictly vanishes for any odd $l$, while
in the latter case it either gets  contribution only from Cooper pairs, $\mathcal{C}=\mathcal{C}_p$ (for even $l$) or from the unpaired fermions   $\mathcal{C}=i\mathcal{C}_u$ (for odd $l$).
 }
\end{table}

The  even/odd effects summarized in Tables \ref{table1} and \ref{table2} originate from  special symmetries that the Hamiltonian~$\mathcal{H}$ in Eq.(\ref{KTV-Ham}) acquires for the particular parameter values $\mu=0$ or $Q=\pm \pi/2$, and that  do not hold for   generic values of $\mu$ and $Q$.
Specifically, for $\mu=0$, $\mathcal{H}$ exhibits the {\it chiral} symmetry, i.e. it is invariant under the (anti-unitary) transformation   
\begin{equation}\label{S-def}
\mathcal{S} c_{j} \mathcal{S}^{-1}  =  (-1)^{j}c^{\dagger}_{j}\hspace{0.5cm}
\Rightarrow\hspace{0.5cm}
\mathcal{S} c_{k} \mathcal{S}^{\dagger}  = c^{\dagger}_{k-\pi} \quad,
\end{equation}
while for  $Q = \pm \pi/2$  the Hamiltonian is {\it inversion}-symmetric, i.e. it is invariant under the following (unitary) spatial inversion   
\begin{equation}\label{I-def}
\mathcal{I}c_{j}\mathcal{I}^{-1} =c_{-j} \hspace{0.5cm}
\Rightarrow 
\hspace{0.5cm}
\mathcal{I}c_{k}\mathcal{I}^{-1}  = c_{-k}\quad.
\end{equation} 
The proof that the symmetries $[\mathcal{H},\mathcal{S}]=0$ and $[\mathcal{H},\mathcal{I}]=0$   imply the above even/odd effects is provided in details in  Appendix \ref{AppB}. Here we limit ourselves to mention that for small values of distance ($l \le 3$), it is also possible to find  analytical expressions for the (non vanishing) correlations in terms of elliptic functions, which are also reported in Appendix \ref{AppB}.

We conclude this subsection by a comment on the two special points of the parameter space, namely  $(Q,\mu)=(\pm \pi/2,0)$, which  were recently analyzed in Ref.\cite{ma2023}   in the regime $\Delta_0>w$, where  off-diagonal long  range order was found in the correlations. Our analysis allows one to obtain additional information about these points. First, we can identify them as {\it high  symmetry points}, where the system exhibits {\it both} chiral and inversion symmetries, $\mathcal{S}$ and~$\mathcal{I}$. Moreover, 
 we can generalize the results obtained in Ref.\cite{ma2023} for the regime $\Delta_0>w$ by observing that, 
because in such a regime the system is always in the gapped phase [Fig.\ref{Fig1}(a)],   normal correlation at {\it arbitrary site distance} $l$ vanishes, $\mathcal{C}(l)\equiv 0\,\, \forall l>0$ (even and odd), and only anomalous correlation exists at such point, as can be deduced from  the intersection of Tables \ref{table1} and \ref{table2}.   Finally, our results   extend the analysis of correlations also to the regime $\Delta_0<w$, where the high symmetry points  always correspond to the gapless phase.   In this case Tables \ref{table1} and \ref{table2} predict that   the normal correlation is  vanishing  for even $l$, while for odd $l$   it only gets contribution from the unpaired fermion sector, $\mathcal{C}=i\mathcal{C}_u$. As we shall see in Sec.\ref{sec6}, $\mathcal{C}_u$ can be related to spin chiral gapless phases in spin models. 
%%%%%%%%%%%%%%%%%%%%%%%%%%%%%
%%%%%%%%%%%%%%%%%%%%%%%%%%%%%
%%%%%%%%%%%%%%%%%%%%%%%%%%%%%
\section{Short distance behavior of correlation functions}
\label{sec4}
We analyze now the  behavior of the  correlation functions (\ref{C-def}) and (\ref{A-def}) at short distance $l$,  for arbitrary values of the parameters $Q$, $\mu$, and $\Delta_0$. Specifically, by the numerically exact evaluation of $\mathcal{C}_p$ and $\mathcal{A}$ in Eqs.(\ref{Cp-k})  and (\ref{A-k}),
we shall analyze the quantities $\left|\mathcal{C}(l)\right|^{2} = \mathcal{C}^{2}_{p}(l) + \mathcal{C}^{2}_{u}(l)$ and $\left|\mathcal{A}(l)\right|^{2}$. The former can be considered as the probability for an electron to hop from site  $j+l$ to $j$, while  the latter corresponds to the probability of creating  a Cooper pair at sites $j$ and $j + l$. 
We shall focus here on the short distance values $l=1$ and $l=2$.

\subsection{The case $l=1$} 
\begin{figure*}
        \includegraphics[scale = 0.45]{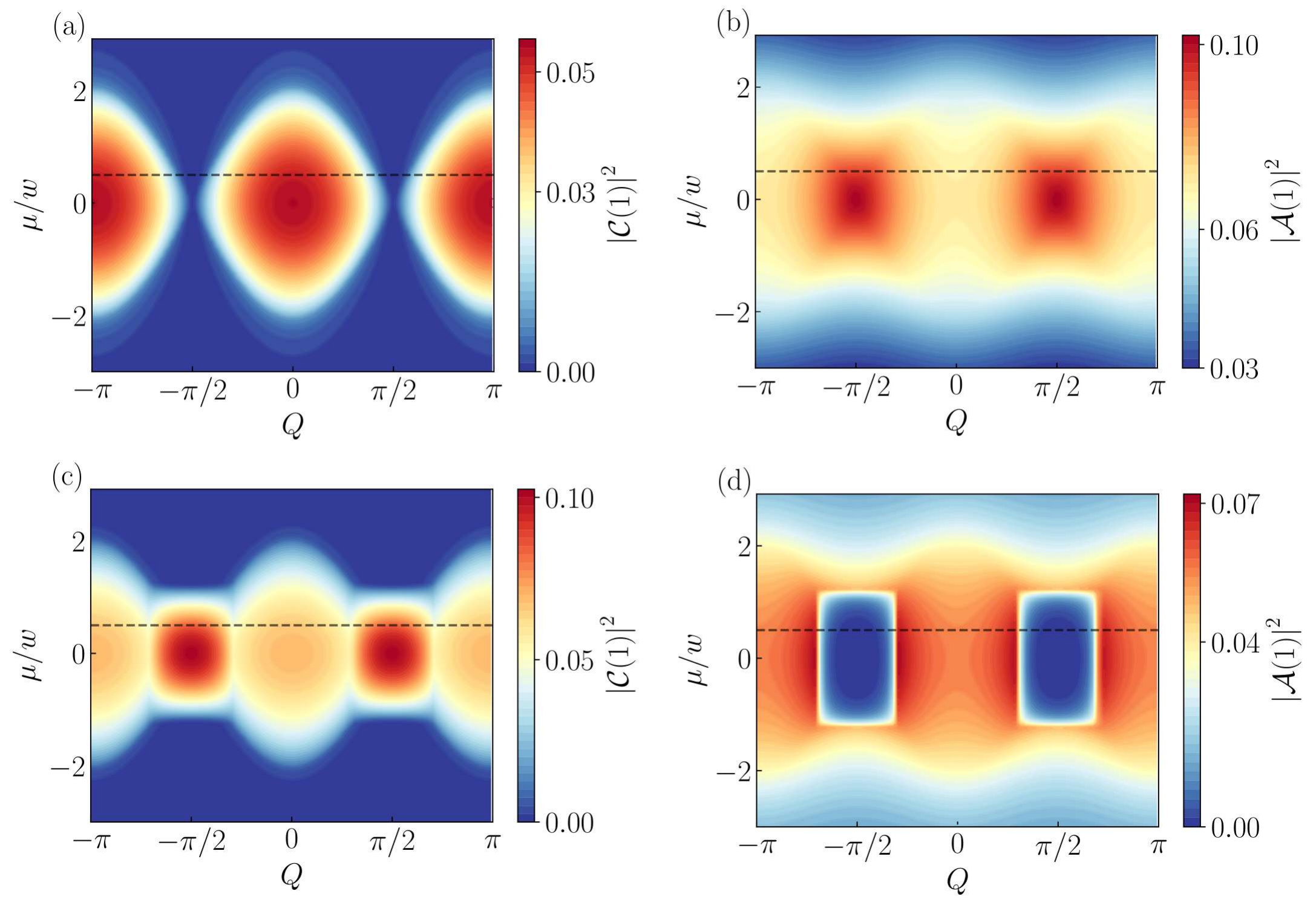}
 		\caption{ \label{Fig3}   Normal and anomalous correlation functions, Eq.(\ref{C-def}) and Eq.(\ref{A-def}), for $l=1$. Contour plots of $|\mathcal{C}(1)|^2$ and $|\mathcal{A}(1)|^2$ as a function of the  Cooper pair wavevector $Q$ and chemical potential $\mu$. Panels (a) and (b) are obtained for  $\Delta_{0} = 1.3 w$. The behavior of $|\mathcal{C}(1)|^2$ in panel (a) reflects the phase diagram of Fig.\ref{Fig1}(a) where two topologically distinct gapped phases exist. Panels (c) and (d) are obtained for $\Delta_{0} = 0.8w$, and both clearly show the emergence of a gapless region as a sharp rectangular area centered around the  high symmetry points $(Q,\mu)=(\pm \pi/2,0)$ [green areas of Fig.\ref{Fig1}(c)]. Horizontal dashed lines identify the cuts at $\mu=0.5 w$, shown in Figs.\ref{Fig4} and  \ref{Fig5}.}
 	\end{figure*}

Let us start by analyzing the case $l=1$. The two quantities $\left|\mathcal{C}(1)\right|^{2}$ and $\left|\mathcal{A}(1)\right|^{2}$ are shown in Fig.\ref{Fig3} as contour plots over the parameter space $(Q, \mu)$, at a fixed value of $\Delta_0$. Specifically, the two upper panels refer to the regime $\Delta_0>w$, where the system exhibits only the two topologically different gapped phases [see Fig.\ref{Fig1}(a)], while the two lower panels of Fig.\ref{Fig3} refer to the regime $\Delta_0<w$, where the gapless phase  appears in the  $(Q, \mu)$ parameter space [green areas of Fig.\ref{Fig1}(c)]. 

\begin{figure}[h]
\centering
\includegraphics[scale = 0.55]{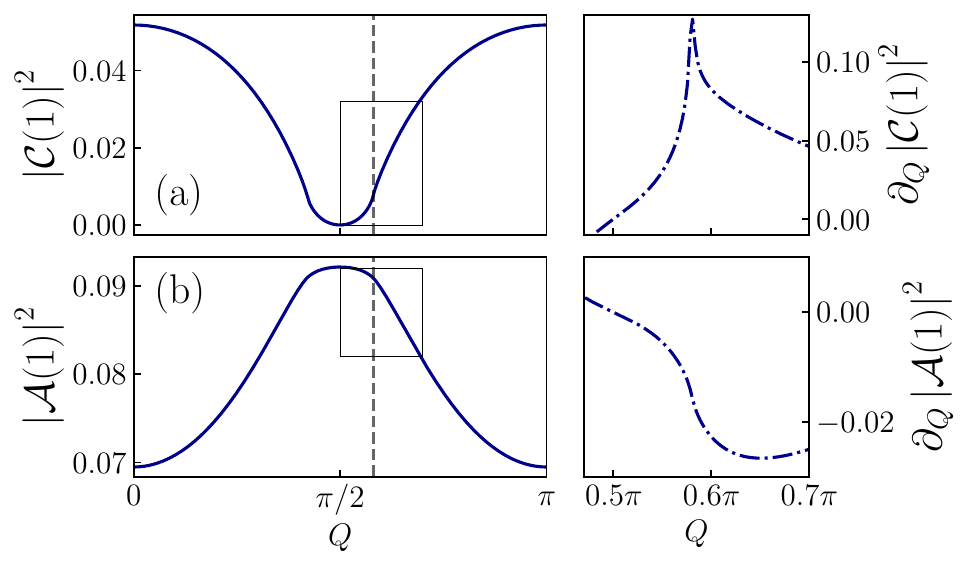}
\caption{\label{Fig4} The squared modulus   of the normal and anomalous correlations $\mathcal{C}(1)$ and $\mathcal{A}(1)$ are plotted as a function of $Q$, for $\Delta_0=1.3 w$. Panels (a) and (b)  represent cuts of Figs.\ref{Fig3}(a) and \ref{Fig3}(b), respectively, at $\mu = 0.5w$.   Right panels   display the $Q$-derivative of the corresponding curve on the left in the range highlighted by rectangles. Vertical dashed lines mark the transition point between the trivial  and the topological gapped phases. }
\end{figure}
\begin{figure}[h]
\includegraphics[scale = 0.55]{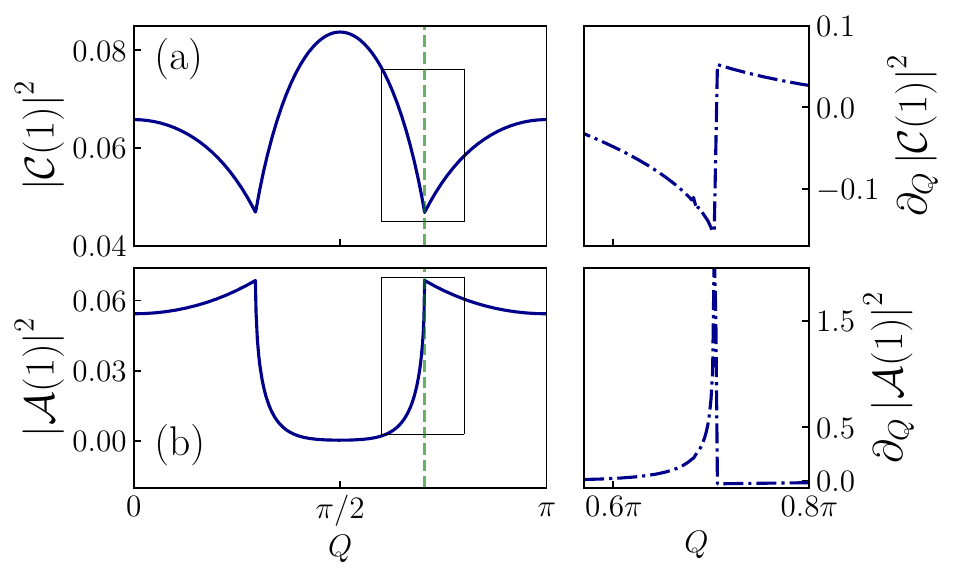}
\caption{\label{Fig5}  The squared modulus   of the normal and anomalous correlations $\mathcal{C}(1)$ and $\mathcal{A}(1)$ are plotted as a function of $Q$, for $\Delta_0=0.8 w$. (a) and (b) panels represent cuts of Fig.\ref{Fig3}(c) and Fig.\ref{Fig3}(d), respectively, at $\mu = 0.5w$.   Right panels   display the $Q$-derivative of the corresponding curve on the left in the range highlighted by rectangles. Vertical dashed lines mark the Lifshitz transition line separating  the gapped from the gapless phases.}
\end{figure}

From Fig.\ref{Fig3}(a), one can see that the contour plot of the squared normal correlation function $|\mathcal{C}(1)|^2$ qualitatively reproduces the phase diagram of Fig.\ref{Fig1}(a), reaching its maximal   values at the center of the gapped topological  phase and being suppressed in the trivial gapped phase. Instead, it would be harder to infer such phase diagram from  the inspection of  the anomalous correlation function $|\mathcal{A}(1)|^2$, shown in Fig.\ref{Fig3}(b). Yet, from such a plot  we deduce that the maximal probability of finding nearest neighbors Cooper pairs is at the  high symmetry points $(Q,\mu)=(\pm \pi/2,0)$, in agreement with the result discussed at the end of Sec.\ref{sec3-A} that  all normal correlation functions    vanish  (at any distance) for such parameter values. 
Focussing now on  the $\Delta_{0} < w$ regime, we observe that {\it both} the normal and the anomalous correlation functions depicted in  Figs.\ref{Fig3}(c) and  \ref{Fig3}(d)  clearly exhibit a rectangular shape, centered around the high symmetry points
$(Q,\mu)=(\pm \pi/2,0)$, 
identifying the gapless region of Fig.\ref{Fig1}(b). It is straightforward to 
 show  that the values of correlations at the  high symmetry points are
\begin{eqnarray}
\left. \mathcal{C}(1)\right|_{\mu=0; Q=\pm \frac{\pi}{2}} &=&i \left. \mathcal{C}_u(1)\right|_{\mu=0; Q=\pm \frac{\pi}{2}}= \nonumber \\
&=&\left\{ 
\begin{array}{lcl}
0 & \mbox{for} & \Delta_0>w \\
\pm \frac{i}{\pi} & \mbox{for} & \Delta_0<w \\
\end{array}\right. \label{C(1)-special}
\end{eqnarray}
and
\begin{equation}
\left. \mathcal{A}(1)\right|_{\mu=0; Q=\pm \frac{\pi}{2}}=\left\{ 
\begin{array}{lcl}
0 &\mbox{for}  & \Delta_0>w \\
- \frac{1}{\pi} & \mbox{for} & \Delta_0<w \\
\end{array}\right. \label{A(1)-special}
\end{equation}

The question we now want to address is whether  the $Q$-dependence of the  bulk  correlation functions enables one to distinguish the two types of transitions, namely the band topology transition (from trivial gapped to topological gapped) and   the Fermi surface topology Lifshitz transition (gapped to gapless). To this purpose,  for each panel in Fig.\ref{Fig3}, we have analysed a horizontal cut at $\mu = 0.5w$.
The cuts of the upper panels (a) and (b) of Fig.\ref{Fig3} are  shown in Fig.\ref{Fig4} [panels   (a) and (b), respectively] and display the behavior of $|\mathcal{C}(1)|^2$ of $|\mathcal{A}(1)|^2$ across the band topology transition, occurring  at the two values $Q=\arccos(\pm \mu/2w)$. One of them  is highlighted as a vertical dashed line. As one can see, the behavior appears to be smooth. 
To have a closer inspection around the transition point, we have focused on the $Q$-range enclosed by boxes, and we have depicted the   derivatives $\partial_Q |\mathcal{C}(1)|^2$ and $\partial_Q|\mathcal{A}(1)|^2$ in the right panels of Fig.\ref{Fig4}. As one can see, while the anomalous correlation has a finite and continuous derivative, the normal correlation  exhibits a divergent derivative at the transition point, related to the {\it direct} closing at $k=\pi$ of the gap between the two bands $E_{+}$ and $E_{-}$. Note, however, that the behavior is the {\it same} on both sides (trivial and topological) of the transition, in agreement with the universality of the  correlation length scaling discussed in Ref.\cite{sigriest_PRB_2017}. Indeed the two sides of the transition can only be distinguished by invoking edge correlation functions in a finite chain \cite{zhou_PRL_2017,chen_PRB_2017,zhou_scirep_2018}.

Let us now analyze the cuts  of  Fig.\ref{Fig3}(c) and Fig.\ref{Fig3}(d), which are shown in panels (a) and (b) of Fig.\ref{Fig5}, and refer to the Lifshitz transition.  
We now observe  clear  cusps appearing in {\it both}   $|\mathcal{C}(1)|^2$ and    $|\mathcal{A}(1)|^{2}$ at the boundaries between gapped and gapless phases, which are determined by the {\it indirect} closing of the gap between the two bands $E_{+}$ and $E_{-}$. The boundary  at the  value      $Q=\pi-\arcsin(\Delta_0/w)$  is highlighted by the vertical dashed line, and the related  {\it discontinuity} in the derivatives is shown by the focus in the right panels of Fig.\ref{Fig5}. 
The comparison between Figs.\ref{Fig4} and \ref{Fig5} shows the difference in the $Q$-dependence of the correlation functions across the two types of transitions. Because of the   presence  of cusps, the Lifshitz transition   has a much sharper evidence than the band topology transition between gapped phases.
The origin of such cusps boils down to the intrinsically different structure of the ground state on the two sides (gapped {\it vs} gapless) of the transition, which affects the correlation functions. Indeed, as observed at the beginning of this section, the integral expressions (\ref{Cp-k}) and (\ref{A-k}) have a twofold dependence on $Q$, namely through the integrand function and through  the integration domain. In the gapped side of the transition  only the former is present, since $S_p \equiv {\rm BZ}$  is $Q$-independent, whereas in the  gapless side  also the latter leads to a finite contribution, since $S_p$ is given by Eqs.(\ref{Sp-gapless}) and (\ref{k-star-pm}) and depends on $Q$. This gives rise to the discontinuity of the correlation function derivatives $\partial_Q|\mathcal{C}(1)|^2$ and $\partial_Q|\mathcal{A}(1)|^2$ shown in the right panels of Fig.\ref{Fig5}.

\begin{figure*}
    \includegraphics[scale = 0.45]{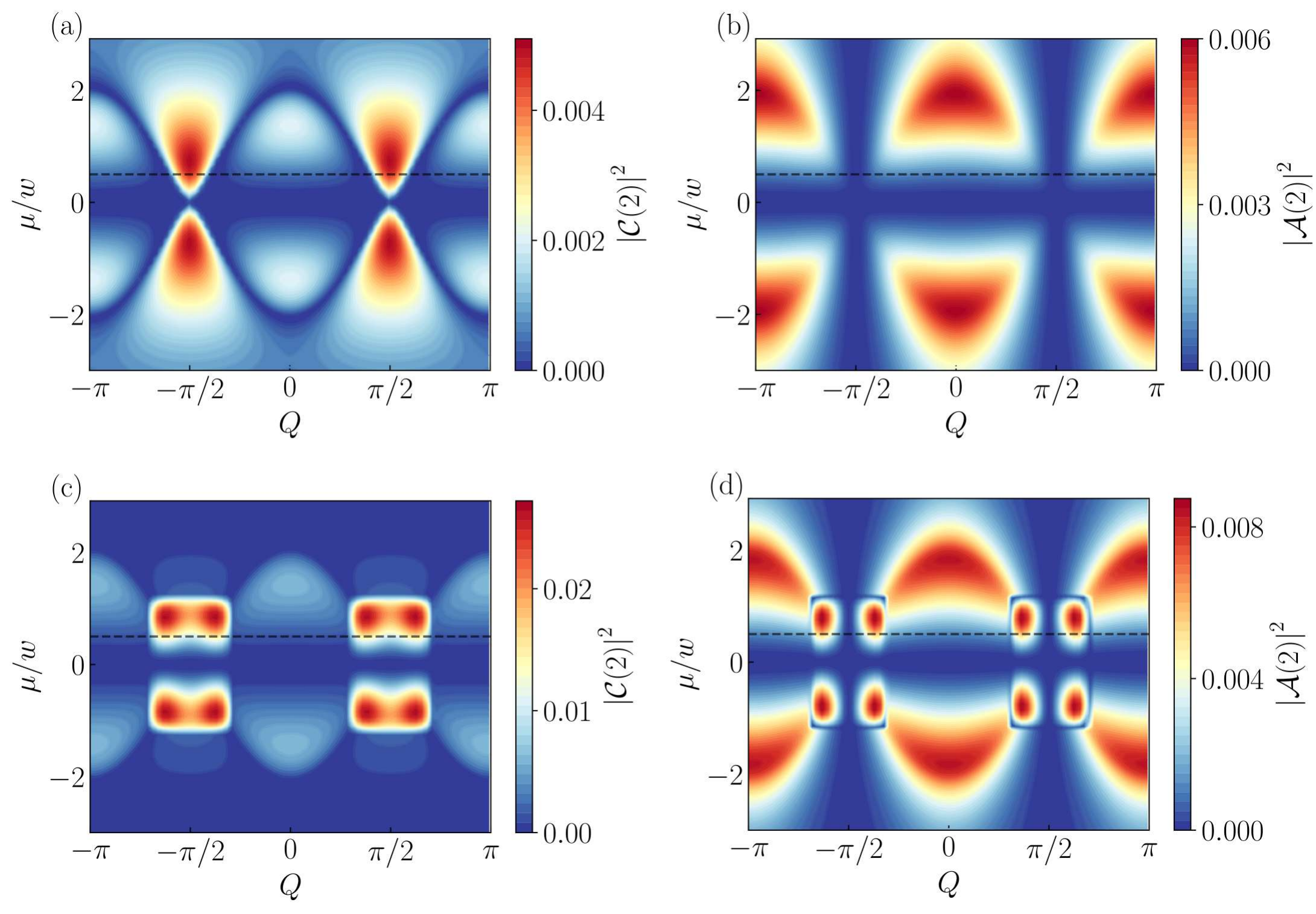}
    \caption{\label{Fig6}  Normal and anomalous correlation functions, Eq.(\ref{C-def}) and Eq.(\ref{A-def}), for $l=2$. Contour plots of $|\mathcal{C}(2)|^2$ and $|\mathcal{A}(2)|^2$ as a function of the Cooper pair wavevector $Q$ and chemical potential $\mu$. 
    Panels (a) and (b) are obtained for  $\Delta_{0} = 1.3 w$, while panels (c) and (d) for $\Delta_{0} = 0.8w$. The behavior of $|\mathcal{C}(2)|^2$ in panel (a) reflects the phase diagram of Fig.\ref{Fig1}(a), while both $|\mathcal{C}(2)|^2$  and $|\mathcal{A}(2)|^2$ in panels (c) and (d)   acquire some local maxima inside the rectangular region identifying the gapless phase.  }
\end{figure*}

\begin{figure}[h]
\includegraphics[scale = 0.55]{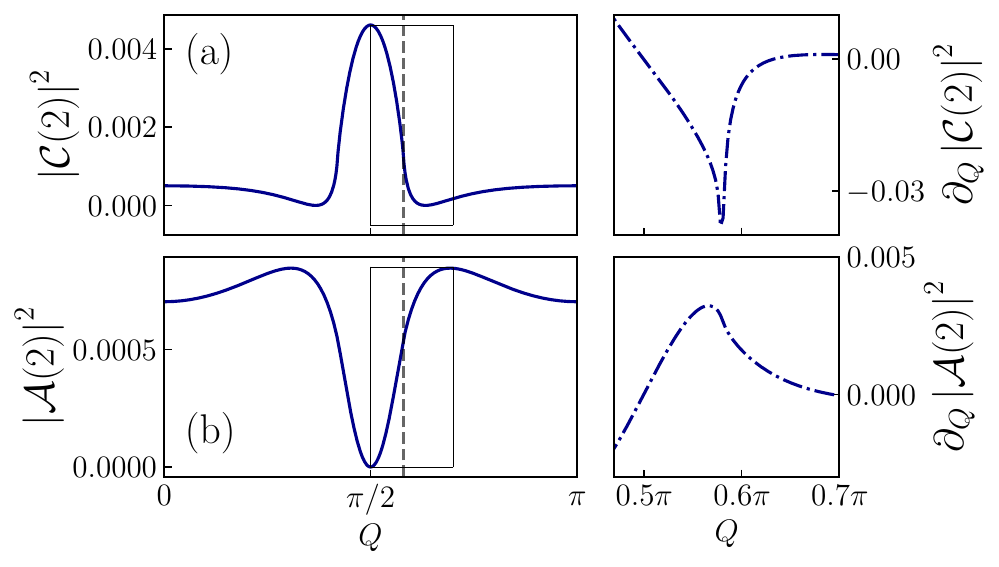}
\caption{\label{Fig7} 
The squared modulus   of the normal and anomalous correlations $\mathcal{C}(2)$ and $\mathcal{A}(2)$ are plotted as a function of $Q$, for $\Delta_0=1.3 w$. Panels (a) and (b)  represent cuts of Figs.\ref{Fig6}(a) and \ref{Fig6}(b), respectively, at $\mu = 0.5w$.   Right panels   display the $Q$-derivative of the corresponding curve on the left in the range highlighted by rectangles. Vertical dashed lines mark the transition point between the trivial  and the topological gapped phases.}
\end{figure}
\begin{figure}[h]
\includegraphics[scale = 0.55]{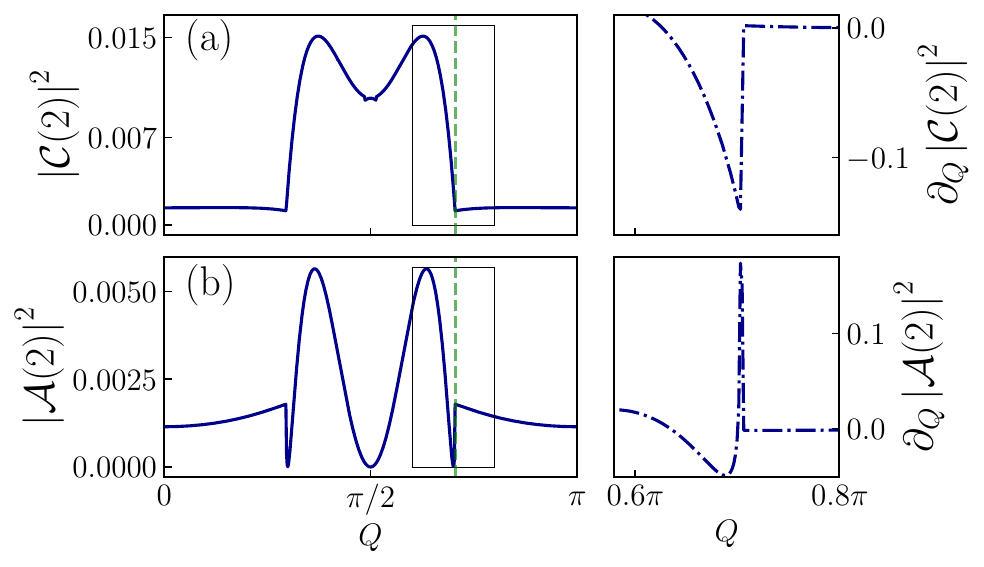}
\caption{\label{Fig8} The squared modulus   of the normal and anomalous correlations $\mathcal{C}(2)$ and $\mathcal{A}(2)$ are plotted as a function of $Q$, for $\Delta_0=0.8 w$. Panels (a) and (b)  represent cuts of Figs.\ref{Fig6}(c) and \ref{Fig6}(d), respectively, at $\mu = 0.5w$.   Right panels   display the $Q$-derivative of the corresponding curve on the left in the range highlighted by rectangles. Vertical dashed lines mark the Lifshitz transition line separating  the gapped from the gapless phases.}
\end{figure}

\subsection{The case $l=2$}
Let us now consider the probabilities $|\mathcal{C}(2)|^2$ and $|\mathcal{A}(2)|^2$  related to next nearest neighbor processes of electron hopping or pair production. These are shown in Fig.\ref{Fig6} as contour plots, where again the upper panels (a) and (b) correspond to   the regime $\Delta_{0} > w$, and the lower panels  (c) and (d) to the regime $\Delta_{0} < w$ where the gapless phase appears. 
 A first striking difference between Fig.\ref{Fig6} ($l=2$) and Fig.\ref{Fig3} ($l=1$) is the even/odd effect, which is   schematically described in Tables \ref{table1} and \ref{table2} and can now be appreciated by inspecting the horizontal line $\mu=0$ and the vertical lines $Q=\pm \pi/2$, respectively. Indeed at $\mu=0$ both $|\mathcal{C}(2)|^2$ and $|\mathcal{A}(2)|^2$ vanish, in striking contrast with $|\mathcal{C}(1)|^2$ and $|\mathcal{A}(1)|^2$. For $Q=\pm \pi/2$ one observes that $|\mathcal{A}(2)|^2$ vanishes, while $|\mathcal{C}(2)|^2$ does not, in agreement with Table \ref{table2}. 
Another striking difference is related to the gapless region. While from the $l=1$ correlations shown in Fig.\ref{Fig3}  the gapless region appears as a uniform rectangular shape, the  $l=2$ correlations in Fig.\ref{Fig6}  reveal  an inner structure with local maxima.
 
 Then, similarly to what was done for $l=1$, we have investigated the difference between two types of transitions by analyzing the $Q$ dependence of $|\mathcal{A}(2)|^2$ and $|\mathcal{C}(2)|^2$. The cuts at $\mu = 0.5w$ of the upper panels (a) and (b) of Fig.\ref{Fig6} are shown in Fig.\ref{Fig7} and are related to the transition between the gapped phases, while the cuts of the lower panels (c) and (d) of Fig. \ref{Fig6} are shown in Fig.\ref{Fig8} and highlight the behavior across the Lifshitz transition.  The  corresponding derivatives around the transition boundaries are shown in the right panels of Figs. \ref{Fig7} and \ref{Fig8}.
 As one can see from  Fig.\ref{Fig7}, $|\mathcal{A}(2)|^2$ and $|\mathcal{C}(2)|^2$ vary smoothly across the band topology transition, while cusps clearly appear in Fig.\ref{Fig8}.  This means that, despite the above mentioned differences between the $l=1$ and the $l=2$ case, the $Q$ dependence of both cases indicates that the gapped to gapless Lifshitz transition is far more detectable than the band topology transition.

\section{Long distance behavior of correlation functions}	
\label{sec5}
We now turn to evaluate the asymptotic behavior of the normal and anomalous correlations $\mathcal{C}(l)$ and $\mathcal{A}(l)$   at long distance $l \gg 1$. The behavior significantly changes from the gapped to the gapless phase. Moreover, even within the gapped phase, different asymptotic behaviors arise  in the regime $\Delta_0<w$.

\subsection{Asymptotic behavior in the gapped phases}
\label{sec5-1}
In the gapped phase (\ref{gapped-regime}), the asymptotic behavior of the correlation functions  can be   obtained by re-expressing Eqs.(\ref{Cp-k}) and (\ref{A-k}) in a different but equivalent form, based on complex analysis. While the technicalities of this procedure are given in the Appendix~\ref{AppC}, here we briefly illustrate its main steps, which will be useful to elucidate the various asymptotic behaviors that  emerge, depending on the parameter ranges.

As observed above, in the gapped phase the unpaired fermion sector $S_u$ is an empty set [see Eq.(\ref{sectors-gapped})], and the
 unpaired fermion contribution  in Eq.(\ref{Cu-k}) therefore vanishes,
 \begin{equation}
\mathcal{C}_u=0 \quad.
 \end{equation}
As the ground state only involves Cooper pairs,   the pair sector $S_p$ coincides with the entire Brillouin zone [see Eq.(\ref{sectors-gapped})]. Thus, by interpreting  $k \in [-\pi;\pi]$ as the phase of a complex number $z=e^{i k}$ that spans over the unit circle, it is possible to recast the correlation functions in the form of integrals   in the complex plane~\cite{ercolessi_NJP_2016,dellanna_PRB_2022}
	\begin{eqnarray}
		\mathcal{C}_{p}(l) &=& - \frac{1}{4\pi}\text{Im}\left\{\oint_{\left|z\right| = 1} dz \frac{z^{l-1} g\left(z\right)}{\sqrt{g^{2}\left(z\right)  - f^{2}\left(z\right)   }}   \right\} \label{Cp-cmplx}\\
		\mathcal{A}(l) &=&  \frac{1}{4\pi}\text{Im}\left\{ \oint_{\left|z\right| = 1} dz \frac{z^{l-1} f\left(z\right)  }{ \sqrt{ g^{2}\left(z\right) - f^{2}\left(z\right)   }  } \right\}, \label{A-cmplx}
	\end{eqnarray}
where the functions $f(z) = \Delta_{0}(z - z^{-1}) $ and $g(z)= w\left(z - z^{-1}\right)\cos{Q} - \mu $    of the complex variable  take the values $f(e^{i k})=\Delta(k)$ and $f(e^{i k})=\xi(k;Q)$ over the unit circle $|z|=1$, respectively. The  denominator of Eqs.(\ref{Cp-cmplx})-(\ref{A-cmplx})  exhibits four branch points,
whose location depends on the specific parameter values $\Delta_0$, $Q$ and $\mu$.
It is possible to show that two of such branch points, which we shall denote as $z^{*}_\pm$, lie  inside the unit circle ($|z^*_\pm|<1$), while the other two lie outside it and are given by $1/z^{*}_\pm$.  Then,   Cauchy  theorem applied to Eqs.(\ref{Cp-cmplx})-(\ref{A-cmplx}) enables one to   rewrite Eqs.(\ref{Cp-cmplx})-(\ref{A-cmplx}) as contour integrals over the branch cuts connecting the inner branch points $z^{*}_\pm$. Thus, it is the location of   $z^{*}_\pm$ that determines the different asymptotic behavior  of the correlation functions for $l \gg 1$. Details are given in the Appendix~\ref{AppC}

With keeping in mind that  here   the parameter conditions (\ref{gapped-regime}) for the gapped phase are assumed to hold, one can  identify three possible configurations for the location of  $z^*_\pm$ in the complex plane, which determine three different types of asymptotic behaviors and are highlighted with different symbols in Fig.\ref{Fig1}.\\

\noindent  {\it   $(a)$ \, $ z^*_\pm$ are real and have opposite signs} \\
When this branch point configuration occurs, the  correlation  functions decay as
 \begin{align}
		\mathcal{C}_{p}(l) \sim& - \frac{1}{2\sqrt{l}}\left(\alpha^{+}_{1}e^{-\kappa_{+}l}    - \left(-1\right)^{l} \alpha^{-}_{1}e^{-\kappa_{-}l}  \right), \label{Cp-asym-(a)-1}\\
		\mathcal{A}(l) \sim& - \frac{1}{2\sqrt{l}}\left(\beta^{+}_{1}e^{-\kappa_{+} l} - \left(-1\right)^{l}\beta^{-}_{1}e^{-\kappa_{-}l}   \right), \label{A-asym-(a)-1}
	\end{align}
where 
\begin{equation}
\kappa_{\pm} =  \ln \frac{1}{\left|z^{*}_{\pm}\right|}  \label{kappa_pm-(a)-1}
\end{equation} 
represent the inverse decay lengths, while the constants $\alpha_1^\pm$ and $\beta_1^\pm$ depend on $|z^*_\pm|$ and are explicitly given in Appendix \ref{AppC} as a function of the parameter values. 
Such a   configuration of   real branch points with opposite signs  occurs  when  the  conditions
\begin{equation}
\left\{ \begin{array}{l}
\Delta_0^2+\frac{\mu^2}{4}-w^2\cos^2{Q}>0 \\
\Delta_0>w|\cos{Q}|
\end{array}
\right. \label{cond-(a)-1}
\end{equation}
are both fulfilled. 
For a given $\Delta_0$ value, the portions of the $Q$-$\mu$ phase diagram where the gapped phase fulfills the additional conditions (\ref{cond-(a)-1}) and the correlation asymptotic   behavior is given by Eqs.(\ref{Cp-asym-(a)-1})-(\ref{A-asym-(a)-1})  are highlighted as circles   ``$\circ$" in  Fig.\ref{Fig1}. 
Note that, in the regime $\Delta_{0}>w$, where  for any value  of $Q$ and $\mu$ the Kitaev chain only exhibits gapped  phases (either   trivial or topological)  [Fig.\ref{Fig1}(a)], the conditions (\ref{cond-(a)-1}) are   fulfilled, whereas for $\Delta_{0}<w$ it only holds in subregions of the phase diagram  [Figs.\ref{Fig1}(b) and \ref{Fig1}(c)].

Various features are noteworthy in the expressions Eqs.(\ref{Cp-asym-(a)-1}) and (\ref{A-asym-(a)-1}). First, they  exhibit a combination of two exponential decays that are further enhanced by the additional  $1/\sqrt{l}$ factor.
Second, the  inverse decay lengthscales are determined by  the moduli $|z^*_\pm|$ of the inner branch points Eq.(\ref{kappa_pm-(a)-1}). Note that, because the inner branch points   $z^*_\pm$   have opposite signs, the magnitude of $|z^*_+|$ and $|z^*_-|$ might be comparable, and {\it both} the exponential terms have to be retained in general.
Finally,  the relative sign $(-1)^l$ between the two terms actually changes when the site distance $l$ alternates from even to odd values. This implies that the two exponential either sum up or mutually suppress, depending on the parity of~$l$.

This effect is particularly striking  for the special  cases $\mu=0$ and $Q=\pm \pi/2$, where  one can now find an analytical asymptotic expression of  the   even/odd effect proven in Sec.\ref{sec3}. Indeed both these special cases correspond  to the configuration where  two inner real branch points stemming from Eqs.(\ref{roots}) are symmetrically placed with respect to the origin, $z^*_\pm=\pm |z^*_\pm|$. In particular, for $\mu=0$ the constants become equal, $\alpha^+_1=\alpha^-_1$ and $\beta^+_1=\beta^-_1$, and the asymptotic expansions   Eqs.(\ref{Cp-asym-(a)-1})-(\ref{A-asym-(a)-1}) reduce to 
\begin{eqnarray}
\lefteqn{\left. \mathcal{C}_{p}(l>0) \right|_{\mu=0}\,\,  \sim   } & & \label{Cp-asym-(a)-1-mu=0}\\
& \sim &  \displaystyle \left\{ \begin{array}{lcl}    \displaystyle - \mbox{sgn}(\cos{Q})  \sqrt{\frac{\Delta_0 w |\cos{Q}|}{\Delta_0^2- w^2 \cos^2{Q}}}\,\, \frac{e^{-\kappa l}}{\sqrt{2\pi l}} &  & l \mbox{ odd} \\  0  &  & l \mbox{ even}  \end{array} \right. \nonumber  
\end{eqnarray}
and
\begin{eqnarray}
\left. \mathcal{A}(l) \right|_{\mu=0}\,\,  \sim    \displaystyle \left\{ \begin{array}{lcl}    \displaystyle  - \sqrt{\frac{\Delta_0 w |\cos{Q}|}{\Delta_0^2- w^2 \cos^2{Q}}}\,\, \frac{e^{-\kappa l}}{\sqrt{2 \pi l}} &  & l \mbox{ odd} \\ 0  &  & l \mbox{ even}  \end{array} \right.  \label{A-asym-(a)-1-mu=0}
\end{eqnarray}
where
\begin{equation}
\kappa=\kappa_\pm=\frac{1}{2} \ln\left( \frac{\Delta_0+w|\cos{Q}|}{\Delta_0-w|\cos{Q}|}\right)
\end{equation}
is the inverse decay length. 
The case shown in Fig.\ref{Fig2}(a) is an example   where the conditions (\ref{cond-(a)-1}) are fulfilled, and the normal correlation decays as described by Eq.(\ref{Cp-asym-(a)-1-mu=0}). By contrast, for the case $Q=\pm \pi/2$, one finds $\alpha^+_1=-\alpha^-_1$ and $\beta^+_1=\beta^-_1$, and the asymptotic expansions   Eqs.(\ref{Cp-asym-(a)-1})-(\ref{A-asym-(a)-1}) acquire the form 
{\small
\begin{eqnarray}
\lefteqn{\left. \mathcal{C}_{p}(l>0) \right|_{Q=\pm \frac{\pi}{2}}\,\,  \sim   } & & \label{Cp-asym-(a)-1-cos(Q)=0}\\
& \sim  &  \displaystyle \left\{ 
\begin{array}{lcl}  0  &  & l \mbox{ odd} \\ & & \\  \displaystyle  \frac{\mu \left( \sqrt{\Delta_0^2+\frac{\mu^2}{4}}-\frac{|\mu|}{2}\right)}{\sqrt{ \left(\Delta^4_0  -\left( \sqrt{\Delta_0^2+\frac{\mu^2}{4}}-\frac{|\mu|}{2}\right)^4\right)}} \, \frac{e^{-\kappa l}}{\sqrt{2 \pi l}}  &  & l \mbox{ even}   
\end{array} \right.   \nonumber
\end{eqnarray}
}
whereas
\begin{eqnarray}
\lefteqn{\left. \mathcal{A}(l) \right|_{Q=\pm\frac{\pi}{2}}\,\,  \sim   } & & \\
& \sim  &  \displaystyle \left\{ \begin{array}{lcl}  0  &  & l \mbox{ odd}  \\ & & \\ \displaystyle   \sqrt{\frac{\Delta_0^2-\left( \sqrt{\Delta_0^2+\frac{\mu^2}{4}}-\frac{|\mu|}{2}\right)^2}{\Delta_0^2+\left( \sqrt{\Delta_0^2+\frac{\mu^2}{4}}-\frac{|\mu|}{2}\right)^2}}
 \, \frac{e^{-\kappa l}}{\sqrt{2 \pi l}}  &  & l \mbox{ even}  \end{array} \right. \nonumber  \label{A-asym-(a)-1-cos(Q)=0}
\end{eqnarray}
with
\begin{equation}
\kappa=\kappa_\pm= \ln \frac{  \sqrt{\Delta_0^2+\frac{\mu^2}{4}}-\frac{|\mu|}{2} }{\Delta_0} \quad.
\end{equation}
\\
\\
\noindent  {\it   $(b) \, z^*_\pm$ are real and have the same sign.} \\
When the two inner branch points $z^*_{-}< z^*_{+}$ have the same sign 
\begin{equation}
\sigma^*=\mbox{sgn}(z^*_{-})=\mbox{sgn}(z^*_{+}) =\mbox{sgn}\left(\mu \cos{Q}\right)\quad, \label{sigma-star-def}
\end{equation}
the   asymptotic behavior of the
 correlation  functions is 
\begin{eqnarray}
		\mathcal{C}_{p}(l) &\sim & - \frac{{\sigma^*}^{l-1}}{2}\, \left\{ \alpha_M\, \frac{e^{- \kappa_{M} l }}{\sqrt{l}}  \,  - \alpha_{m}\, \frac{e^{- \kappa_{m} l }}{\sqrt{l}} \right\}   \label{Cp-asym-(a)-2}\\
		\mathcal{A}(l) &\sim&  - \frac{{\sigma^*}^{l-1}}{2}\, \left\{ \beta_M\, \frac{e^{- \kappa_{M} l }}{\sqrt{l}}  \,  - \beta_{m}\, \frac{e^{- \kappa_{m} l }}{\sqrt{l}} \right\}  \quad,\label{A-asym-(a)-2}
\end{eqnarray}
where the   values of the constants $\alpha_{M/m}$ and $\beta_{M/m}$ are given in  Appendix  \ref{AppC}, while  the inverse decay lengths are  
\begin{eqnarray}
\kappa_M &=&  \ln \frac{1}{\mbox{max}(|z^*_+|,|z^*_-|)}\\
\kappa_m &=&  \ln \frac{1}{\mbox{min}(|z^*_+|,|z^*_-|)} \quad.
\end{eqnarray}
Differently from   the case (a) [see Eqs.(\ref{Cp-asym-(a)-1})-(\ref{A-asym-(a)-1})],  the two exponential terms in Eqs.(\ref{Cp-asym-(a)-2})-(\ref{A-asym-(a)-2}) do not compete with a $l$-dependent relative factor.  
Because the two branch points have the same sign, two limiting situations can occur. If  $\kappa_M\ll \kappa_m$, the leading asymptotic term is dictated by $\kappa_M$, whereas if $\kappa_M \simeq  \kappa_m$ the two terms give comparable contributions.
The   case $(b)$  occurs if and only if  the  relations
\begin{equation}
\left\{ \begin{array}{l}
\Delta_0^2+\frac{\mu^2}{4}-w^2\cos^2{Q}>0 \\
\Delta_0<w|\cos{Q}|
\end{array}
\right. \label{cond-(a)-2}
\end{equation}
are both fulfilled, in addition to   the gapped phase conditions (\ref{gapped-regime}). 
In the $Q$-$\mu$ phase diagram shown in Fig.\ref{Fig1}, the sub-regions   where Eq.(\ref{cond-(a)-2}) holds in the gapped phases are highlighted by dots  ``$\cdot $".\\

\noindent {\it $(c)$ $z^*_\pm=x^*\pm i y^*$ are a complex conjugate pair.}  \\
A direct inspection of the branch points (\ref{roots}) shows that they can exhibit an imaginary part if and only if 
\begin{equation}
\Delta_0^2+\frac{\mu^2}{4}-w^2\cos^2{Q}<0   \quad.
\label{cond-(b)}
\end{equation}
Such condition describes the area enclosed by an ellipse in the variables $\cos{Q}$ and $\mu$, and is identified in Figs.\ref{Fig1}(b) and (c) by the cyan regions with horizontal lines ``$-$",  centered around the origin $(Q,\mu)=(0,0)$ and around $(Q,\mu)=(0,\pm \pi)$. Note that the pole configuration (c) can only occur for $\Delta_0<w$ and within the gapped topological phase.  

In this case, the asymptotic behavior of the correlation functions  exhibits an exponential decay  that is combined with an oscillatory behavior~\cite{barouch1970}. The suppression   is characterized by {\it one} decay length, related to the real part $x^*$ of the the branch points,  
\begin{eqnarray}
x^*&=& \displaystyle \frac{\frac{\mu}{2}\, \mbox{sgn}(\cos{{Q}})}{w|\cos{Q}|+\Delta_0} \quad, \label{x*-(b)}  
\end{eqnarray}
while the period of the oscillatory behavior  is determined by their imaginary part 
\begin{eqnarray}
y^*  =  \frac{\sqrt{w^2\cos^2{Q}-\Delta_0^2-\frac{\mu^2}{4}}}{w|\cos{Q}|+\Delta_0} \quad. \label{y*-(b)}  
\end{eqnarray}
In this case the inner branch points $z^*_\pm=x^*\pm i y^*$ are straightforwardly given by $Z_{1,2}$ in Eq.(\ref{roots}).

While an asymptotic expression of the correlations cannot be obtained for arbitrary values, an analytical result can be obtained in the most interesting regime $|x^*|\ll y^*$, where the imaginary part dominates over the real part. This corresponds to the relevant case where the period of the oscillations is short compared to the decay length, and the oscillations become appreciable. In this case one obtains  
\begin{equation}
\mathcal{C}_p(l)    \sim     \displaystyle  \frac{e^{- \kappa_+ \,l }}{\sqrt{l}} \left( \alpha^s_3 \sin[q l]+\alpha^c_3 \cos[q l]\right)\label{Cp-asym-(b)}
\end{equation}
while   
\begin{equation}
\mathcal{A}(l)    \sim     \displaystyle \frac{e^{- \kappa_+ \,l }}{\sqrt{l}} \left( \beta^s_3 \sin[q l]+\beta^c_3 \cos[q l]\right) \quad. \label{A-asym-(b)}
\end{equation}
Here,
\begin{eqnarray}
\kappa_{+} &=& \ln\frac{1}{|z^*_{+}|}   \label{kappa+-(b)-def}\\
q &=&  \arccos  \left(  \frac{\frac{\mu}{2}\, \mbox{sgn}(\cos{Q})}{\sqrt{w^2 \cos^2{Q}-\Delta_0^2}}\right) \label{q-def}
\end{eqnarray}
represent the decay of the exponential suppression and the period of the oscillatory terms, respectively. The expression of the constants $\alpha^{c/s}_3$ and $\beta^{c/s}_3$ are given in Appendix \ref{AppC}. It is straightforward to check that, for either $\mu=0$ or $Q=\pm \pi/2$, the even/odd effect is recovered. \\

We conclude this subsection by two remarks  about the regime  $\Delta_0<w$, which is the most realistic one in implementations of the Kitaev chain. Firstly, the conditions  (\ref{cond-(a)-2}) and (\ref{cond-(b)}) related to   gapped phases, along with the conditions (\ref{gapless-regime}) for the gapless phase, identify two wavevectors
\begin{eqnarray}
Q^* &=& \arcsin(\Delta_0/w) \label{Q-star-def} \\
{Q}_0 &=& \arccos(\Delta_0/w) \quad. \label{Q0-def} 
\end{eqnarray}
As can be seen from   panels (b) and (c) of Fig.\ref{Fig1}, the former determines the boundaries $Q^*<|Q|<\pi-Q^*$ of the gapless region, while $Q_0$ specifies the boundary between the two different asymptotic behaviors ``$\circ$" and  ``$\cdot$" of the correlation functions in the   gapped region. In turn,   Eqs.(\ref{Q-star-def})-(\ref{Q0-def}) imply that there exists a special value  $\Delta^*_0=w/\sqrt{2}\simeq 0.71\,w $ of  $\Delta_0$ that determines the relative order between $Q^*$ and $Q_0$. Indeed in Fig.\ref{Fig1}(b), which refers to the range  $\Delta^*_0<\Delta_0<w$, one has $Q_0<Q^*$. In this case $Q_0$ also represents the $Q$-boundary of the striped elliptic region. However, in Fig.\ref{Fig1}(c), which refers to the range  $\Delta_0<\Delta^*_0$, the gapped elliptic region is cut by the onset of the gapless phase, and its $Q$-boundaries are determined by $Q^*$ instead. \\
The second remark is that the striped elliptic region can be given a twofold interpretation. On the one side, it is the sub-portion of the topological gapped phase where the exponential decay of the long distance correlation functions is combined with an oscillatory behavior. On the other hand, it can be seen as the set of parameter values that are connected to the gapless region through $Q$. Indeed for {\it any} values of $\Delta_0<w$ and of chemical potential $\mu$, if the ground state is within the elliptic region for  $Q=0$ (no current flows), by increasing  $Q$ with keeping $\mu$ and $\Delta_0$ constant, the system will eventually enter the gapless phase. This is not the case for other parameter points of the topological gapped phase.

%%%%%%%%%%%%%%%%%%%%%%%%%%%%%%%%%%%
%%%%%%%%%%%%%%%%%%%%%%%%%%%%%%%%%%%
%%%%%%%%%%%%%%%%%%%%%%%%%%%%%%%%%%%
\subsection{Asymptotic behavior in the gapless phase}
 \label{sec5-2}
	We now turn to determine the behavior of the normal and anomalous correlations $\mathcal{C}(l)$ and $\mathcal{A}(l)$ for large distance, $l \gg 1$, in the gapless phase.
	As mentioned above, in the gapless phase,  emerging when the parameters fulfill Eq.(\ref{gapless-regime}), the  current carrying ground state (\ref{G-gen}) is characterized   by both Cooper pairs and unpaired fermions (electron and holes).

 As observed above, the unpaired fermion contribution~$\mathcal{C}_u$ to  the normal correlation $\mathcal{C}(l)$ can be evaluated exactly at arbitrary values of parameters, see Eq.(\ref{Cu-exact}), and exhibits for $l \gg 1$  a power decay as $ \sim 1/l$, with   oscillations characterized by two spatial frequencies dictated by $k^*_\pm$ [see Eq.(\ref{k-star-pm})]. In contrast,  the   integrals (\ref{Cp-k}) and (\ref{A-k}) yielding the Cooper pair contribution $\mathcal{C}_p$ and the anomalous correlation function $\mathcal{A}$  cannot be computed analytically and  an asymptotic expansion must be determined. We note, however, that   the   approach adopted   to derive the asymptotic expansion in  the gapped case, where $k$ is treated as the angle of a complex number describing a circle in the complex plane, is not straightforwardly applicable to the gapless case. This is because the $k$-domain $S_p$ appearing in the integrals  Eqs.(\ref{Cp-k}) and (\ref{A-k}) does not coincide with the entire {\rm BZ}, and the   angle $k$~spans only disconnected arcs, rather than a closed circle.
 Nevertheless,  an asymptotic expansion of such integrals can be computed with the method of the stationary phase. Details of these calculations are given in  the Appendix~\ref{AppD}. To leading order,  the Cooper pair contribution of the normal correlator is found to acquire  the form
	\begin{eqnarray}
		\mathcal{C}^{}_p(l) & \sim&  - \frac{1}{2\pi} \, \frac{1}{l} \left\{ {F}_{-}(|k^{*}_{-}|) \sin{(|k^{*}_{-}|l)}\right. 		\label{Cp-asym-gapless}\\
		& & \hspace{1cm} -\left.(-1)^l {F}_{+}(|k^{*}_{+}|) \sin{(|k^{*}_{+}|l)}, \right\},
	 \nonumber 
	\end{eqnarray}
	which is similar to the 
 exact contribution Eq.(\ref{Cu-exact}) from the unpaired fermions, 
 while the anomalous correlation behaves as 
	\begin{eqnarray}
			\mathcal{A}^{}(l) &\sim & \frac{1}{2\pi}\frac{1}{l}\left\{ {G}_{-}\left(\left|k^{*}_{-}\right|\right) \cos\left( k^{*}_{-} l\right)        \right.	\label{A-asym-gapless}\\
			 & & \hspace{1cm} -\left. \left(-1\right)^{l}{G}_{+}\left(\left|k^{*}_{+}\right|\right) \cos\left( k^{*}_{+}  l\right)  \right\}
	 \nonumber 
	\end{eqnarray}
	where
	\begin{equation} 
	{F}_{\pm} (k^{*}_{\pm}) =  \frac{\xi_{\pm} (k^{*}_{\pm} )}{\sqrt{ \xi^{2}_{\pm}(|k^{*}_{\pm}| ) + |\Delta(k^{*}_{\pm})|^{2}}},
	\end{equation}
	and
	\begin{equation}
	{G}_{\pm} (  k^{*}_{\pm}) =   \frac{ |\Delta(k^{*}_{\pm})|   }{\sqrt{ \xi^{2}_{\pm}(k^{*}_{\pm} ) + |\Delta(k^{*}_{\pm})|^{2}} }.
	\end{equation}
    with $\xi_{\pm}( k^{*}_{\pm} ) = 2w\cos(k^{*}_{\pm})\cos{Q} \pm \mu$.\\
	 
Equations (\ref{Cp-asym-gapless}) and (\ref{A-asym-gapless}), together with (\ref{Cu-exact}), show that, in   contrast with the exponential  decay  $\sim e^{-\kappa l}/\sqrt{l}$ obtained in the gapped phase,   in the gapless phase correlation functions   always exhibit an {\it algebraic} decay $\sim 1/l$, combined with  spatial oscillations characterized by {\it two} periods given by $2\pi/|k^*_\pm|$, whose dependence on the parameters in given by Eq.(\ref{k-star-pm}). \\

Before concluding this subsection, we note that, for $\mu=0$ or for $Q=\pm \pi/2$ the two periods become equal $|k^*_{+}|=|k^*_{-}|$, and the expressions given above acquire a simpler form. Specifically, for $\mu=0$  one has $k^{*}_{Q} \doteq |k^*_+|=|k^*_{-}| =\arcsin(w|\cos{Q}|/\sqrt{w^2- \Delta_0^2})$, and Eq.(\ref{Cp-asym-gapless}) reduces to
\begin{equation}
\left. \mathcal{C}_p(l)\right|_{\mu=0}  \sim  \left\{ 
 \begin{array}{lcl}
\displaystyle    0  & & l \,\, \mbox{even} \\ & & \\ -  \frac{\sin{(|k^{*}_{Q}| l)}}{\pi\,l} \,\, \mbox{sgn}(\cos{Q}) \sqrt{\frac{w^2 \sin^2{Q}-\Delta_0^2}{w|\sin{Q}|}}  & & l \,\, \mbox{odd} \end{array}\right.
\end{equation} 
and
\begin{equation}
\left. \mathcal{A}(l)\right|_{\mu=0}  \sim  \left\{ 
 \begin{array}{lcl}
\displaystyle    0  & & l \,\, \mbox{even} \\ & & \\  \frac{\cos{(k^{*}_{Q}  l)}}{\pi\,l} \frac{\Delta_0 }{  w  |\sin{Q}|  }  & & l \,\, \mbox{odd} \end{array}\right.
\end{equation} 
In contrast, for
 $Q=\pm \pi/2$ one has $k^*_\mu=|k^*_{+}|=|k^*_{-}|=\arcsin(|\mu|/2\sqrt{w^2-\Delta_0^2})$. Then, from Eq.(\ref{Cu-exact}) one finds
 \begin{equation} 
\left. \mathcal{C}_u(l) \right|_{Q=\pm \frac{\pi}{2}} \equiv \left\{ 
\begin{array}{lcl} 
0 & & l \, \mbox{even}\\ & & \\
\mp \frac{\cos(k^*_\mu l)}{\pi\,l}    & & l \,\, \mbox{odd} 
\end{array}
\right.  \label{corr-norm-unpaired-cos(phi)=0-ris}
\end{equation}
whereas from Eqs.(\ref{Cp-asym-gapless})   and Eq.(\ref{A-asym-gapless}) one obtains  
\begin{equation}
\left. \mathcal{C}_p(l) \right|_{Q=\pm \frac{\pi}{2}} \sim  \left\{ 
 \begin{array}{lcl}
\displaystyle    \mbox{sgn}(\mu)  \,  \sqrt{  1   -\frac{\Delta^{2}_{0}}{w^2}     }  \, \, \frac{\sin{(k^*_\mu l)}}{\pi\,l} & & l \,\, \mbox{even} \\ & & \\ 0 & & l \,\, \mbox{odd} \end{array}\right.
\end{equation} 
and
\begin{equation}
\left. \mathcal{A}(l) \right|_{Q=\pm \frac{\pi}{2}} \sim  \left\{ 
 \begin{array}{lcl}
\displaystyle  0  & & l \,\, \mbox{even} \\ & & \\ \mbox{sgn}(\mu)  \,  \frac{\Delta_{0}}{w}  \, \, \frac{\cos{(k^*_\mu l)}}{\pi\,l}  & & l \,\, \mbox{odd} \end{array}\right.
\end{equation} 
respectively. This is the asymptotic behavior exhibited by  the anomalous correlation function in Fig.\ref{Fig2}(b).

\section{Connection with a XY spin chain with Dzyaloshinskii-Moriya interaction}
\label{sec6}
 In this section we discuss the relation between the Kitaev chain with spatial modulation of the superconducting order parameter and spin chain models. As is well known, 1D models of spinless fermions can be mapped onto   spin models through the Jordan-Wigner transformation~\cite{jw_1928}. For the conventional 1D Kitaev chain without superconducting modulation ($Q=0$), the mapping returns a  XY spin model under a transverse field. It is worth recalling that,    even though in the thermodynamic limit the first-quantized version of the fermionic and spin models share the same single-particle eigenvalues and eigenfunctions,   the  physical nature of the many-particle ground states is not equivalent. In particular, while the 1D Kitaev chain exhibits topological order and two topologically distinct phases sharing the same symmetries,  the 1D spin model exhibits conventional order \cite{greiter2014,pan2023}. Indeed, the gapped trivial and topological phases of the Kitaev chain correspond to the gapped paramagnetic (PM) and ferromagnetic (FM) phases in the XY spin chain, respectively.

    By applying to  Eq.(\ref{KTV-Ham})  the following Jordan-Wigner representation of the fermionic operators  
    \begin{equation}\label{JW}
    \begin{array}{lcl}
    c^{\dagger}_{j} &=& \displaystyle e^{iQj} \sigma^{+}_{j}\prod^{j-1}_{n = 1} \left(- \sigma^{z}_{n}\right), \\  c_{j} &=& \displaystyle e^{-iQj}\prod^{j - 1}_{n = 1} \left(-\sigma^{z}_{n}\right) \sigma^{-}_{j},
    \end{array}
    \end{equation}
    where $\sigma_j^z$ and  $\sigma^{\pm}_{j} = (\sigma^{x}_{j} \pm i \sigma^{y}_{j})/2$ are spin component operators  at the $j$-th site, one obtains  the following  spin model Hamiltonian  
    \begin{eqnarray}
    \mathcal{H}_s &=&  \frac{1}{2}\sum_{j}\left[-\mu \sigma^{z}_{j}  + {J}_{x} (Q)\, \sigma^{x}_{j}\, \sigma^{x}_{j+1}   +  {J}_{y} (Q) \, \sigma^{y}_{j} \sigma^{y}_{j+1}  +\right. \nonumber \\
    & & \hspace{1cm} -   \left. {D} (Q ) \left(  \sigma^{x}_{j}\sigma^{y}_{j+1}  -  \sigma^{y}_{j}\sigma^{x}_{j+1}  \right)   \right] \quad. \label{H-spin}
    \end{eqnarray}
    Here  the coupling constants
    \begin{eqnarray} 
  J_{x,y}(Q)&=& w \cos{Q}\pm \Delta_{0} \label{Jxy-def} \\ {D}(Q) &=& w\sin(Q)\label{D(Q)-def}
    \end{eqnarray}
    are independent of the site $j$ because of the phase factors $e^{\pm iQj}$ introduced in Eq.(\ref{JW}). 
  For $Q=0$ one recovers from Eq.(\ref{H-spin}) the customary XY-model, where $\Delta_0$ acts as an  anisotropy parameter for the exchange couplings $J_{x,y}$, while $\mu$ plays the role of a transverse field along~$z$. The spatial modulation wavevector~$Q$ of the Kitaev chain gives rise to two effects. Firstly, it renormalizes the exchange coupling constants  $J_{x,y}(Q)$ through $w \rightarrow w \cos{Q}$ and, similarly to renormalization of the tunneling amplitude in the fermionic model, it modifies the boundaries between the gapped PM and FM phases.  
  The second effect of $Q$ is to introduce  the term in the second line of Eq.(\ref{H-spin}), characterized by the coupling constant $D(Q)$ in Eq.(\ref{D(Q)-def}), and known as the Dzyaloshinskii-Moriya  interaction (DMI)~\cite{dzyaloshinsky_1958,moriya_1960}. 
  In magnetic systems, this coupling originates from the interplay of broken inversion symmetry and spin-orbit interaction and, despite being typically small, it  can give rise to interesting    chiral magnetic orders such as spin spirals and skyrmions \cite{bogdanov2020, fert2017, wiesendanger2016,mahdavifar2024}. In particular, it can  lead to a gapless chiral phase, where 
  the chirality operator
  \begin{equation}\label{kappa_j-def}
  \kappa_j=\sigma^{x}_{j}\sigma^{y}_{j+1} -\sigma^{y}_{j}\sigma^{x}_{j+1}
  \end{equation}
  exhibits a finite long-range order \cite{hikihara2001, mahdavifar2024, roy2019}. 
  
The correlation functions of the spin model Eq.(\ref{H-spin}) are determined by the interplay between the   above two effects, which are both controlled by the parameter $Q$.   By exploiting the inverse Jordan-Wigner transformation
    \begin{equation}\label{JW-inv}
    \begin{array}{lcl}
    \sigma^{+}_{j} &=& \displaystyle e^{-iQj}  c^{\dagger}_{j}\prod^{j-1}_{n = 1}\left(1 - 2c^{\dagger}_{n}c_{n}\right) \\
    \sigma^{-}_{j} &=& \displaystyle e^{iQj} \prod^{j- 1}_{n = 1}\left(1 - 2c^{\dagger}_{n}c_{n}\right)c_{j} \\
    \sigma^{z}_{j} &=& \displaystyle 2c^{\dagger}_{j}c_{j} - 1 
    \end{array}\quad,
    \end{equation}
 spin-spin correlations can be expressed in terms of the fermionic correlations. In particular,   for nearest neighbors correlations one finds
 \begin{equation}\label{Spin-corr-1}
  \begin{array}{lcl}
  \langle \sigma^+_j \sigma^+_{j+1} \rangle &=& \langle \sigma^-_j \sigma^-_{j+1} \rangle= \mathcal{A}(1)  \\
    \langle \sigma^+_j \sigma^-_{j+1} \rangle &=& \langle \sigma^{-}_{j} \sigma^{+}_{j+1} \rangle^*= \mathcal{C}(1)  
    \end{array}\quad,
  \end{equation}
%  \begin{eqnarray}
%  \langle \sigma^+_j \sigma^+_{j+1} \rangle &=& \mathcal{A}(1) \label{S-1} \\
%  \langle \sigma^-_j \sigma^-_{j+1} \rangle &=& -e^{iQ(2j+1)}\langle c_{j}c_{j+1} \rangle \label{S-2}\\
%    \langle \sigma^+_j \sigma^-_{j+1} \rangle &=& e^{iQ}\langle c^{\dagger}c_{j+1} \rangle \label{S-3}\\
%    \langle \sigma^{-}_{j} \sigma^{+}_{j+1} \rangle & = & -e^{-iQ}\langle c_{j}c^{\dagger}_{j+1} \rangle. \label{S-4}
%  \end{eqnarray}
which in turn straightforwardly imply the connection between spin orders and   the paired and unpaired contributions to fermion correlations, namely
\begin{eqnarray}
  \langle \sigma^{x}_{j}\sigma^{x}_{j+1} + \sigma^{y}_{j}\sigma^{y}_{j+1} \rangle &=& 4\mathcal{C}_{p}(1)  \\
   \langle \sigma^{x}_{j}\sigma^{x}_{j+1} - \sigma^{y}_{j}\sigma^{y}_{j+1} \rangle &=&  4\mathcal{A}(1)\\
  \langle  \sigma^{x}_{j}\sigma^{y}_{j+1}   + \sigma^{y}_{j}\sigma^{x}_{j+1}    \rangle &=& 0\quad, 
\end{eqnarray}
and
\begin{eqnarray}
\kappa=  \langle\kappa_j\rangle=    \langle \sigma^{x}_{j}\sigma^{y}_{j+1} - \sigma^{y}_{j}\sigma^{x}_{j+1} \rangle &=& -4\mathcal{C}_{u}(1)\quad. \label{chiral-unp}
\end{eqnarray}

\begin{figure}
    \includegraphics[scale = 0.65]{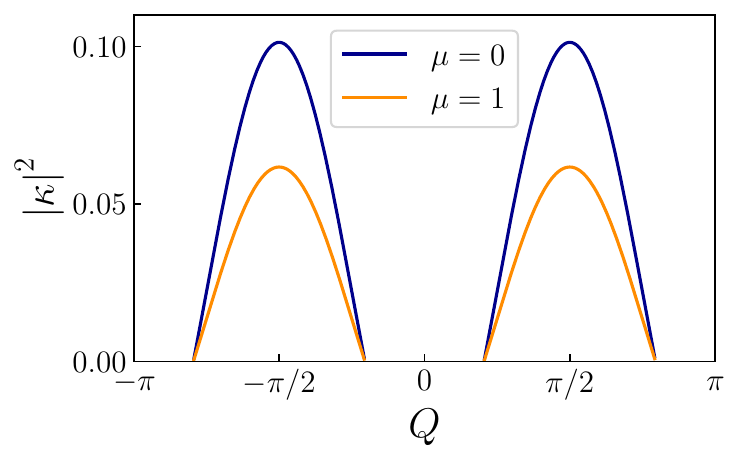}
    \caption{\label{Fig9}  The  squared modulus of the chiral order parameter~$\kappa$ in Eq.(\ref{chiral-unp})   is plotted as a function of $Q$, at fixed $\Delta_0=0.6w$ (anisotropy parameter), for two values of~$\mu$ (magnetic field). The maximum of chiral order occurs at $Q=\pm\pi/2$, its maximal value depends on $\mu$ and the onset of the chiral phase ($\kappa \neq 0$) is determined by the range $Q^*<|Q|<\pi-Q^*$, and only depends on the anisotropy parameter $\Delta_0$ [see Eq.(\ref{Q-star-def})] and not on $\mu$.}
\end{figure}

In particular,  Eq.(\ref{chiral-unp}) establishes   that the gapless chiral phase in the XY model, $   \kappa \neq 0$, corresponds to a gapless superconducting phase in the fermionic model, where unpaired fermions appear in the ground state. The results obtained in the previous section about the Kitaev chain can now be interpreted in terms of the spin model Eq.(\ref{H-spin}). In particular, Eq.(\ref{gapless-regime}) identifies the parameter regimes where such a chiral phase exists, while the exact result Eq.(\ref{Cu-exact}) returns the expectation value  $\kappa=\langle\kappa_j\rangle$. From Fig.\ref{Fig9}, which shows $|\kappa |^2$ as a function of $Q$, one can see that its maximal value is always reached at $Q=\pm \pi/2$, for any value of the magnetic field $\mu$. Yet, the value of $|\kappa |^2$  in such maxima depends on $\mu$, with $\mu=0$ corresponding to the global maximum. Furthermore,  the boundaries of the chiral phase, given by $Q^*<|Q|<\pi-Q^*$,   only depend  on the anisotropy parameter $\Delta_0$   through   Eq.(\ref{Q-star-def}) and are {\it independent} of the magnetic field $\mu$. 

As far as non-local spin correlations at  arbitrary distance $l$ are concerned, their evaluation through fermionic correlation requires to account for the string operators appearing in the inverse Jordan-Wigner transformation~(\ref{JW-inv})
\begin{eqnarray}
  \langle \sigma^+_j \sigma^+_{j+l} \rangle &=& e^{-iQ(2j +l)}\langle c^{\dagger}_{j} \prod^{l-1}_{n = j}\left( 1 - 2c^{\dagger}_{n}c_{n}\right)c^{\dagger}_{j+l} \rangle \label{Sl1}\\
  \langle \sigma^-_j \sigma^-_{j+l} \rangle &=& e^{iQ(2j +l)}\langle  c_{j} \prod^{l -.1}_{n = j} \left(1 - 2c^{\dagger}_{n}c_{n}\right)c_{j+l} \rangle \label{Sl2}\\
    \langle \sigma^+_j \sigma^-_{j+l} \rangle &=& e^{-iQl}\langle c^{\dagger}_{j} \prod^{l-1}_{n = j}\left( 1 - 2c^{\dagger}_{n}c_{n}\right)c_{j+l} \rangle   \label{Sl3}\\
    \langle \sigma^{-}_{j}\sigma^{+}_{j + l} \rangle &=& e^{iQl}\langle  c_{j} \prod^{l-1}_{n =j}\left(1 - 2c^{\dagger}_{n}c_{n}\right)c_{j+l}\rangle. \label{Sl4}
\end{eqnarray}
The right hand side of Eqs.(\ref{Sl1})-(\ref{Sl4}) can be computed by applying Wick's theorem    and expressing spin correlations at a given distance $l$ as combination  of products of the fermionic two-point correlation  functions $\mathcal{C}(m)$ and $\mathcal{A}(m)$ determined in the previous section, for various $m$.

%%%%%%%%%%%%%%%%%%%%%%%%%%%%%%%%%%%%%%%

%%%%%%%%%%%%%%%%%%%%%%%%%%
%%%%%%%%%%%%%%%%%%%%%%%%%%
%%%%%%%%%%%%%%%%%%%%%%%%%%
    \section{Summary and Conclusions}
    \label{sec-conclusions} 

In this article we have investigated  the correlation functions of the 1D Kitaev chain model in the presence of a spatially modulated phase of its  superconducting order parameter, which describes a
$p$-wave topological superconductor crossed by an electrical current. It has recently been shown that,  depending on  the parameters $w$ (single particle hopping parameter), $Q$ (wavevector of the superconducting phase modulation), $\Delta_0$ (magnitude of the superconducting order parameter) and  $\mu$ (chemical potential), the model exhibits two types of topological transitions. Indeed, in addition  to   the band topological phase transition separating the topologically trivial and nontrivial gapped phases, a Fermi surface Lifshitz transition between gapped and gapless phases can arise. 
 Here, we have found that such a  rich scenario emerging in the presence of a current flow leads to various interesting effects in the normal and anomalous correlation functions, which have not been reported so far. 

First, we have shown that in the special cases $\mu=0$ and $Q=\pm \pi/2$ the model acquires chiral and inversion symmetries, respectively, which are otherwise broken for generic values of $Q$ and $\mu$. These symmetries cause  an even/odd effect  in both the normal and the anomalous correlation functions, $\mathcal{C}(l)$ and $\mathcal{A}(l)$,  which turn out to strictly vanish at either even or odd values of the site distance $l$, measured in units of the lattice spacing.

Then, we have shown that the difference between the band topology and the Lifshitz  transitions can be signalled by analyzing the behavior of the bulk correlation functions   at short distance ($l=1,2$) as a function of  the modulation wavevector $Q$. Across the band topology transition the anomalous correlation function $\mathcal{A}$ turns out to behave  very smoothly in $Q$ and is not very informative about the transition, while the normal  correlation function $\mathcal{C}$ signals the transition only through a divergence of its $Q$-derivative. This is a  consequence of the {\it direct} closing of the gap at $k=0$ or $k=\pi$ across the two topologically distinct gapped phases, and the related divergence of the correlation length is the same on both sides of the transition, in agreement with the universality of the correlation length scaling discussed in Ref.\cite{sigriest_PRB_2017}.   In contrast, across the Lifshitz transition {\it both}   correlations functions $\mathcal{C}$ and $\mathcal{A}$  exhibit {\it sharp cusps}, which reflect     discontinuity jumps in their derivative. We have shown that  a jump  is the hallmark of the {\it indirect}  closing of the gap at the Lifshitz transition and the appearance of unpaired fermions  in the ground state.

Furthermore, we have been able 
to determine  the asymptotic behavior   of   $\mathcal{C}(l)$ and $\mathcal{A}(l)$ for long distance  $l \gg 1$. 
In particular, in the gapped phase we have found that the correlation functions can exhibit three types of exponential decay. Indeed,  depending on the parameter range, $\mathcal{C}(l)$ and $\mathcal{A}(l)$ can acquire the  form of  i)  a linear combination of two exponential decays  with a relative sign $(-1)^l$ alternating with the distance $l$, or ii) two exponential decays  without any  alternating sign, or iii)  one single exponential decay with additional spatial oscillations. These three types of behavior  are identified  by the symbols ``$\circ$",  ``$\cdot$" and  ``$-$", respectively, in the cyan and grey regions of Fig.\ref{Fig1}. Moreover, we have found that in  the gapless phase,  correlations decay algebraically as  $\sim l^{-1}$, with an additional spatially oscillatory behavior characterized by two periods related to $Q$, $\mu/w$, and $\Delta_0/w$.

Finally, we have shown that the  Kitaev chain with superconducting phase modulation can be mapped in a XY spin model, where $Q$ controls both the exchange coupling constants (\ref{Jxy-def}) and the strength (\ref{D(Q)-def}) of a  Dzyaloshinskii-Moriya interaction. We have argued that the gapless superconducting phase of the Kitaev chain, where unpaired fermions appear in the ground state, can be interpreted as the spin chiral phase of the XY model, whose order parameter reaches its maximal values around for $Q=\pm \pi/2$.
We have  used our results about fermionic correlations to evaluate   also spin correlations between neighboring sites.\\

{\it Implementations.}
Before concluding, we would like to briefly discuss some possible  setups where the results obtained in this paper could be applied.  
At the moment, there are mainly two promising setups for the realization of topological superconductivity. The first one is semiconductor nanowires with strong spin-orbit coupling, such as InSb and InAs, proximitized by a  superconducting layer (e.g., Al or Nb) and exposed to a longitudinal magnetic field~\cite{kouwenhoven2012,furdyna2012,heiblum2012,kouwenhoven2018,yacoby2014,yu2021non}.  The second one is ferromagnetic atom chains deposited on a superconducting film~\cite{yazdani2014,tang2016,loss-meyer_2016}.  
Scanning tunneling microscopy has been proposed as a technique to measure local   correlation functions in magnetic atom  chains~\cite{choy2011,nadjperge2013}, while spatial correlations in nanowires have been probed by x-ray scattering~\cite{wu_APL_2008,zheng_2021}.
Transport measurements  are also closely related to correlation functions. 
Indeed the current $I= 2 \text{e} w \, \text{Im} \, \langle c^{\dagger}_{j}c_{j+1} \rangle  /\hbar$ can be expressed in terms of the normal correlation function $\mathcal{C}(1)$ as  
\begin{equation}
I =   \frac{2\text{e}w}{\hbar}\left(\cos(Q)\mathcal{C}_{u}(1) - \sin(Q)\mathcal{C}_{p}(1)\right) \quad.
\end{equation}
In particular, we would like to outline a connection between  our results and the recent prediction that  the electrical current through a topological superconductor  exhibits cusps as a function of $Q$~\cite{FFF2024}. 
On the one hand, the cusps in the current can now be interpreted as a straightforward consequence of the correlation singularity across the Lifshitz transition. On the other hand, because our findings show that such cusps are a general hallmark of such type of transition, their signature is expected to be observable in other correlation functions as well, like the anomalous correlation $\mathcal{A}$. The latter  can be extracted from Andreev reflection spectroscopy~\cite{miyoshi_2005,gonnelli_2011} and non-local correlations are accessible via  crossed Andreev reflection and cross correlation measurements~\cite{beckmann_2004,das2012,he_2014,poschl2022}.

Moreover, because our 
 findings  can also be interpreted in terms of spin chain models, we mention that  nuclear magnetic resonance techniques enable one to determine correlation  functions in magnetic systems even in out of equilibrium conditions~\cite{cappellaro_PRL_108}.
Finally, spin models with adjustable spin-spin interactions can also be implemented with ions confined in a linear Paul trap, which can be manipulated using lasers. This approach allows  both collective and individual control over ion spins through laser interactions, and enables one to access  single-shot measurements of spin correlations \cite{richerme2014, jurcevic2014}.  \\\\

%%%%%%%%%%%%%%%%%%%%%%%%%%%%%%%%%%%%%
%%%%%%%    ACKNOWLEDGMENTS  %%%%%%%%%
%%%%%%%%%%%%%%%%%%%%%%%%%%%%%%%%%%%%%
\acknowledgments
F.G.M.C. acknowledges financial support from ICSC Centro Nazionale di Ricerca in High-Performance Computing, Big Data, and Quantum Computing (Spoke 7), Grant No. CN00000013, funded by European Union NextGeneration EU. \,\, F.D. acknowledges financial support  from the TOPMASQ project, CUP E13C24001560001, funded by the  National Quantum Science and Technology Institute (Spoke 5), Grant No. PE00000023, funded by the European Union – NextGeneration EU. Fruitful discussions with Lorenzo Rossi are also greatly acknowledged.

%%%%%%%%%%%%%%%%%%%%%%%%%%%%%%%%% 
%%%%%%%    APPENDICES   %%%%%%%%%
%%%%%%%%%%%%%%%%%%%%%%%%%%%%%%%%% 
\appendix

	\section{\label{appendixA} Derivation of the real space correlations functions}
 In this Appendix we provide some details about the evaluation of the correlation functions (\ref{C-def}) and (\ref{A-def}) given in the main text. By re-expressing the real space operators through their Fourier modes, $c_j=N_s^{-1/2} \sum_{k \in {\rm BZ}} e^{i k j} c_k$, one can  rewrite Eqs.(\ref{C-def}) and (\ref{A-def}) as
\begin{eqnarray} 
		\mathcal{C}(j_2-j_1) &=& e^{i Q (j_2-j_1) }\langle  c^{\dagger}_{j_1}c^{}_{j_2} \rangle =  \nonumber \\
  &=& \frac{e^{i Q (j_2-j_1) }}{N_s} \sum_{k_1,k_2} e^{-i k_1 j_1} e^{i k_2 j_2} \langle c^\dagger_{k_1} c^{}_{k_2} \rangle = \nonumber \\
  &=& \frac{1}{N_{s}} \sum_{k,k^\prime}e^{-i\left(k j_{1} - k'j_{2}\right)} \langle c^{\dagger}_{k-Q}\,c^{}_{k'-Q} \rangle, \label{normal_V1-k} 
	\end{eqnarray}
 and
 \begin{eqnarray}
     \mathcal{A}(j_2-j_1) 
 &=&e^{-i Q (j_2+j_1) } \langle  c^{\dagger}_{j_{1}}c^{\dagger}_{j_{2}} \rangle \nonumber \\
  &=& e^{-i Q (j_2+j_1) } \frac{1}{N_s} \sum_{k_1,k_2} e^{-i k_1 j_1} e^{-i k_2 j_2} \langle c^\dagger_{k_1} c^\dagger_{k_2} \rangle = \nonumber \\
  &=&\frac{1}{N_{s}} \sum_{k,k'}e^{-i\left(k j_{1} - k'j_{2}\right)} \langle c^{\dagger}_{k-Q}c^{\dagger}_{-k'-Q} \rangle,
		\label{anomalous_V1-k}
	\end{eqnarray}
respectively.
	The correlations   $\langle c^{\dagger}_{k-Q}c^{}_{k'-Q} \rangle$,  and    $ \langle c^{\dagger}_{k-Q}c^{\dagger}_{-k'-Q} \rangle$ appearing in Eqs.(\ref{normal_V1-k})-(\ref{anomalous_V1-k}) can now be computed by inverting  Eqs.(\ref{Bogoliubov-quasi}) in favor of the $c^{}_{k-Q}, c^\dagger_{k-Q}$ operators
\begin{equation}
\label{c-intermsof-gamma-old}
 \left\{ \begin{array}{lcl}
  c^{}_{k -Q} &= & u_{Q}(k)\gamma^{}_{k-Q}    -v^*_{Q}(k)\, \gamma^\dagger_{-k-Q}  \\	    c^{\dagger}_{k-Q} &= & u_{Q}(k) \gamma^{\dagger}_{k-Q}  -v_{Q}(k)\gamma^{}_{-k -Q}  
\end{array}  \right.   
\end{equation}
and by exploiting the 
  action of the $\gamma$-Bogolubov quasi-particles onto the current carrying ground state $|G(Q)\rangle$ given in Eq.(\ref{G-gen}). Such an action depends on   the specific Brillouin sector $S_p$, $S_e$ or $S_h$ where the $k$-wavevector is located, namely  $\gamma_{k-Q}\ket{G(Q)}=0$ for $k \in S_{+} \equiv S_p \cup S_h$ and $\gamma^\dagger_{k-Q}\ket{G(Q)}=0$ for $k \in S_{-} \equiv S_e$. Therefore, one has to consider the following cases 
\\
\noindent {1. \,$k,k' \in S_{p}$}
		\begin{equation}
			\begin{array}{ccl}
				\langle \gamma^{\dagger}_{k-Q}\gamma_{k'-Q} \rangle &=& \langle \gamma^{\dagger}_{k-Q}\gamma^{\dagger}_{-k'-Q} \rangle =\\
				 & = &  
				\langle \gamma_{-k-Q}\gamma_{k'-Q} \rangle = 0 \\
				\langle \gamma_{-k-Q}\gamma^{\dagger}_{-k'-Q} \rangle & = & \delta_{k,k'}
			\end{array}
		\end{equation}

\noindent {2. $ k \in S_{h}, \, -k \in S_{e}$ }
		\begin{equation}
			\begin{array}{l}
				\langle \gamma^{\dagger}_{k-Q}\gamma_{k'-Q} \rangle  =  
				\langle \gamma^{\dagger}_{k-Q}\gamma^{\dagger}_{-k'-Q} \rangle   =  \\
				=\langle \gamma_{-k-Q}\gamma_{k'-Q} \rangle   =
				\langle \gamma_{-k-Q}\gamma^{\dagger}_{-k'-Q} \rangle  = 0
			\end{array}
		\end{equation}
\noindent {3. $k \in S_{e}, \, -k \in S_{h}$}	
	\begin{equation}
			\begin{array}{ccl}
				\langle \gamma^{\dagger}_{k-Q}\gamma^{}_{k'-Q} \rangle &=& 
    \langle \gamma^{}_{-k-Q}\gamma^{\dagger}_{-k'-Q} \rangle = \delta_{k,k'}    \\
				\langle \gamma^{\dagger}_{k-Q}\gamma^{\dagger}_{-k'-Q} \rangle & = &  
				\langle \gamma^{}_{-k-Q}\gamma^{}_{k'-Q} \rangle = 0 \\			
			\end{array}
		\end{equation}	
Exploiting  the above results one obtains the correlations in momentum space 
	\begin{equation}
		\langle c^{\dagger}_{k-Q}c_{k'-Q} \rangle  = \left\{
		\begin{array}{ccc}
			\left|v_Q(k)\right|^{2} \delta_{k,k'} && k \in S_{p}\\
			0         && k \in S_{h} \\
			\delta_{k,k'} && k \in S_{e}
		\end{array}\right.,
        \label{A4}
	\end{equation}
	\begin{equation}
		\langle c_{-k-Q}c^{\dagger}_{-k'-Q} \rangle = \left\{
		\begin{array}{ccc}
			\left|u_Q(k)\right|^{2}\delta_{k,k'} && k \in S_{p}\\
			0        && k \in S_{h}\\
			\delta_{k,k'} && k \in S_{e}
		\end{array}\right.,
        \label{A5}
	\end{equation}
	\begin{equation}
		\langle c_{-k-Q}c_{k'-Q} \rangle =   
		\left\{\begin{array}{ccc}
			- u_Q(k)v^{*}_Q(k)\delta_{k,k'} &&    k \in S_{p}\\
			0   &&   k \in S_{h}\\
			0   &&   k \in S_{e}
		\end{array}\right.,
        \label{A6}
	\end{equation}
	\begin{equation}
		\langle c^{\dagger}_{k-Q}c^{\dagger}_{-k'-Q} \rangle =   
		\left\{\begin{array}{ccc}
			- u_Q(k)v_Q(k)\delta_{k,k'} &&    k \in S_{p}\\
			0   &&   k \in S_{h}\\
			0   &&   k \in S_{e}
		\end{array}\right.,
        \label{A7}
	\end{equation}
	where $u_Q(k)$ and $v_Q(k)$ are given by Eq.(\ref{u_v}). Replacing Eqs.(\ref{A4})-(\ref{A5})-(\ref{A6})-(\ref{A7})  in Eq.(\ref{normal_V1-k}) and Eq.(\ref{anomalous_V1-k}), and denoting $l=j_2-j_1$, it is straightforward to obtain
    \begin{eqnarray}
   \mathcal{C}(l)  
    &=& \frac{1}{N_{s}} \left( \sum_{k \in S_{p}} e^{ik l} \left|v_Q(k)\right|^{2}    +   \sum_{k \in S_{e}} e^{i k l} \right)  \label{Cp-k-discrete}
\\
    \mathcal{A}(l)&=& -\frac{1}{N_{s}} \sum_{k \in S_{p}}  e^{ ik l} u_Q(k)v_Q(k)  \label{A-k-discrete}
    \end{eqnarray}
Substituting Eq.(\ref{u_v}) into Eq.(\ref{A-k-discrete}), and taking the thermodynamic limit $N_s \rightarrow \infty$, one obtains Eq.(\ref{A-k}) of the main text. Moreover, exploiting the mirror symmetries $S_e \rightarrow S_h$ and $S_p \rightarrow S_p$  under $k\rightarrow -k$,   one can rewrite Eq.(\ref{Cp-k-discrete}) as
\begin{eqnarray}
\lefteqn{\mathcal{C}(l)  
     = } & &  \\
     &=&\frac{1}{2N_{s}} \left( \sum_{k \in {\rm BZ}} \cos(k l) -\sum_{k \in S_p}\frac{\xi(k;Q)}{h(k;Q)}    +    2 \sum_{k \in S_{e}} \sin(kl)    \right)    \nonumber
    \end{eqnarray}
whence Eq.(\ref{Cp-k}) and (\ref{Cu-k}) of the main text are obtained by recalling that $l \neq 0$ is assumed, and by taking the thermodynamic limit $N_s \rightarrow \infty$.
 
%%%%%%%%%%%%%%%%%%%%%%%%%%%%%%%%%
%%%%%%%    Appendix B    %%%%%%%%
%%%%%%%%%%%%%%%%%%%%%%%%%%%%%%%%%
\section{Details about correlation functions in the cases $\mu=0$ and $Q=\pm \pi/2$}
\label{AppB}
In this Appendix we prove the even/odd effect occurring for $\mu=0$ or $Q=\pm \pi/2$ and discussed in Sec.\ref{sec3}. Moreover, we provide the analytical expressions of the non-vanishing correlation functions at some values of $l$. 

We start by proving that the special symmetries acquired by the Hamiltonian (\ref{KTV-Ham}) for these parameter values imply   the vanishing of some correlation functions. To this purpose, the first straightforward step is to realize, as mentioned in the main text, that, by applying the (anti-unitary) chiral transformation   (\ref{S-def}) to the Hamiltonian~(\ref{KTV-Ham}), one obtains  $\mathcal{S}\mathcal{H}\mathcal{S}^\dagger=\mathcal{H}$ when $\mu=0$. Similarly,  when $Q=\pm \pi/2$,   one has $\mathcal{I}\mathcal{H}\mathcal{I}^\dagger=\mathcal{H}$, where the (unitary) inversion transformation $\mathcal{I}$ is defined in Eq.(\ref{I-def}). The second step is to show that the 
 ground state (\ref{G-gen}) is also an eigenstate of $\mathcal{S}$ (for $\mu=0$) and of $\mathcal{I}$ (for $Q=\pm \pi/2 $), both in the gapped and in the gapless phase, which justifies why the even/odd effect is robust across the Lifshitz transition.  To this purpose, we observe that, by rewriting the Hamiltonian in $k$-space [see Eq.(\ref{KTV-Ham-k})], both the $\mathcal{S}$ and the $\mathcal{I}$  symmetry separately imply the relation 
\begin{equation}\label{sym-HSP}
    E_\pm(k)=E_\pm(\pi-k)
\end{equation}
for   the two bands (\ref{EigenV}).  As a consequence of Eq.(\ref{sym-HSP}), each of the three sectors $S_p$, $S_e$ and $S_h$, defined through the Eqs.(\ref{values_eta})-(\ref{spectral-asymmetry}) and entering the expression (\ref{G-gen}) of the ground state, becomes symmetric under $k \rightarrow \pi-k$ for either $\mu=0$ or $Q=\pm \pi/2$. 
Moreover, the fermionic vacuum $|0\rangle$ 
transforms as $\mathcal{I}|0\rangle=|0\rangle$ and $\mathcal{S}|0\rangle=|F\rangle$, where $|F\rangle=\prod_{k \in {\rm BZ}}c^\dagger_k|0\rangle$ is the completely filled state, as straightforwardly follows by applying $\mathcal{I}$ or $\mathcal{S}$ on the left of the relation $c^{}_k|0\rangle=0\,\, \forall k \in{\rm BZ}$ characterizing the vacuum. By exploiting the action of $\mathcal{S}$ or $\mathcal{I}$ on   $|0\rangle$ and  the expression of the coefficients (\ref{u_v}) appearing in the ground state (\ref{G-gen}) of $\mathcal{H}$, it is easy to show that   $|G(Q)\rangle$    also exhibits the above symmetries. Specifically, for $\mu=0$, one has $\mathcal{S}|G(Q)\rangle =\pm |G(Q)\rangle$, where the sign depends on the specific parameter values, while for $Q=\pm \pi/2$ one has $\mathcal{I}|G(Q)\rangle =|G(Q)\rangle$.

We now have all the ingredients to prove that the even/odd effects originate from symmetry arguments. Let us first consider the case $\mu=0$ and prove that both the normal and the anomalous correlation function vanish for even $l$, as sketched in Table~\ref{table1}. Indeed we observe that the expectation value of a second-quantized operator ${O}$   on the $\mathcal{S}$-symmetric ground state $|G(Q)\rangle$ is
\begin{eqnarray}
\langle G|O G\rangle &=& \langle G|O^\dagger G\rangle^* =\langle \mathcal{S} G|\mathcal{S} {O}^\dagger   G\rangle = \nonumber \\
&=& \langle \mathcal{S} G|\mathcal{S} {O}^\dagger \mathcal{S}^\dagger |\mathcal{S} G\rangle = \nonumber \\
&=& \langle  G|\mathcal{S} {O}^\dagger \mathcal{S}^\dagger |  G\rangle  \quad,
\end{eqnarray}
where we have exploited the antiunitarity of $\mathcal{S}$. Now, the property $\langle G|O|G\rangle=0$  straightforwardly follows for any operator  ${O}$   fulfilling $\mathcal{S} {O}^\dagger \mathcal{S}^\dagger=-{O}$. This is the case for ${O}=c^\dagger_j c^{}_{j+l}$ and for ${O}=c^\dagger_j c^\dagger_{j+l}$ for any even $l$. Since these are precisely the operators appearing in the normal and anomalous correlation functions (\ref{C-def})-(\ref{A-def}), the above symmetry argument explains the vanishing of the normal and anomalous correlations functions appearing in Table~\ref{table1}. 

Let us now turn to the case $Q= \pi/2$, and prove  that the real part of the normal correlation function vanishes  for any odd $l$, while anomalous correlation function vanishes for even $l$, as sketched in Table\ref{table2}. Indeed, for the normal correlation function (\ref{C-def}) one has for any odd $l$
\begin{eqnarray}
\mathcal{C}(l)&=& i^l \,  \langle G|c^\dagger_j c^{}_{j+l}| G\rangle \nonumber \\
&=&    i^l\,  \langle G| \mathcal{I}^\dagger c^\dagger_{-j} c^{}_{-j-l} \mathcal{I}|  G\rangle  =\nonumber \\
&=&   i^l\, \langle \mathcal{I} G|c^\dagger_{-j} c^{}_{-j-l}  |\mathcal{I}  G\rangle = \nonumber \\
&=&  i^l\,\langle  G| c^\dagger_{-j-l}  c^{}_{-j}  |  G\rangle^* =  \nonumber \\
&=& (-1)^{l} \,\, \left(i^l \langle   G| c^\dagger_{-j-l}  c^{}_{-j}  |  G\rangle \right)^*= \nonumber \\
&=&- \,\, \mathcal{C}^*(l) \quad,\label{sym-on-C-due-to-I}
\end{eqnarray}
where we have exploited   the $\mathcal{I}$-symmetry of the ground state $|G\rangle$ and  the property that $\mathcal{C}(l)$ only depends on the site distance $l$. From Eq.(\ref{sym-on-C-due-to-I}) we deduce that $\mathcal{C}$ can only be purely imaginary at odd $l$. In particular, because its imaginary part originates from the unpaired fermions  [see Eq.(\ref{C-Re-Im})], $\mathcal{C}\equiv 0$ in the gapped phase.
Considering now the anomalous correlation function~(\ref{A-def}), one has
\begin{eqnarray}
\mathcal{A}(l)&=& (-1)^j \, (-i)^l\langle G|c^\dagger_j c^\dagger_{j+l}| G\rangle= \nonumber \\
&=& (-1)^j \, (-i)^l \langle G| \mathcal{I}^\dagger c^\dagger_{-j} c^\dagger_{-j-l} \mathcal{I}|  G\rangle  =\nonumber \\
&=& (-1)^j \, (-i)^l \langle \mathcal{I} G|c^\dagger_{-j} c^\dagger_{-j-l}  |\mathcal{I}  G\rangle = \nonumber \\
&=& -(-1)^j \, (-i)^l \langle   G| c^\dagger_{-j-l}  c^\dagger_{-j}  |  G\rangle =  \nonumber \\
&=& (-1)^{l+1} \,\, (-1)^j\, i^l \langle   G| c^\dagger_{-j-l}  c^\dagger_{-j}  |  G\rangle = \nonumber \\
&=&(-1)^{l+1} \,\, \mathcal{A}(l) \quad.\label{sym-on-A-due-to-I}
\end{eqnarray}
 From  Eq.(\ref{sym-on-A-due-to-I}) we deduce that the anomalous correlation vanishes for any even $l$, as described by Table \ref{table2}. A quite similar argument holds for $Q=-\pi/2$.

So far, we have used the symmetry arguments to prove that correlation functions vanish depending on the parity of $l$. We now turn to provide the analytically exact expression for some of the non-vanishing correlations functions.
Let us first consider the case $\mu=0$ and analyze first the normal correlation function  $\mathcal{C}(l)$.
We start by analyzing the  unpaired fermion contribution $\mathcal{C}_u(l)$, which vanishes in the  gapped phase, $\mathcal{C}_u(l) \equiv 0$, while in the gapless phase is exactly given    by Eq.(\ref{Cu-exact}). We now note that, for $\mu=0$, the two wavevectors in Eq.(\ref{k-star-pm}) share the same magnitude, which we can denote as  
\begin{equation}
k^{*}_{Q}=|k^*_{\pm}(\mu=0)| =\arcsin\left(\frac{|\cos{Q}|}{\sqrt{1-\frac{\Delta_0^2}{w^2}}}\right)   \quad. 
\end{equation}
Thus, from Eq.(\ref{Cu-exact})
one obtains 
\begin{eqnarray}
\left.  \mathcal{C}_u(l) \right|_{\mu=0}  =  - 
  \frac{\text{sgn}(Q)}{2\pi}\frac{1}{l}\left( 1 - (-1)^{l}  \right) \cos(k^{*}_{Q}l)\quad,
\label{Cu-exact-mu=0}
\end{eqnarray}
 where we  also recover  that $\mathcal{C}_u(l)$ vanishes for even values of $l$.
Turning now to the pair contribution $\mathcal{C}_p$, we observe   
 that in its general expression   Eq.(\ref{Cp-k})  both the integrand function  and the integration domain $S_p$ are symmetric under $k \rightarrow -k$, implying that this expression  can be rewritten as an integral  over the positive-$k$ values of $S_p$ only. Focussing first  on the gapless phase, where the $S_p$ domain is given by Eq.(\ref{Sp-gapless}),  one can rewrite Eq.(\ref{Cp-k})  as
\begin{eqnarray}
\left.  \mathcal{C}_p(l) \right|_{\mu=0}
    &=& -\frac{\cos{Q}}{\pi} \left\{ \int_{0}^{k^{*}_{Q}}  \frac{\cos(k l)\, w \cos{k} }{h(k;Q)}\, dk  \,+\, \right. \nonumber \\
    & & \left. \hspace{0.5cm} +\int_{\pi-k^{*}_{Q}}^{\pi}  \frac{\cos(k l)\, w \cos{k} }{h(k;Q)}\, dk \right\} \,\,. \nonumber  
\end{eqnarray}
The band symmetry relation (\ref{sym-HSP}) that holds for $\mu=0$ in turns implies that $h(k)=h(\pi-k)$. 
Thus, changing $k \rightarrow \pi-k$ in the second integral, and exploiting  $\cos(\pi l -kl)=(-1)^l \cos(k l)$, one finds
\begin{eqnarray}
\left.  \mathcal{C}_p(l) \right|_{\mu=0}
    = -  \frac{1-(-1)^l}{2} \, \mathcal{Q}^n_p(l;k^*_Q) \quad, \label{Cp-mu=0-pre} 
\end{eqnarray}
where we have introduced 
\begin{eqnarray}
\mathcal{Q}^n_p(l;\alpha)  \doteq    \frac{\mbox{sgn}(\cos {Q})}{\pi}  \int_{0}^{\alpha} dk\, \frac{    \cos k  \, \cos(k l)}{\sqrt{  \cos^2k  + \delta_{Q} \sin^2 k}} \quad,   \label{QR-norm-def}
\end{eqnarray}
with
\begin{equation}
    \delta_{Q}=\frac{\Delta_0^2}{w^2 \cos^2{Q}}  \quad.
    \label{delta_Q-def}
\end{equation}
A similar argument applies to the evaluation of the anomalous correlation function $\mathcal{A}(l)$ in Eq.(\ref{A-k}), and leads to conclude that
\begin{eqnarray}
\left.  \mathcal{A}(l) \right|_{\mu=0}
    = -  \frac{1-(-1)^l}{2} \, \mathcal{Q}^a(l;k^*_Q) \quad, \label{A-mu=0-pre} 
\end{eqnarray}
where 
\begin{eqnarray}
\mathcal{Q}^a(l;\alpha)  \doteq   \displaystyle \frac{\sqrt{\delta_\phi}}{\pi}  \int_{0}^{\alpha} dk\, \frac{    \sin k  \, \sin(k l)}{\sqrt{  \cos^2k  + \delta_\phi \sin^2 k}}  \quad.  \label{QR-anom-def}
\end{eqnarray}
From Eqs.(\ref{Cp-mu=0-pre}) and (\ref{A-mu=0-pre})  we   recover  that the normal and anomalous correlations vanish  for  even $l$,  while for odd $l$ they are given by (minus) the quantities $\mathcal{Q}^n_p$ and $\mathcal{Q}^a$, respectively. Interestingly, these quantities can be given  analytically exact expressions in terms of elliptic functions of the first and second kind, $F$ and $E$. Here we limit ourselves to provide the   expressions  for $l=1$ in the gapless phase, namely  
\begin{equation}
\mathcal{C}_{p}(1) = \frac{\text{sgn}(\cos{Q})}{\pi} \left(  \frac{  \delta_{Q}F(k^{*}_{Q}; 1- \delta_{Q}) - E(k^{*}_{Q}; 1 - \delta_{Q})    }{1- \delta_{Q}}   \right)
\end{equation}
and
\begin{equation}
\mathcal{A}(1) = \frac{\sqrt{\delta_{Q}}}{\pi}\left(  \frac{  E(k^{*}_{Q}; 1 - \delta_{Q}) - F(k^{*}_{Q}; 1 - \delta_{Q})   }{ 1 - \delta_{Q}   }    \right)
\end{equation}
For the gapped phase, the corresponding results are obtained from the above formulas by replacing $k^*_Q \rightarrow \pi/2$ and $k^*_\mu \rightarrow \pi/2$, thereby obtaining expressions in terms of the complete elliptic integrals $K$ and $E$.\\

Let us now turn to the  case $Q=\pm \pi/2$. Following similar arguments and denoting 
\begin{equation}
k^{*}_{\mu}=|k^*_{+}\left(Q=\pm \frac{\pi}{2}\right)|=   \arcsin\left(\frac{|\mu|}{2 \sqrt{w^2-\Delta_0^2}}\right) \quad, 
\end{equation}
 one finds for the gapless phase
\begin{eqnarray}
\left.  \mathcal{C}_p(l) \right|_{Q=\pm \frac{\pi}{2}}
    =    \frac{1+(-1)^l}{2} \, \mathcal{R}^n_p(l;k^*_\mu) \quad, \label{Cp-cos(Q)=0-pre} 
\end{eqnarray}
and  
\begin{eqnarray}
\left.  \mathcal{A}(l) \right|_{Q=\pm \frac{\pi}{2}}
    =    \frac{1-(-1)^l}{2} \, \mathcal{R}^a(l;k^*_\mu)  \label{A-cos(Q)=0-pre} \quad.
\end{eqnarray}
Here we have introduced  
\begin{eqnarray}
\label{R-norm-def}
\mathcal{R}^n_{p}(l;\alpha)   \doteq   \frac{\mbox{sgn}(\mu)}{ \pi}   \int_{0}^{\alpha} dk\, \frac{\cos(k l)}{\sqrt{1+4\delta_\mu \sin^2 k}} 
\end{eqnarray}
and
\begin{eqnarray}
\label{R-anom-def}
\mathcal{R}^a(l;\alpha)  \doteq \displaystyle \frac{2\sqrt{\delta_{\mu}}}{\pi}     \int_{0}^{\alpha} dk\, \frac{  \sin{k} \, \sin(k l)}{\sqrt{1+4\delta_{\mu} \sin^2 k}}
\end{eqnarray}
with
\begin{equation}
    \delta_{\mu}=\frac{\Delta_0^2}{\mu^2} \quad. 
    \label{delta_Q-def}
\end{equation}
From Eq.(\ref{Cp-cos(Q)=0-pre}) we recover that  $\mathcal{C}_p$, i.e. the real part   of the normal correlation, vanishes for odd $l$, in agreement with the symmetry-based argument given above. Moreover, Eq.(\ref{A-cos(Q)=0-pre}) implies  that   $\mathcal{A}$  vanishes for even $l$, in both the gapped and gapless phase.  Also in this case, the quantities $\mathcal{R}^n_p$ and $\mathcal{R}^a$ in Eqs.(\ref{R-norm-def}) and (\ref{R-anom-def}) can be given an exact expression for small $l$, namely
\begin{eqnarray}
\mathcal{C}_p(2) = \frac{\text{sgn}(\mu)}{\pi}\left(  \frac{(1 + 2\delta_{\mu}) F(  k^{*}_{\mu} ; -4\delta_{\mu})   }{   2 \delta_{\mu}}      -\frac{ E(k^{*}_{\mu};-4\delta_{\mu}) }{ 2\delta_{\mu}  }\right) \nonumber \\
\end{eqnarray}
and
\begin{equation}
\mathcal{A}(1) = \frac{\sqrt{\delta_{\mu}}}{\pi}\left(\frac{F(k^{*}_{\mu}; -4\delta_{\mu}  )     -   E(k^{*}_{\mu}; -4\delta_{\mu})}{2\delta_{\mu}}\right)\quad.
\end{equation}
Again, for the gapped phase the same formulas (\ref{Cp-cos(Q)=0-pre}) and (\ref{A-cos(Q)=0-pre})  apply, upon replacing $k^*_\mu \rightarrow \pi/2$.

%%%%%%%%%%%%%%%%%%%%%%%%%%%%%%%%
\section{Asymptotic Expansion  of the correlations functions in the gapped phase}
\label{AppC} 
In this Appendix we provide details of the derivation of results about the  asymptotic behavior of the correlation functions at long distance ($l \gg 1$) given in Section \ref{sec5-1}. The functions appearing in Eqs.(\ref{Cp-cmplx}) and (\ref{A-cmplx}) exhibit
  four branch points at $z=Z_j$ given by  
\begin{equation}
\begin{array}{l}
Z_{1,2} = \displaystyle \frac{\frac{\mu}{2}\, \mbox{sgn}(\cos{Q}) \pm \sqrt{\Delta_0^2+\frac{\mu^2}{4}-w^2\cos^2{Q}}}{w|\cos{Q}|+\Delta_0}  \\  \\
Z_{3,4} = \displaystyle \frac{\frac{\mu}{2}\, \mbox{sgn}(\cos{Q}) \pm \sqrt{\Delta_0^2+\frac{\mu^2}{4}-w^2\cos^2{Q}}}{w|\cos{Q}|-\Delta_0} 
\end{array}\quad, \label{roots}
\end{equation}
whose location depends on the paramters $\Delta_0$, $\mu$ and $Q$. Note that the case $|\mu|=2w |\cos{Q}|$
is ruled out from the gapped phase parameter conditions~(\ref{gapped-regime}), since in such a case two of the branch points (\ref{roots}) coalesce into a pole on the circle at either $z=+ 1$ or $z=-1$. This corresponds to the direct closing of the superconducting gap at $k=0$ or $k=\pi$, and   identifies  the separatrix between the two topologically distinct gapped phases, as observed above. Avoiding this singularity and focussing on the gapped phases, two branch points, which we shall denote as  $z^{**}_\pm$,  lie outside the unit circle ($|z^{**}_\pm|>1$), whereas, as observed in the main text, the two inner  branch points ($|z^{*}_\pm|<1$) can have three different configurations in the complex plane, namely (1) both   real and with opposite sign; (2) both real and with equal signs; (3)   complex conjugate pair. 
These are illustrated in Fig.\ref{Fig10}, and determine the three different types of asymptotic behavior in the gapped phase. 
By applying Cauchy theorem to Eqs.(\ref{Cp-cmplx}) and (\ref{A-cmplx}), one can rewrite
\begin{eqnarray}
\mathcal{C}_p(l) &=& + \frac{1}{4\pi} {\rm Im} \left\{   \oint_{b. cut}  dz    \frac{z^{l-1}\,  g(z)}{\mathcal{D}(z)}   \right\} \label{Cp-cmplx-bc}\\
\mathcal{A}(l) &=& - \frac{1}{4\pi} {\rm Im} \left\{   \oint_{b.cut}  dz   \frac{z^{l-1}\,  \,  f(z)  }{\mathcal{D}(z)}  \right\} \label{A-cmplx-bc}
\end{eqnarray}
where $\mathcal{D}(z)=\sqrt{g^2(z)-f^2(z)}$, and "b.cut" denotes the   inner branch cuts, taken   {\it clockwise}, as also shown in Fig.\ref{Fig10}. Here we have exploited the fact that  the integrals along infinitesimally small circles around the branch cuts vanish.
One can now determine the values of the denominator $\mathcal{D}(z)$    appearing in Eqs.(\ref{Cp-cmplx-bc})-(\ref{A-cmplx-bc}) along the branch cuts, analyzing the three possible configurations of the branch points $z^*_\pm$.\\

\noindent (a): {\it  $z^*_\pm$ real  with opposite sign}\\
This configuration occurs when the conditions (\ref{cond-(a)-1}) are fulfilled. 
In this case, the outer branch points are given by
$z^{**}_{+}=1/z^*_{+} >0$ and $z^{**}_{-}=1/z^*_{-} <0$. The values of $\mathcal{D}(z)$ along the inner branch cut in Fig.\ref{Fig10}(a) are 
\begin{equation}
\left\{\begin{array}{lcl}
\mathcal{D}(x \pm i\epsilon) =  \pm i D_{1}(x) & & x>0\\
\mathcal{D}(x \pm i\epsilon) = \mp i D_{1}(x) & & x<0  
\end{array}\right. \label{square-root-values-(a)-1} 
\end{equation}
where
\begin{eqnarray}
\lefteqn{D_{1}(x)  =  \sqrt{\frac{\Delta_0^2-w^2\cos^2{Q}}{x^2}} \times } \label{D(x)-def-1a}   \\
& & \times \sqrt{(z^*_{+}-x) \,   (x-z^*_{-})\,   (z^{**}_{+}-x)\,   \, (x-z^{**}_{-}) }   \nonumber 
\end{eqnarray}
Inserting Eqs.(\ref{square-root-values-(a)-1})-(\ref{D(x)-def-1a}) into Eq.(\ref{Cp-cmplx-bc}) one obtains
\begin{widetext}
\begin{eqnarray}
\mathcal{C}_p(l)&=&\displaystyle - \frac{1}{2\pi}  \, \frac{1}{\sqrt{\frac{\Delta_0^2}{w^2}- \cos^2{Q}}}  \, \left\{ \int_{0}^{z_{+}^*}      \,dx  \frac{x^{(l-1)}}{\sqrt{z^*_{+}-x }}  f^{n,+}_{reg}(x)  -(-1)^l \int_{0}^{|z_{-}^*|}      \,dx  \frac{x^{(l-1)}}{\sqrt{|z^*_{-}|-x}}
 f^{n,-}_{reg}(x) \right\}   \label{Cp-cmplx-pre-(a)-1} 
\end{eqnarray}
\end{widetext}
where
\begin{eqnarray}
f^{n,+}_{reg}(x)&=&\frac{(1+x^2 ) \cos{{Q}}-\frac{\mu}{w} x}{ \sqrt{  \,\,   (|z^*_{-}|+x)\,   (\frac{1}{z^{*}_{+}}-x)\,   \, (\frac{1}{|z^{*}_{-}|}+x)   }} \label{fn+_reg-def} \\
f^{n,-}_{reg}(x)&=& \frac{(1+x^2 ) \cos{{Q}}+\frac{\mu}{w} x}{\sqrt{ \,(z^*_{+}+x) \,  \,   (\frac{1}{z^{*}_{+}}+x)\,   \, (\frac{1}{|z^{*}_{-}|}-x) }  }   \label{fn-_reg-def} 
\end{eqnarray}
In the first integral the term $x^{l-1}$ ranges from 0 to $(z^*_{+})^{l-1}$, which is exponentially small for $l \rightarrow \infty$ and strongly suppresses the contribution from $f^{n,+}_{reg}(x)$ away from $x \simeq z^*_{+}$. Similarly, in the second integral, the contribution from $f^{n,-}_{reg}(x)$ away from $x \simeq |z^*_{-}|$ is negligible.
Therefore,  in the $l \gg 1$ limit one can approximate Eq. (\ref{Cp-cmplx-pre-(a)-1}) as 
\begin{widetext}
\begin{figure*}
		\centering
    \includegraphics[scale = 0.7]{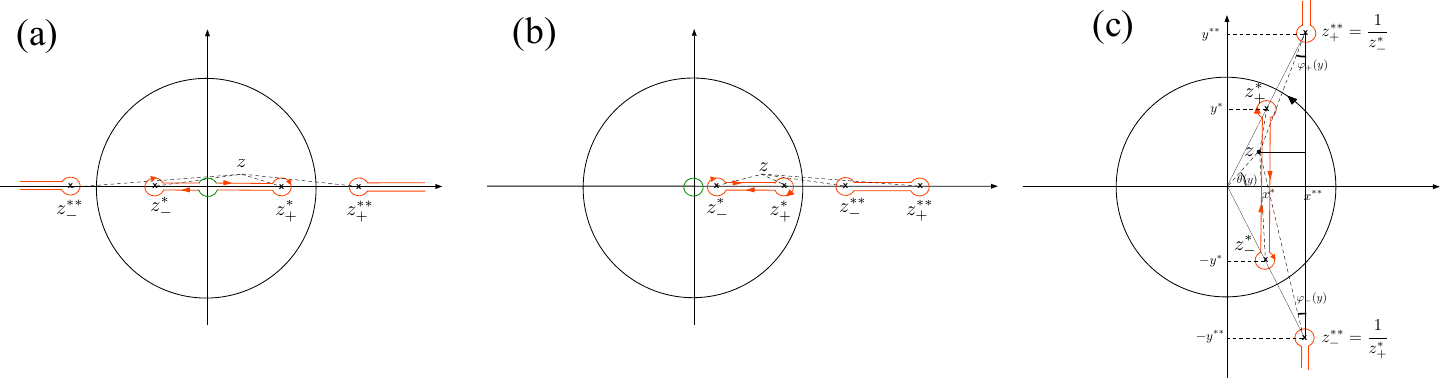}
    \caption{\label{Fig10}  The three different configurations of the inner branch  points   $z^*_\pm$ of the  denominator $\mathcal{D}(z)=\sqrt{g^2(z)-f^2(z)}$  appearing in Eqs.(\ref{Cp-cmplx-bc})-(\ref{A-cmplx-bc}). The branch cuts are highlighted in red.}
\end{figure*}
\begin{eqnarray}
\mathcal{C}_p(l)& \sim   &\displaystyle - \frac{1}{2\pi}  \, \frac{1}{\sqrt{\frac{\Delta_0^2}{w^2}- \cos^2{Q}}}   \, \left\{ f^{n,+}_{reg}(z^{*}_{+}) \int_{0}^{z_{+}^*}       \frac{x^{(l-1)}}{\sqrt{z^*_{+}-x }}   \,dx \,  -(-1)^l f^{n,-}_{reg}(|z^{*}_{-}|) \int_{0}^{|z_{-}^*|}   \frac{x^{(l-1)}}{\sqrt{|z^*_{-}|-x}}      \,dx \right\}   
\end{eqnarray}
\end{widetext}
A similar expression can be obtained for $\mathcal{A}(l)$ inserting Eq.(\ref{square-root-values-(a)-1}) into Eq.(\ref{A-cmplx-bc}).
Using the identity
\begin{equation}
\int_{0}^{a}     \frac{x^{(l-1)}}{\sqrt{a -x}}  \,dx   =    a^{l-\frac{1}{2}}\, \sqrt{\pi} \frac{\Gamma(l)}{\Gamma \left(l+\frac{1}{2}\right)} \simeq  \sqrt{\frac{\pi}{a\,l} }\, a^{l}\,  \label{ide-int-Gamma}
\end{equation}
where we have exploited $\Gamma(l+\frac{1}{2}) \sim \Gamma(l) \, l^{1/2}$ for $l\rightarrow \infty$, one obtains Eqs.(\ref{Cp-asym-(a)-1}) and (\ref{A-asym-(a)-1}), where 
\begin{align}
		\alpha^{\pm}_{1} =& \frac{w  A^{\pm}_{1}}{\sqrt{\pi}\sqrt{\Delta^{2}_{0} - w^{2}\cos^{2}\left(Q\right) }}, \\
		\beta^{\pm}_{1} = & \frac{\Delta_{0} B^{\pm}_{1}}{\sqrt{\pi}\sqrt{\Delta^{2}_{0} - w^{2}\cos^{2}\left(Q\right)}} 
	\end{align}
	with 
	\begin{align}
		A^{\pm}_{1} =& \frac{\left(1 - \left|z^{*}_{\pm}\right|^{2}\right) \cos\left(Q\right) \mp \frac{\mu}{w}\left|z^{*}_{\pm}\right| }{ \sqrt{ \left|z^{*}_{\pm}\right| \left( \left|z^{*}_{+}\right| + \left|z^{*}_{-}\right|       \right) \left(  \frac{1}{\left|z^{*}_{\mp}\right|}  + \left|z^{*}_{\pm}\right|   \right) \left( \frac{1}{\left|z^{*}_{\pm}\right|}  - \left|z^{*}_{\pm}\right|    \right)   }        } \\
		B^{\pm}_{1} =& \frac{1- \left|z^{*}_{\pm}\right|^{2}}{ \sqrt{ \left|z^{*}_{\pm}\right| \left( \left|z^{*}_{+}\right| + \left|z^{*}_{-}\right|       \right) \left(  \frac{1}{\left|z^{*}_{\mp}\right|}  + \left|z^{*}_{\pm}\right|   \right) \left( \frac{1}{\left|z^{*}_{\pm}\right|}  - \left|z^{*}_{\pm}\right|    \right)   }        }.
	\end{align}

\noindent (b): {\it $z^*_\pm$ real with the same sign}\\
This configuration occurs when the parameters fulfill Eqs.(\ref{cond-(a)-2}). In this case  one has $
z^{**}_{+}=1/z^*_{-}$ and $z^{**}_{-}=1/z^*_{+}$. 
The values of $\mathcal{D}(z)$ along the inner branch cut in Fig.\ref{Fig10}(b) are 
\begin{equation}
\mathcal{D}( x \pm i\epsilon  ) =  \pm i \, \sigma^*\, D_{2}(x)  \label{square-root-values-(a)-2} 
\end{equation}
where $\sigma^*$ is given by Eq.(\ref{sigma-star-def}), and
\begin{eqnarray}
\lefteqn{D_{2}(x) =  \sqrt{
\frac{w^2\cos^2{Q}-\Delta_0^2}{{x^*}^2+y^2  } } \times } & & \nonumber \\
& & \times \sqrt{(z^*_{+}-x) \,   (x-z^*_{-})\,   (z^{**}_{+}-x)\,   \, (z^{**}_{-}-x) }\quad. \label{D(x)-def-(a)-2}
\end{eqnarray}
Inserting Eqs.(\ref{square-root-values-(a)-2})-(\ref{D(x)-def-(a)-2}) into Eq.(\ref{Cp-cmplx-bc}), and denoting
\begin{equation}
\begin{array}{lcl}
x_M&=& \mbox{max}(|z^*_+|,|z^*_-|) \\
x_m&=& \mbox{min}(|z^*_+|,|z^*_-|)
\end{array}
\end{equation}
one obtains
\begin{eqnarray}
\lefteqn{\mathcal{C}_p(l) = \displaystyle  \frac{1}{2\pi} \frac{{\sigma^*}^{l-1}}{\sqrt{ \cos^2{Q}-\frac{\Delta_0^2}{w^2}} } \, \times} & &\label{Cp-complex-1b-pre} \\
&  & \int_{x_m}^{x_M} \! \! \! dx  \frac{x^{l-1}\,  \left((1+x^2 ) \cos{{Q}}-\frac{\mu}{w} x \, \sigma^*\right)    }{ \sqrt{\,(x_M-x) \,   (x-x_m)\,   (\frac{1}{x_M}-x)\,  ( \frac{1}{x_m}-x)    }}   \nonumber 
\end{eqnarray}
A similar expression is obtained for $\mathcal{A}(l)$. Applying the same approximation strategy as in the case (1) for $l \gg 1$, one obtains Eqs.(\ref{Cp-asym-(a)-2}) and  (\ref{A-asym-(a)-2}), where
\begin{eqnarray}
\alpha_{M} &=& \frac{1}{\sqrt{ \cos^2{Q}-\frac{\Delta_0^2}{w^2}} }   \, \frac{(1+x_M^2 ) \cos{{Q}}-\frac{\mu}{w} x_M \sigma^*    }{\sqrt{  \pi \,   (x_M-x_m)      \, ( \frac{1}{x_m}-x_M) \,(1-x_M^2)\,    }} \nonumber \\
& & \label{AM-def-(a)-2}\\
\alpha_{m} &=& \frac{1}{\sqrt{ \cos^2{Q}-\frac{\Delta_0^2}{w^2}} }   \, \frac{(1+x_m^2 ) \cos{{Q}}-\frac{\mu}{w} x_m \sigma^*    }{\sqrt{  \pi \,   (x_M-x_m)      \, ( \frac{1}{x_M}-x_m) \,(1-x_m^2)\,    }}\nonumber \\
& & \label{Am-def-(a)-2} \\
\beta_{M} &=& \frac{\frac{\Delta_0}{w}}{\sqrt{ \cos^2{Q}-\frac{\Delta_0^2}{w^2}} }   \, \sqrt{  \frac{1-x_M^2        }{\pi \,   (x_M-x_m)      \, ( \frac{1}{x_m}-x_M)      }} \nonumber \\
& & \label{AM-def-(a)-2}\\
\beta_{m} &=& \frac{\frac{\Delta_0}{w}}{\sqrt{ \cos^2{Q}-\frac{\Delta_0^2}{w^2}} }   \, \sqrt{  \frac{ 1-x_m^2 }{\pi \,   (x_M-x_m)      \, ( \frac{1}{x_M}-x_m)  }}\nonumber \\
& & \label{Am-def-(a)-2} 
\end{eqnarray}
\noindent (c): {\it $z^*_\pm=x^*\pm i y^*$ a complex conjugate pair} \\
This configuration occurs in the parameter range (\ref{cond-(b)}). In this case $z^{**}_{+}=1/z^*_{-}$ and $   z^{**}_{-}=1/z^*_{+}$, as displayed in Fig.\ref{Fig10}(c). The real and imaginary parts of the inner roots are given in Eqs.(\ref{x*-(b)})-(\ref{y*-(b)}) of the main text, while for the outer roots $z^{**}_\pm= x^{**}\pm i y^{**}$,  one has
\begin{eqnarray}
x^{**}&=& \displaystyle \frac{\frac{\mu}{2}\, \mbox{sgn}(\cos{{Q}})}{w|\cos{{Q}}|-\Delta_0} \\ y^{**}&=&  \frac{\sqrt{w^2\cos^2{Q}-\Delta_0^2-\frac{\mu^2}{4}}}{w|\cos{{Q}}|-\Delta_0}  \quad.
\end{eqnarray}
The values of $\mathcal{D}(z)$ along the branch cut are given by
\begin{equation}
\mathcal{D}(x^* \pm \epsilon+i y)=\pm\, D_{3}(y) e^{-i ( \Delta \varphi(y)+ \theta(y) )} \label{square-root-f-g-values-(b)} 
\end{equation}
where
\begin{eqnarray}
 D_{3}(y) &=& \sqrt{
\frac{w^2\cos^2{Q}-\Delta_0^2}{{x^*}^2+y^2  } } \times \nonumber \\
& & \times \sqrt{\,\rho_{+}(y) \, \,   \rho_{-}(y)\,    \, r_{+}(y)\,  \, r_{-}(y)     
 }  \label{square-root-values-(b)} 
 \end{eqnarray}
with
\begin{eqnarray}
\rho_{\pm}(y) &=&   y^* \mp y  \,  \\
r_{\pm}(y) &=& \, \,  \sqrt{(x^*-x^{**})^2+(y\mp y^{**})^2} \, \label{rpm(y)-def}  \\
|z(y)| &=& \sqrt{{x^*}^2+y^2} \\
\theta(y) &=& \mbox{arg}(x^*+ i y) \,\in [-\pi, \pi]
  \label{theta(y)-def}\\
\Delta \varphi(y) &=&   \frac{1}{2} \left\{\arctan \left( \frac{x^{**}-x^*}{y^{**}-y}\right)-\right. \nonumber \\
& & \left. \hspace{1cm} \arctan \left( \frac{x^{**}-x^*}{y^{**}+y}\right) \right)\label{varphi(y)-def}
\end{eqnarray}
Inserting Eqs.(\ref{square-root-f-g-values-(b)})-(\ref{square-root-values-(b)}) into Eq.(\ref{Cp-cmplx-bc}) one obtains
\begin{widetext}
\begin{eqnarray} 
\mathcal{C}_p(l) &=& \displaystyle    \frac{1}{\pi}   \frac{1}{\sqrt{\cos^2{Q}-\frac{\Delta_0^2}{w^2}}}  \,   \int_{0}^{y^*}     \frac{|z(y)|^{l}}{\sqrt{
({y^*}^2-y^2) \,   \, r_{+}(y)\,  \, r_{-}(y)     
 }} \,\, \times     \nonumber  \\ &  & \nonumber\\
&  &\hspace{1cm} \displaystyle    \times \left\{     \sin\left[\Delta\varphi(y)+l \, \theta(y) \right]   \,   \cos{{Q}} \,(|z(y)|-|z(y)|^{-1})\sin{\theta(y)}      - \right.  \nonumber\\
& &\displaystyle    \hspace{1.2cm} \left. -\cos\left[\Delta\varphi(y)+l \, \theta(y) \right]    \left(     \cos{{Q}} \,(|z(y)|+|z(y)|^{-1})\cos{\theta(y)}   -\frac{\mu}{w} \right)   
\right\} \,dy  = \nonumber\\
&=& \int_{0}^{y^*}  \frac{({x^*}^2+y^2)^{\frac{l-1}{2}}}{\sqrt{y^*-y}} \, \left\{ F^n_{s}(y) \sin[l \theta(y)]+F^n_{c}(y) \cos[l \theta(y)] \right\}   \, dy 
\label{Cpasym-(b)-pre1}
\end{eqnarray} 
where
\begin{eqnarray} 
F^n_{s}(y)   &=&   \frac{1}{\pi}   \frac{1}{\sqrt{\cos^2{Q}-\frac{\Delta_0^2}{w^2}}}  \,     \frac{1}{\sqrt{
(y^* +y) \,   \, r_{+}(y)\,  \, r_{-}(y)     
 }}  \times   \nonumber   \\
&  &   \times \left\{     \cos\left[\Delta\varphi(y)  \right]  \,   \cos{{Q}} \,\,(|z(y)|^2-1)\sin{\theta(y)}  +  \sin\left[\Delta\varphi(y)\right]    \left(     \cos{{Q}} \,(|z(y)|^2+1 )\cos{\theta(y)}   -\frac{\mu}{w} \right)   
\right\} \label{Fsnorm-def}\\
%%%%%%%%%
F^n_{c}(y)   &=&   \frac{1}{\pi}   \frac{1}{\sqrt{\cos^2{Q}-\frac{\Delta_0^2}{w^2}}}  \,     \frac{1}{\sqrt{
(y^* +y) \,   \, r_{+}(y)\,  \, r_{-}(y)     
 }}  \times   \nonumber   \\
&  & \,\,  \times \left\{     \sin\left[\Delta\varphi(y)  \right]  \,   \cos{{Q}} \,\,(|z(y)|^2-1)\sin{\theta(y)}     - \cos\left[\Delta\varphi(y)\right]    \left(     \cos{{Q}} \,(|z(y)|^2+1 )\cos{\theta(y)}   -\frac{\mu}{w} \right)   
\right\} \label{Fcnorm-def} 
\end{eqnarray} 
\end{widetext}
A similar expression can be obtained for $\mathcal{A}(l)$. In the regime $|x^*| \ll y^*$, we observe that, for $l \gg 1$, the function $({x^*}^2+y^2)^{\frac{l-1}{2}}$    suppresses  exponentially in $l$  the contribution from $F^{n}_{c/s}(y)$  away from $y \simeq y^*$. Therefore, we can approximate Eq.(\ref{Cpasym-(b)-pre1}) as
\begin{eqnarray} 
\mathcal{C}_p(l)     
 &\sim & \left\{  {F}^n_{s}(y^*) \sin[l  \theta^*]+ {F}^n_{c}(y^*) \cos[l  \theta^*] \right\}\,\, \times \nonumber \\
 & & \,\times \int_{0}^{y^*}  \frac{({x^*}^2+y^2)^{\frac{l-1}{2}}}{\sqrt{y^*-y}} \,    \, dy   
\label{Cpasym-(b)-pre3}
\end{eqnarray}
where
\begin{equation}
\theta^*=\theta(y^*)=\mbox{arg}(z^*_{+}) \quad.
\end{equation}
Moreover,  again for $|x^*| \ll y^*$, the integral  appearing in Eq.(\ref{Cpasym-(b)-pre3}) can be fairly well approximated as
\begin{eqnarray}
\int_{0}^{y^*}  \frac{({x^*}^2+y^2)^{\frac{l-1}{2}}}{\sqrt{y^*-y}} \,    \, dy   
&\sim  &     ({x^*}^2+{y^*}^2)^{\frac{l}{2}}\,   \sqrt{\frac{\pi}{{y^*}\,l} }\, = \nonumber \\
&=& |z^*_{+}|^{l}\,   \sqrt{\frac{\pi}{{y^*}\,l} }\label{approx-unknown}
\end{eqnarray}
Inserting Eq.(\ref{approx-unknown}) into Eq.(\ref{Cpasym-(b)-pre3}) and proceeding in a similar way for $\mathcal{A}(l)$, one obtains Eqs.(\ref{Cp-asym-(b)}) and (\ref{A-asym-(b)}), where
\begin{widetext}
\begin{eqnarray}
%%%%%%%%
\alpha^s_3    &=&   \frac{1}{\sqrt{2 \pi}}   \frac{1}{\sqrt{\cos^2{Q}-\frac{\Delta_0^2}{w^2}}}  \,     \frac{1}{\sqrt{
 {y^*}^2   \,   \, r_{+}(y^*)\,  \, r_{-}(y^*)     
 }}   \times \left\{     \cos\left[\Delta\varphi(y^*)  \right]  \,   \cos{{Q}} \,\,(|z^*_{+}|^2-1)\sin{\theta^*}  + \right. \nonumber \\
& & \hspace{7cm} \left. +\sin\left[\Delta\varphi(y^*)\right]    \left(     \cos{{Q}} \,(|z^*_{+}|^2+1 )\cos{\theta^*}   -\frac{\mu}{w} \right)   
\right\}  \label{Abs-def}\\
\alpha^c_3    &=&   \frac{1}{\sqrt{2\pi}}   \frac{1}{\sqrt{\cos^2{Q}-\frac{\Delta_0^2}{w^2}}}  \,     \frac{1}{\sqrt{
{y^*}^2 \,   \, r_{+}(y^*)\,  \, r_{-}(y^*)     
 }}  \times     \left\{     \sin\left[\Delta\varphi(y^*)  \right]  \,   \cos{{Q}} \,\,(|z^*_{+}|^2-1)\sin{\theta^*}     - \right. \nonumber \\
& & \hspace{7cm} \left. -\cos\left[\Delta\varphi(y^*)\right]    \left(     \cos{{Q}} \,\left( \left| z^*_{+} \right|^{2}+1 \right)\cos{\theta^*}   -\frac{\mu}{w} \right)   
\right\} \label{Abc-def} \\
%%%%%%%%%
\beta^s_3 &=& -\frac{1}{\sqrt{ 2\pi \, }} \frac{\frac{\Delta_0}{w} }{\sqrt{ \cos^2{Q}-\frac{\Delta_0^2}{w^2}}}\frac{1}{\sqrt{
  {y^*}^2  \,   \, r_{+}(y^*)\,  \, r_{-}(y^*)     
 } }  \times    \nonumber \\
 & & \hspace{1.2cm} \times \left\{     \cos\left[\Delta\varphi(y^*)  \right]  \,(|z^*_{+}|^2+1) \, \sin{\theta^*}    +\sin\left[\Delta\varphi(y^*)\right]        \,(|z^*_{+}|^2-1 )\cos{\theta^*}    
\right\} \label{Bbs-def} \\
\beta^c_3 &=& -\frac{1}{\sqrt{ 2\pi \, }} \frac{\frac{\Delta_0}{w} }{\sqrt{ \cos^2{Q}-\frac{\Delta_0^2}{w^2}}}\frac{1 }{\sqrt{
  {y^*}^2  \,   \, r_{+}(y^*)\,  \, r_{-}(y^*)     
 } }   \times   \nonumber  \\
 & & \hspace{1.2cm} \times \left\{     \sin\left[\Delta\varphi(y^*)  \right]  \,    \,\,(|z^*_{+}|^2+1)\sin{\theta^*}     -  \cos\left[\Delta\varphi(y^*)\right]      \,(|z^*_{+}|^2-1 )\cos{\theta^*}      
\right\}  \label{Bbc-def}
\end{eqnarray}
and $r_\pm(y)$ and $\Delta \varphi(y)$ are given in Eqs.(\ref{rpm(y)-def}) and (\ref{varphi(y)-def}), respectively.
\end{widetext}
%%%%%%%%%%%%%%%%%%%%%%%%%%%%%%%%%
\section{Asymptotic expansions of the correlation functions in the Gapless Phase}
\label{AppD}
    In this Appendix, we provide details about the derivation of the asymptotic expansions  of the correlation functions found in the gapless phase, and given  in Sec.\ref{sec5-2} [see Eqs.(\ref{Cp-asym-gapless}) and (\ref{A-asym-gapless})]. 
    For definiteness, we shall provide the derivation for $\mathcal{C}_p$, as the one for $\mathcal{A}$ follows along the same lines.

    As mentioned in Appendix \ref{AppB}, recalling that in the gapless phase  the $S_p$ domain is given by Eq.(\ref{Sp-gapless}) and exploiting the symmetry under $k \rightarrow -k$ of both the integrand function in Eq.(\ref{Cp-k})  and of the integration domain $S_p$, one can rewrite the Eq.(\ref{Cp-k}) as 
    \begin{eqnarray}
    \mathcal{C}_{p}&=& - \frac{1}{2\pi} \intop^{\left|k^{*}_{-}\right|}_{0} dk \cos(kl)\frac{\xi(k;Q,\mu)}{h(k;Q,\mu)} \nonumber\\
    & & - \frac{1}{2\pi} \intop^{\pi}_{\pi - \left|k^{*}_{+}\right|}  dk \cos(kl)\frac{\xi(k;Q,\mu)}{h(k;Q,\mu)} 
    \end{eqnarray}
    By applying a change of variable  $k  \rightarrow \pi - k$  in the second term, one obtains 
    \begin{equation}
    \mathcal{C}_{p}(l) = -\frac{1}{2\pi} \text{Re}\left\{\mathcal{I}_{-} - (-1)^{l}\mathcal{I}_{+}\right\}, \label{Cp-gapless-asym-pre}
    \end{equation}
    where
    \begin{equation}
    \mathcal{I}_{\pm} = \intop^{\left|k^{*}_{\pm}\right|}_{0} dk \, e^{ikl}{F}_{\pm}(k),
    \label{Integrals_c}
    \end{equation}
    and
    \begin{align}
    {F}_{\pm}(k) =& \frac{\xi_{\pm}(k;Q,  \mu)}{h_{\pm}(k;Q, \mu)} \nonumber \\
    =& \frac{  2w\cos(k)\cos(Q)  \pm \mu      }{  \sqrt{ ( 2w \cos(k)\cos(Q) \pm \mu     )^{2}   + \left| \Delta(k)\right|^{2}     }    } \quad.
    \end{align}
    For $l \gg 1$, one can apply the   stationary phase method to Eq.(\ref{Cp-gapless-asym-pre}), obtaining
    \begin{equation}
    \mathcal{I}_{\pm} \sim \sum_{n = 0} \frac{(-1)^{n}}{(i l)^{n+1}} \left\{ {F}^{(n)}_{\pm}\left(\left|k^{*}_{\pm}\right|\right)e^{i\left|k^{*}_{\pm}\right|l} - {F}^{(n)}_{\pm}(0)  \right\}
    \end{equation}
    To  leading order in the asymptotic expansion  ($n = 0$) one obtains the Eq.(\ref{Cp-asym-gapless}) given in the main text. Following the very same lines, one also obtains the  asymptotic expansion  Eq.(\ref{A-asym-gapless}) given for the anomalous correlation function.

	\bibliography{Biblio}

\begin{thebibliography}{86}%
\makeatletter
\providecommand \@ifxundefined [1]{%
 \@ifx{#1\undefined}
}%
\providecommand \@ifnum [1]{%
 \ifnum #1\expandafter \@firstoftwo
 \else \expandafter \@secondoftwo
 \fi
}%
\providecommand \@ifx [1]{%
 \ifx #1\expandafter \@firstoftwo
 \else \expandafter \@secondoftwo
 \fi
}%
\providecommand \natexlab [1]{#1}%
\providecommand \enquote  [1]{``#1''}%
\providecommand \bibnamefont  [1]{#1}%
\providecommand \bibfnamefont [1]{#1}%
\providecommand \citenamefont [1]{#1}%
\providecommand \href@noop [0]{\@secondoftwo}%
\providecommand \href [0]{\begingroup \@sanitize@url \@href}%
\providecommand \@href[1]{\@@startlink{#1}\@@href}%
\providecommand \@@href[1]{\endgroup#1\@@endlink}%
\providecommand \@sanitize@url [0]{\catcode `\\12\catcode `\$12\catcode
  `\&12\catcode `\#12\catcode `\^12\catcode `\_12\catcode `\%12\relax}%
\providecommand \@@startlink[1]{}%
\providecommand \@@endlink[0]{}%
\providecommand \url  [0]{\begingroup\@sanitize@url \@url }%
\providecommand \@url [1]{\endgroup\@href {#1}{\urlprefix }}%
\providecommand \urlprefix  [0]{URL }%
\providecommand \Eprint [0]{\href }%
\providecommand \doibase [0]{https://doi.org/}%
\providecommand \selectlanguage [0]{\@gobble}%
\providecommand \bibinfo  [0]{\@secondoftwo}%
\providecommand \bibfield  [0]{\@secondoftwo}%
\providecommand \translation [1]{[#1]}%
\providecommand \BibitemOpen [0]{}%
\providecommand \bibitemStop [0]{}%
\providecommand \bibitemNoStop [0]{.\EOS\space}%
\providecommand \EOS [0]{\spacefactor3000\relax}%
\providecommand \BibitemShut  [1]{\csname bibitem#1\endcsname}%
\let\auto@bib@innerbib\@empty
%</preamble>
\bibitem [{\citenamefont {Wen}(2004)}]{wen-book}%
  \BibitemOpen
  \bibfield  {author} {\bibinfo {author} {\bibfnamefont {X.-G.}\ \bibnamefont
  {Wen}},\ }\href@noop {} {\emph {\bibinfo {title} {Quantum {F}ield {T}heory of
  {M}any-{B}ody {S}ystems: From the {O}rigin of {S}ound to an {O}rigin of
  {L}ight and {E}lectrons}}}\ (\bibinfo  {publisher} {Oxford university
  press},\ \bibinfo {year} {2004})\BibitemShut {NoStop}%
\bibitem [{\citenamefont {Chen}\ \emph {et~al.}(2010)\citenamefont {Chen},
  \citenamefont {Gu},\ and\ \citenamefont {Wen}}]{wen_PRB_2010}%
  \BibitemOpen
  \bibfield  {author} {\bibinfo {author} {\bibfnamefont {X.}~\bibnamefont
  {Chen}}, \bibinfo {author} {\bibfnamefont {Z.-C.}\ \bibnamefont {Gu}},\ and\
  \bibinfo {author} {\bibfnamefont {X.-G.}\ \bibnamefont {Wen}},\ }\bibfield
  {title} {\bibinfo {title} {Local unitary transformation, long-range quantum
  entanglement, wave function renormalization, and topological order},\ }\href
  {https://doi.org/10.1103/PhysRevB.82.155138} {\bibfield  {journal} {\bibinfo
  {journal} {Phys. Rev. B}\ }\textbf {\bibinfo {volume} {82}},\ \bibinfo
  {pages} {155138} (\bibinfo {year} {2010})}\BibitemShut {NoStop}%
\bibitem [{\citenamefont {Nussinov}\ and\ \citenamefont
  {Ortiz}(2009)}]{ortiz-review}%
  \BibitemOpen
  \bibfield  {author} {\bibinfo {author} {\bibfnamefont {Z.}~\bibnamefont
  {Nussinov}}\ and\ \bibinfo {author} {\bibfnamefont {G.}~\bibnamefont
  {Ortiz}},\ }\bibfield  {title} {\bibinfo {title} {A symmetry principle for
  topological quantum order},\ }\href
  {https://doi.org/https://doi.org/10.1016/j.aop.2008.11.002} {\bibfield
  {journal} {\bibinfo  {journal} {Ann. Phys. (NY)}\ }\textbf {\bibinfo {volume}
  {324}},\ \bibinfo {pages} {977} (\bibinfo {year} {2009})}\BibitemShut
  {NoStop}%
\bibitem [{\citenamefont {Hasan}\ and\ \citenamefont {Kane}(2010)}]{hasan2010}%
  \BibitemOpen
  \bibfield  {author} {\bibinfo {author} {\bibfnamefont {M.~Z.}\ \bibnamefont
  {Hasan}}\ and\ \bibinfo {author} {\bibfnamefont {C.~L.}\ \bibnamefont
  {Kane}},\ }\bibfield  {title} {\bibinfo {title} {Colloquium: Topological
  insulators},\ }\href {https://doi.org/10.1103/RevModPhys.82.3045} {\bibfield
  {journal} {\bibinfo  {journal} {Rev. Mod. Phys.}\ }\textbf {\bibinfo {volume}
  {82}},\ \bibinfo {pages} {3045} (\bibinfo {year} {2010})}\BibitemShut
  {NoStop}%
\bibitem [{\citenamefont {Qi}\ and\ \citenamefont
  {Zhang}(2011)}]{zhang-review}%
  \BibitemOpen
  \bibfield  {author} {\bibinfo {author} {\bibfnamefont {X.-L.}\ \bibnamefont
  {Qi}}\ and\ \bibinfo {author} {\bibfnamefont {S.-C.}\ \bibnamefont {Zhang}},\
  }\bibfield  {title} {\bibinfo {title} {Topological insulators and
  superconductors},\ }\href {https://doi.org/10.1103/RevModPhys.83.1057}
  {\bibfield  {journal} {\bibinfo  {journal} {Rev. Mod. Phys.}\ }\textbf
  {\bibinfo {volume} {83}},\ \bibinfo {pages} {1057} (\bibinfo {year}
  {2011})}\BibitemShut {NoStop}%
\bibitem [{\citenamefont {Chiu}\ \emph {et~al.}(2016)\citenamefont {Chiu},
  \citenamefont {Teo}, \citenamefont {Schnyder},\ and\ \citenamefont
  {Ryu}}]{chiu2016}%
  \BibitemOpen
  \bibfield  {author} {\bibinfo {author} {\bibfnamefont {C.-K.}\ \bibnamefont
  {Chiu}}, \bibinfo {author} {\bibfnamefont {J.~C.~Y.}\ \bibnamefont {Teo}},
  \bibinfo {author} {\bibfnamefont {A.~P.}\ \bibnamefont {Schnyder}},\ and\
  \bibinfo {author} {\bibfnamefont {S.}~\bibnamefont {Ryu}},\ }\bibfield
  {title} {\bibinfo {title} {Classification of topological quantum matter with
  symmetries},\ }\href {https://doi.org/10.1103/RevModPhys.88.035005}
  {\bibfield  {journal} {\bibinfo  {journal} {Rev. Mod. Phys.}\ }\textbf
  {\bibinfo {volume} {88}},\ \bibinfo {pages} {035005} (\bibinfo {year}
  {2016})}\BibitemShut {NoStop}%
\bibitem [{\citenamefont {Ringel}\ and\ \citenamefont
  {Kraus}(2011)}]{kraus_PRB_2011}%
  \BibitemOpen
  \bibfield  {author} {\bibinfo {author} {\bibfnamefont {Z.}~\bibnamefont
  {Ringel}}\ and\ \bibinfo {author} {\bibfnamefont {Y.~E.}\ \bibnamefont
  {Kraus}},\ }\bibfield  {title} {\bibinfo {title} {Determining topological
  order from a local ground-state correlation function},\ }\href
  {https://doi.org/10.1103/PhysRevB.83.245115} {\bibfield  {journal} {\bibinfo
  {journal} {Phys. Rev. B}\ }\textbf {\bibinfo {volume} {83}},\ \bibinfo
  {pages} {245115} (\bibinfo {year} {2011})}\BibitemShut {NoStop}%
\bibitem [{\citenamefont {Chen}\ \emph {et~al.}(2017)\citenamefont {Chen},
  \citenamefont {Legner}, \citenamefont {R\"uegg},\ and\ \citenamefont
  {Sigrist}}]{sigriest_PRB_2017}%
  \BibitemOpen
  \bibfield  {author} {\bibinfo {author} {\bibfnamefont {W.}~\bibnamefont
  {Chen}}, \bibinfo {author} {\bibfnamefont {M.}~\bibnamefont {Legner}},
  \bibinfo {author} {\bibfnamefont {A.}~\bibnamefont {R\"uegg}},\ and\ \bibinfo
  {author} {\bibfnamefont {M.}~\bibnamefont {Sigrist}},\ }\bibfield  {title}
  {\bibinfo {title} {Correlation length, universality classes, and scaling laws
  associated with topological phase transitions},\ }\href
  {https://doi.org/10.1103/PhysRevB.95.075116} {\bibfield  {journal} {\bibinfo
  {journal} {Phys. Rev. B}\ }\textbf {\bibinfo {volume} {95}},\ \bibinfo
  {pages} {075116} (\bibinfo {year} {2017})}\BibitemShut {NoStop}%
\bibitem [{\citenamefont {Chen}\ and\ \citenamefont
  {Schnyder}(2019)}]{schnyder_2019}%
  \BibitemOpen
  \bibfield  {author} {\bibinfo {author} {\bibfnamefont {W.}~\bibnamefont
  {Chen}}\ and\ \bibinfo {author} {\bibfnamefont {A.~P.}\ \bibnamefont
  {Schnyder}},\ }\bibfield  {title} {\bibinfo {title} {Universality classes of
  topological phase transitions with higher-order band crossing},\ }\href
  {https://doi.org/10.1088/1367-2630/ab2a2d} {\bibfield  {journal} {\bibinfo
  {journal} {N. J. Phys.}\ }\textbf {\bibinfo {volume} {21}},\ \bibinfo {pages}
  {073003} (\bibinfo {year} {2019})}\BibitemShut {NoStop}%
\bibitem [{\citenamefont {Castelnovo}\ and\ \citenamefont
  {Chamon}(2008)}]{chamon_PRB_2008}%
  \BibitemOpen
  \bibfield  {author} {\bibinfo {author} {\bibfnamefont {C.}~\bibnamefont
  {Castelnovo}}\ and\ \bibinfo {author} {\bibfnamefont {C.}~\bibnamefont
  {Chamon}},\ }\bibfield  {title} {\bibinfo {title} {Quantum topological phase
  transition at the microscopic level},\ }\href
  {https://doi.org/10.1103/PhysRevB.77.054433} {\bibfield  {journal} {\bibinfo
  {journal} {Phys. Rev. B}\ }\textbf {\bibinfo {volume} {77}},\ \bibinfo
  {pages} {054433} (\bibinfo {year} {2008})}\BibitemShut {NoStop}%
\bibitem [{\citenamefont {Abasto}\ \emph {et~al.}(2008)\citenamefont {Abasto},
  \citenamefont {Hamma},\ and\ \citenamefont {Zanardi}}]{zanardi_PRA_2008}%
  \BibitemOpen
  \bibfield  {author} {\bibinfo {author} {\bibfnamefont {D.~F.}\ \bibnamefont
  {Abasto}}, \bibinfo {author} {\bibfnamefont {A.}~\bibnamefont {Hamma}},\ and\
  \bibinfo {author} {\bibfnamefont {P.}~\bibnamefont {Zanardi}},\ }\bibfield
  {title} {\bibinfo {title} {Fidelity analysis of topological quantum phase
  transitions},\ }\href {https://doi.org/10.1103/PhysRevA.78.010301} {\bibfield
   {journal} {\bibinfo  {journal} {Phys. Rev. A}\ }\textbf {\bibinfo {volume}
  {78}},\ \bibinfo {pages} {010301(R)} (\bibinfo {year} {2008})}\BibitemShut
  {NoStop}%
\bibitem [{\citenamefont {Eriksson}\ and\ \citenamefont
  {Johannesson}(2009)}]{johannesson_PRA_2009}%
  \BibitemOpen
  \bibfield  {author} {\bibinfo {author} {\bibfnamefont {E.}~\bibnamefont
  {Eriksson}}\ and\ \bibinfo {author} {\bibfnamefont {H.}~\bibnamefont
  {Johannesson}},\ }\bibfield  {title} {\bibinfo {title} {Reduced fidelity in
  topological quantum phase transitions},\ }\href
  {https://doi.org/10.1103/PhysRevA.79.060301} {\bibfield  {journal} {\bibinfo
  {journal} {Phys. Rev. A}\ }\textbf {\bibinfo {volume} {79}},\ \bibinfo
  {pages} {060301(R)} (\bibinfo {year} {2009})}\BibitemShut {NoStop}%
\bibitem [{\citenamefont {Chen}\ and\ \citenamefont {Li}(2010)}]{li_PRA_2010}%
  \BibitemOpen
  \bibfield  {author} {\bibinfo {author} {\bibfnamefont {Y.-X.}\ \bibnamefont
  {Chen}}\ and\ \bibinfo {author} {\bibfnamefont {S.-W.}\ \bibnamefont {Li}},\
  }\bibfield  {title} {\bibinfo {title} {Quantum correlations in topological
  quantum phase transitions},\ }\href
  {https://doi.org/10.1103/PhysRevA.81.032120} {\bibfield  {journal} {\bibinfo
  {journal} {Phys. Rev. A}\ }\textbf {\bibinfo {volume} {81}},\ \bibinfo
  {pages} {032120} (\bibinfo {year} {2010})}\BibitemShut {NoStop}%
\bibitem [{\citenamefont {Tibaldi}\ \emph {et~al.}(2023)\citenamefont
  {Tibaldi}, \citenamefont {Magnifico}, \citenamefont {Vodola},\ and\
  \citenamefont {Ercolessi}}]{ercolessi_scipost_2023}%
  \BibitemOpen
  \bibfield  {author} {\bibinfo {author} {\bibfnamefont {S.}~\bibnamefont
  {Tibaldi}}, \bibinfo {author} {\bibfnamefont {G.}~\bibnamefont {Magnifico}},
  \bibinfo {author} {\bibfnamefont {D.}~\bibnamefont {Vodola}},\ and\ \bibinfo
  {author} {\bibfnamefont {E.}~\bibnamefont {Ercolessi}},\ }\bibfield  {title}
  {\bibinfo {title} {Unsupervised and supervised learning of interacting
  topological phases from single-particle correlation functions},\ }\href
  {https://doi.org/10.21468/SciPostPhys.14.1.005} {\bibfield  {journal}
  {\bibinfo  {journal} {SciPost Physics}\ }\textbf {\bibinfo {volume} {14}},\
  \bibinfo {pages} {005} (\bibinfo {year} {2023})}\BibitemShut {NoStop}%
\bibitem [{\citenamefont {Kitaev}(2001)}]{kitaev2001}%
  \BibitemOpen
  \bibfield  {author} {\bibinfo {author} {\bibfnamefont {A.~Y.}\ \bibnamefont
  {Kitaev}},\ }\bibfield  {title} {\bibinfo {title} {Unpaired majorana fermions
  in quantum wires},\ }\href {https://doi.org/10.1070/1063-7869/44/10S/S29}
  {\bibfield  {journal} {\bibinfo  {journal} {Phys. Usp.}\ }\textbf {\bibinfo
  {volume} {44}},\ \bibinfo {pages} {131} (\bibinfo {year} {2001})}\BibitemShut
  {NoStop}%
\bibitem [{\citenamefont {Kitaev}(2003)}]{kitaev2003}%
  \BibitemOpen
  \bibfield  {author} {\bibinfo {author} {\bibfnamefont {A.~Y.}\ \bibnamefont
  {Kitaev}},\ }\bibfield  {title} {\bibinfo {title} {Fault-tolerant quantum
  computation by anyons},\ }\href
  {https://doi.org/https://doi.org/10.1016/S0003-4916(02)00018-0} {\bibfield
  {journal} {\bibinfo  {journal} {Ann. Phys. (NY)}\ }\textbf {\bibinfo {volume}
  {303}},\ \bibinfo {pages} {2} (\bibinfo {year} {2003})}\BibitemShut {NoStop}%
\bibitem [{\citenamefont {Alicea}(2012)}]{alicea2012}%
  \BibitemOpen
  \bibfield  {author} {\bibinfo {author} {\bibfnamefont {J.}~\bibnamefont
  {Alicea}},\ }\bibfield  {title} {\bibinfo {title} {New directions in the
  pursuit of {Majorana} fermions in solid state systems},\ }\href
  {https://doi.org/10.1088/0034-4885/75/7/076501} {\bibfield  {journal}
  {\bibinfo  {journal} {Rep. Prog. Phys.}\ }\textbf {\bibinfo {volume} {75}},\
  \bibinfo {pages} {076501} (\bibinfo {year} {2012})}\BibitemShut {NoStop}%
\bibitem [{\citenamefont {Aguado}(2017)}]{aguado2017}%
  \BibitemOpen
  \bibfield  {author} {\bibinfo {author} {\bibfnamefont {R.}~\bibnamefont
  {Aguado}},\ }\bibfield  {title} {\bibinfo {title} {Majorana quasiparticles in
  condensed matter},\ }\href {https://doi.org/10.1393/ncr/i2017-10141-9}
  {\bibfield  {journal} {\bibinfo  {journal} {La Rivista del Nuovo Cimento}\
  }\textbf {\bibinfo {volume} {40}},\ \bibinfo {pages} {523} (\bibinfo {year}
  {2017})}\BibitemShut {NoStop}%
\bibitem [{\citenamefont {Lian}\ \emph {et~al.}(2018)\citenamefont {Lian},
  \citenamefont {Sun}, \citenamefont {Vaezi}, \citenamefont {Qi},\ and\
  \citenamefont {Zhang}}]{zhang2018}%
  \BibitemOpen
  \bibfield  {author} {\bibinfo {author} {\bibfnamefont {B.}~\bibnamefont
  {Lian}}, \bibinfo {author} {\bibfnamefont {X.-Q.}\ \bibnamefont {Sun}},
  \bibinfo {author} {\bibfnamefont {A.}~\bibnamefont {Vaezi}}, \bibinfo
  {author} {\bibfnamefont {X.-L.}\ \bibnamefont {Qi}},\ and\ \bibinfo {author}
  {\bibfnamefont {S.-C.}\ \bibnamefont {Zhang}},\ }\bibfield  {title} {\bibinfo
  {title} {Topological quantum computation based on chiral {Majorana}
  fermions},\ }\href {https://doi.org/10.1073/pnas.1810003115} {\bibfield
  {journal} {\bibinfo  {journal} {Proc. Nat. Acad. Sci.}\ }\textbf {\bibinfo
  {volume} {115}},\ \bibinfo {pages} {10938} (\bibinfo {year}
  {2018})}\BibitemShut {NoStop}%
\bibitem [{\citenamefont {Beenakker}(2020)}]{beenakker2020}%
  \BibitemOpen
  \bibfield  {author} {\bibinfo {author} {\bibfnamefont {C.}~\bibnamefont
  {Beenakker}},\ }\bibfield  {title} {\bibinfo {title} {Search for non-abelian
  majorana braiding statistics in superconductors},\ }\href
  {https://doi.org/10.21468/SciPostPhysLectNotes.15} {\bibfield  {journal}
  {\bibinfo  {journal} {SciPost Physics Lecture Notes}\ ,\ \bibinfo {pages}
  {015}} (\bibinfo {year} {2020})}\BibitemShut {NoStop}%
\bibitem [{\citenamefont {Das~Sarma}(2023)}]{dassarma2023}%
  \BibitemOpen
  \bibfield  {author} {\bibinfo {author} {\bibfnamefont {S.}~\bibnamefont
  {Das~Sarma}},\ }\bibfield  {title} {\bibinfo {title} {In search of
  {Majorana}},\ }\href
  {https://doi.org/https://doi.org/10.1038/s41567-022-01900-9} {\bibfield
  {journal} {\bibinfo  {journal} {Nat. Phys.}\ }\textbf {\bibinfo {volume}
  {19}},\ \bibinfo {pages} {165} (\bibinfo {year} {2023})}\BibitemShut
  {NoStop}%
\bibitem [{\citenamefont {Aghaee}\ \emph {et~al.}(2023)\citenamefont {Aghaee}
  \emph {et~al.}}]{aghaee2023}%
  \BibitemOpen
  \bibfield  {author} {\bibinfo {author} {\bibfnamefont {M.}~\bibnamefont
  {Aghaee}} \emph {et~al.} (\bibinfo {collaboration} {Microsoft Quantum}),\
  }\bibfield  {title} {\bibinfo {title} {{InAs}-{Al} hybrid devices passing the
  topological gap protocol},\ }\href
  {https://doi.org/10.1103/PhysRevB.107.245423} {\bibfield  {journal} {\bibinfo
   {journal} {Phys. Rev. B}\ }\textbf {\bibinfo {volume} {107}},\ \bibinfo
  {pages} {245423} (\bibinfo {year} {2023})}\BibitemShut {NoStop}%
\bibitem [{\citenamefont {Fu}\ and\ \citenamefont {Kane}(2008)}]{fu-kane2008}%
  \BibitemOpen
  \bibfield  {author} {\bibinfo {author} {\bibfnamefont {L.}~\bibnamefont
  {Fu}}\ and\ \bibinfo {author} {\bibfnamefont {C.~L.}\ \bibnamefont {Kane}},\
  }\bibfield  {title} {\bibinfo {title} {Superconducting proximity effect and
  {Majorana} fermions at the surface of a topological insulator},\ }\href
  {https://doi.org/10.1103/PhysRevLett.100.096407} {\bibfield  {journal}
  {\bibinfo  {journal} {Phys. Rev. Lett.}\ }\textbf {\bibinfo {volume} {100}},\
  \bibinfo {pages} {096407} (\bibinfo {year} {2008})}\BibitemShut {NoStop}%
\bibitem [{\citenamefont {Fu}\ and\ \citenamefont {Kane}(2009)}]{fu-kane2009}%
  \BibitemOpen
  \bibfield  {author} {\bibinfo {author} {\bibfnamefont {L.}~\bibnamefont
  {Fu}}\ and\ \bibinfo {author} {\bibfnamefont {C.~L.}\ \bibnamefont {Kane}},\
  }\bibfield  {title} {\bibinfo {title} {Josephson current and noise at a
  superconductor/quantum-spin-{Hall}-insulator/superconductor junction},\
  }\href {https://doi.org/10.1103/PhysRevB.79.161408} {\bibfield  {journal}
  {\bibinfo  {journal} {Phys. Rev. B}\ }\textbf {\bibinfo {volume} {79}},\
  \bibinfo {pages} {161408(R)} (\bibinfo {year} {2009})}\BibitemShut {NoStop}%
\bibitem [{\citenamefont {Lutchyn}\ \emph {et~al.}(2010)\citenamefont
  {Lutchyn}, \citenamefont {Sau},\ and\ \citenamefont
  {Das~Sarma}}]{lutchyn2010}%
  \BibitemOpen
  \bibfield  {author} {\bibinfo {author} {\bibfnamefont {R.~M.}\ \bibnamefont
  {Lutchyn}}, \bibinfo {author} {\bibfnamefont {J.~D.}\ \bibnamefont {Sau}},\
  and\ \bibinfo {author} {\bibfnamefont {S.}~\bibnamefont {Das~Sarma}},\
  }\bibfield  {title} {\bibinfo {title} {Majorana fermions and a topological
  phase transition in semiconductor-superconductor heterostructures},\ }\href
  {https://doi.org/https://doi.org/10.1103/PhysRevLett.105.077001} {\bibfield
  {journal} {\bibinfo  {journal} {Phys. Rev. Lett.}\ }\textbf {\bibinfo
  {volume} {105}},\ \bibinfo {pages} {077001} (\bibinfo {year}
  {2010})}\BibitemShut {NoStop}%
\bibitem [{\citenamefont {Oreg}\ \emph {et~al.}(2010)\citenamefont {Oreg},
  \citenamefont {Refael},\ and\ \citenamefont {von Oppen}}]{oreg2010}%
  \BibitemOpen
  \bibfield  {author} {\bibinfo {author} {\bibfnamefont {Y.}~\bibnamefont
  {Oreg}}, \bibinfo {author} {\bibfnamefont {G.}~\bibnamefont {Refael}},\ and\
  \bibinfo {author} {\bibfnamefont {F.}~\bibnamefont {von Oppen}},\ }\bibfield
  {title} {\bibinfo {title} {Helical liquids and {Majorana} bound states in
  quantum wires},\ }\href {https://doi.org/10.1103/PhysRevLett.105.177002}
  {\bibfield  {journal} {\bibinfo  {journal} {Phys. Rev. Lett.}\ }\textbf
  {\bibinfo {volume} {105}},\ \bibinfo {pages} {177002} (\bibinfo {year}
  {2010})}\BibitemShut {NoStop}%
\bibitem [{\citenamefont {Choy}\ \emph {et~al.}(2011)\citenamefont {Choy},
  \citenamefont {Edge}, \citenamefont {Akhmerov},\ and\ \citenamefont
  {Beenakker}}]{choy2011}%
  \BibitemOpen
  \bibfield  {author} {\bibinfo {author} {\bibfnamefont {T.-P.}\ \bibnamefont
  {Choy}}, \bibinfo {author} {\bibfnamefont {J.~M.}\ \bibnamefont {Edge}},
  \bibinfo {author} {\bibfnamefont {A.~R.}\ \bibnamefont {Akhmerov}},\ and\
  \bibinfo {author} {\bibfnamefont {C.~W.~J.}\ \bibnamefont {Beenakker}},\
  }\bibfield  {title} {\bibinfo {title} {Majorana fermions emerging from
  magnetic nanoparticles on a superconductor without spin-orbit coupling},\
  }\href {https://doi.org/10.1103/PhysRevB.84.195442} {\bibfield  {journal}
  {\bibinfo  {journal} {Phys. Rev. B}\ }\textbf {\bibinfo {volume} {84}},\
  \bibinfo {pages} {195442(R)} (\bibinfo {year} {2011})}\BibitemShut {NoStop}%
\bibitem [{\citenamefont {Nadj-Perge}\ \emph {et~al.}(2013)\citenamefont
  {Nadj-Perge}, \citenamefont {Drozdov}, \citenamefont {Bernevig},\ and\
  \citenamefont {Yazdani}}]{nadjperge2013}%
  \BibitemOpen
  \bibfield  {author} {\bibinfo {author} {\bibfnamefont {S.}~\bibnamefont
  {Nadj-Perge}}, \bibinfo {author} {\bibfnamefont {I.~K.}\ \bibnamefont
  {Drozdov}}, \bibinfo {author} {\bibfnamefont {B.~A.}\ \bibnamefont
  {Bernevig}},\ and\ \bibinfo {author} {\bibfnamefont {A.}~\bibnamefont
  {Yazdani}},\ }\bibfield  {title} {\bibinfo {title} {Proposal for realizing
  {Majorana} fermions in chains of magnetic atoms on a superconductor},\ }\href
  {https://doi.org/10.1103/PhysRevB.88.020407} {\bibfield  {journal} {\bibinfo
  {journal} {Phys. Rev. B}\ }\textbf {\bibinfo {volume} {88}},\ \bibinfo
  {pages} {020407(R)} (\bibinfo {year} {2013})}\BibitemShut {NoStop}%
\bibitem [{\citenamefont {Braunecker}\ and\ \citenamefont
  {Simon}(2013)}]{simon2013}%
  \BibitemOpen
  \bibfield  {author} {\bibinfo {author} {\bibfnamefont {B.}~\bibnamefont
  {Braunecker}}\ and\ \bibinfo {author} {\bibfnamefont {P.}~\bibnamefont
  {Simon}},\ }\bibfield  {title} {\bibinfo {title} {Interplay between classical
  magnetic moments and superconductivity in quantum one-dimensional conductors:
  Toward a self-sustained topological majorana phase},\ }\href
  {https://doi.org/10.1103/PhysRevLett.111.147202} {\bibfield  {journal}
  {\bibinfo  {journal} {Phys. Rev. Lett.}\ }\textbf {\bibinfo {volume} {111}},\
  \bibinfo {pages} {147202} (\bibinfo {year} {2013})}\BibitemShut {NoStop}%
\bibitem [{\citenamefont {Pientka}\ \emph {et~al.}(2013)\citenamefont
  {Pientka}, \citenamefont {Glazman},\ and\ \citenamefont {von
  Oppen}}]{glazman2013}%
  \BibitemOpen
  \bibfield  {author} {\bibinfo {author} {\bibfnamefont {F.}~\bibnamefont
  {Pientka}}, \bibinfo {author} {\bibfnamefont {L.~I.}\ \bibnamefont
  {Glazman}},\ and\ \bibinfo {author} {\bibfnamefont {F.}~\bibnamefont {von
  Oppen}},\ }\bibfield  {title} {\bibinfo {title} {Topological superconducting
  phase in helical {Shiba} chains},\ }\href
  {https://doi.org/10.1103/PhysRevB.88.155420} {\bibfield  {journal} {\bibinfo
  {journal} {Phys. Rev. B}\ }\textbf {\bibinfo {volume} {88}},\ \bibinfo
  {pages} {155420(R)} (\bibinfo {year} {2013})}\BibitemShut {NoStop}%
\bibitem [{\citenamefont {Vazifeh}\ and\ \citenamefont
  {Franz}(2013)}]{franz2013}%
  \BibitemOpen
  \bibfield  {author} {\bibinfo {author} {\bibfnamefont {M.~M.}\ \bibnamefont
  {Vazifeh}}\ and\ \bibinfo {author} {\bibfnamefont {M.}~\bibnamefont
  {Franz}},\ }\bibfield  {title} {\bibinfo {title} {Self-organized topological
  state with {Majorana} fermions},\ }\href
  {https://doi.org/10.1103/PhysRevLett.111.206802} {\bibfield  {journal}
  {\bibinfo  {journal} {Phys. Rev. Lett.}\ }\textbf {\bibinfo {volume} {111}},\
  \bibinfo {pages} {206802} (\bibinfo {year} {2013})}\BibitemShut {NoStop}%
\bibitem [{\citenamefont {Heimes}\ \emph {et~al.}(2014)\citenamefont {Heimes},
  \citenamefont {Kotetes},\ and\ \citenamefont {Sch\"on}}]{kotetes2014}%
  \BibitemOpen
  \bibfield  {author} {\bibinfo {author} {\bibfnamefont {A.}~\bibnamefont
  {Heimes}}, \bibinfo {author} {\bibfnamefont {P.}~\bibnamefont {Kotetes}},\
  and\ \bibinfo {author} {\bibfnamefont {G.}~\bibnamefont {Sch\"on}},\
  }\bibfield  {title} {\bibinfo {title} {Majorana fermions from shiba states in
  an antiferromagnetic chain on top of a superconductor},\ }\href
  {https://doi.org/10.1103/PhysRevB.90.060507} {\bibfield  {journal} {\bibinfo
  {journal} {Phys. Rev. B}\ }\textbf {\bibinfo {volume} {90}},\ \bibinfo
  {pages} {060507(R)} (\bibinfo {year} {2014})}\BibitemShut {NoStop}%
\bibitem [{\citenamefont {Kraus}\ \emph {et~al.}(2012)\citenamefont {Kraus},
  \citenamefont {Diehl}, \citenamefont {Zoller},\ and\ \citenamefont
  {Baranov}}]{kraus-zoller_2012}%
  \BibitemOpen
  \bibfield  {author} {\bibinfo {author} {\bibfnamefont {C.~V.}\ \bibnamefont
  {Kraus}}, \bibinfo {author} {\bibfnamefont {S.}~\bibnamefont {Diehl}},
  \bibinfo {author} {\bibfnamefont {P.}~\bibnamefont {Zoller}},\ and\ \bibinfo
  {author} {\bibfnamefont {M.~A.}\ \bibnamefont {Baranov}},\ }\bibfield
  {title} {\bibinfo {title} {Preparing and probing atomic majorana fermions and
  topological order in optical lattices},\ }\href
  {https://doi.org/10.1088/1367-2630/14/11/113036} {\bibfield  {journal}
  {\bibinfo  {journal} {New Journal of Physics}\ }\textbf {\bibinfo {volume}
  {14}},\ \bibinfo {pages} {113036} (\bibinfo {year} {2012})}\BibitemShut
  {NoStop}%
\bibitem [{\citenamefont {B{\"u}hler}\ \emph {et~al.}(2014)\citenamefont
  {B{\"u}hler}, \citenamefont {Lang}, \citenamefont {Kraus}, \citenamefont
  {M{\"o}ller}, \citenamefont {Huber},\ and\ \citenamefont
  {B{\"u}chler}}]{buchler_2014}%
  \BibitemOpen
  \bibfield  {author} {\bibinfo {author} {\bibfnamefont {A.}~\bibnamefont
  {B{\"u}hler}}, \bibinfo {author} {\bibfnamefont {N.}~\bibnamefont {Lang}},
  \bibinfo {author} {\bibfnamefont {C.~V.}\ \bibnamefont {Kraus}}, \bibinfo
  {author} {\bibfnamefont {G.}~\bibnamefont {M{\"o}ller}}, \bibinfo {author}
  {\bibfnamefont {S.~D.}\ \bibnamefont {Huber}},\ and\ \bibinfo {author}
  {\bibfnamefont {H.-P.}\ \bibnamefont {B{\"u}chler}},\ }\bibfield  {title}
  {\bibinfo {title} {Majorana modes and p-wave superfluids for fermionic atoms
  in optical lattices},\ }\href {https://doi.org/10.1038/ncomms5504} {\bibfield
   {journal} {\bibinfo  {journal} {Nat. Comm.}\ }\textbf {\bibinfo {volume}
  {5}},\ \bibinfo {pages} {4504} (\bibinfo {year} {2014})}\BibitemShut
  {NoStop}%
\bibitem [{\citenamefont {Fraxanet}\ \emph {et~al.}(2021)\citenamefont
  {Fraxanet}, \citenamefont {Bhattacharya}, \citenamefont {Grass},
  \citenamefont {Rakshit}, \citenamefont {Lewenstein},\ and\ \citenamefont
  {Dauphin}}]{lewenstein_2021}%
  \BibitemOpen
  \bibfield  {author} {\bibinfo {author} {\bibfnamefont {J.}~\bibnamefont
  {Fraxanet}}, \bibinfo {author} {\bibfnamefont {U.}~\bibnamefont
  {Bhattacharya}}, \bibinfo {author} {\bibfnamefont {T.}~\bibnamefont {Grass}},
  \bibinfo {author} {\bibfnamefont {D.}~\bibnamefont {Rakshit}}, \bibinfo
  {author} {\bibfnamefont {M.}~\bibnamefont {Lewenstein}},\ and\ \bibinfo
  {author} {\bibfnamefont {A.}~\bibnamefont {Dauphin}},\ }\bibfield  {title}
  {\bibinfo {title} {Topological properties of the long-range {K}itaev chain
  with aubry-andr\'e-harper modulation},\ }\href
  {https://doi.org/10.1103/PhysRevResearch.3.013148} {\bibfield  {journal}
  {\bibinfo  {journal} {Phys. Rev. Res.}\ }\textbf {\bibinfo {volume} {3}},\
  \bibinfo {pages} {013148} (\bibinfo {year} {2021})}\BibitemShut {NoStop}%
\bibitem [{\citenamefont {Mourik}\ \emph {et~al.}(2012)\citenamefont {Mourik},
  \citenamefont {Zuo}, \citenamefont {Frolov}, \citenamefont {Plissard},
  \citenamefont {Bakkers},\ and\ \citenamefont
  {Kouwenhoven}}]{kouwenhoven2012}%
  \BibitemOpen
  \bibfield  {author} {\bibinfo {author} {\bibfnamefont {V.}~\bibnamefont
  {Mourik}}, \bibinfo {author} {\bibfnamefont {K.}~\bibnamefont {Zuo}},
  \bibinfo {author} {\bibfnamefont {S.}~\bibnamefont {Frolov}}, \bibinfo
  {author} {\bibfnamefont {S.}~\bibnamefont {Plissard}}, \bibinfo {author}
  {\bibfnamefont {E.~P. A.~M.}\ \bibnamefont {Bakkers}},\ and\ \bibinfo
  {author} {\bibfnamefont {L.}~\bibnamefont {Kouwenhoven}},\ }\bibfield
  {title} {\bibinfo {title} {Signatures of {Majorana} fermions in hybrid
  superconductor-semiconductor nanowire devices},\ }\href
  {https://doi.org/10.1126/science.1222360} {\bibfield  {journal} {\bibinfo
  {journal} {Sci.}\ }\textbf {\bibinfo {volume} {336}},\ \bibinfo {pages}
  {1003—1007} (\bibinfo {year} {2012})}\BibitemShut {NoStop}%
\bibitem [{\citenamefont {Rokhinson}\ \emph {et~al.}(2012)\citenamefont
  {Rokhinson}, \citenamefont {Liu},\ and\ \citenamefont
  {Furdyna}}]{furdyna2012}%
  \BibitemOpen
  \bibfield  {author} {\bibinfo {author} {\bibfnamefont {L.~P.}\ \bibnamefont
  {Rokhinson}}, \bibinfo {author} {\bibfnamefont {X.}~\bibnamefont {Liu}},\
  and\ \bibinfo {author} {\bibfnamefont {J.~K.}\ \bibnamefont {Furdyna}},\
  }\bibfield  {title} {\bibinfo {title} {The fractional ac {Josephson} effect
  in a semiconductor--superconductor nanowire as a signature of majorana
  particles},\ }\href {https://doi.org/10.1038/nphys2429} {\bibfield  {journal}
  {\bibinfo  {journal} {Nat. Phys.}\ }\textbf {\bibinfo {volume} {8}},\
  \bibinfo {pages} {795} (\bibinfo {year} {2012})}\BibitemShut {NoStop}%
\bibitem [{\citenamefont {Das}\ \emph {et~al.}(2012{\natexlab{a}})\citenamefont
  {Das}, \citenamefont {Ronen}, \citenamefont {Most}, \citenamefont {Oreg},
  \citenamefont {Heiblum},\ and\ \citenamefont {Shtrikman}}]{heiblum2012}%
  \BibitemOpen
  \bibfield  {author} {\bibinfo {author} {\bibfnamefont {A.}~\bibnamefont
  {Das}}, \bibinfo {author} {\bibfnamefont {Y.}~\bibnamefont {Ronen}}, \bibinfo
  {author} {\bibfnamefont {Y.}~\bibnamefont {Most}}, \bibinfo {author}
  {\bibfnamefont {Y.}~\bibnamefont {Oreg}}, \bibinfo {author} {\bibfnamefont
  {M.}~\bibnamefont {Heiblum}},\ and\ \bibinfo {author} {\bibfnamefont
  {H.}~\bibnamefont {Shtrikman}},\ }\bibfield  {title} {\bibinfo {title}
  {Zero-bias peaks and splitting in an {Al}--{InAs} nanowire topological
  superconductor as a signature of {Majorana} fermions},\ }\href
  {https://doi.org/10.1038/nphys2479} {\bibfield  {journal} {\bibinfo
  {journal} {Nat. Phys.}\ }\textbf {\bibinfo {volume} {8}},\ \bibinfo {pages}
  {887} (\bibinfo {year} {2012}{\natexlab{a}})}\BibitemShut {NoStop}%
\bibitem [{\citenamefont {G{\"u}l}\ \emph {et~al.}(2018)\citenamefont
  {G{\"u}l}, \citenamefont {Zhang}, \citenamefont {Bommer}, \citenamefont
  {de~Moor}, \citenamefont {Car}, \citenamefont {Plissard}, \citenamefont
  {Bakkers}, \citenamefont {Geresdi}, \citenamefont {Watanabe}, \citenamefont
  {Taniguchi} \emph {et~al.}}]{kouwenhoven2018}%
  \BibitemOpen
  \bibfield  {author} {\bibinfo {author} {\bibfnamefont {{\"O}.}~\bibnamefont
  {G{\"u}l}}, \bibinfo {author} {\bibfnamefont {H.}~\bibnamefont {Zhang}},
  \bibinfo {author} {\bibfnamefont {J.~D.}\ \bibnamefont {Bommer}}, \bibinfo
  {author} {\bibfnamefont {M.~W.}\ \bibnamefont {de~Moor}}, \bibinfo {author}
  {\bibfnamefont {D.}~\bibnamefont {Car}}, \bibinfo {author} {\bibfnamefont
  {S.~R.}\ \bibnamefont {Plissard}}, \bibinfo {author} {\bibfnamefont {E.~P.}\
  \bibnamefont {Bakkers}}, \bibinfo {author} {\bibfnamefont {A.}~\bibnamefont
  {Geresdi}}, \bibinfo {author} {\bibfnamefont {K.}~\bibnamefont {Watanabe}},
  \bibinfo {author} {\bibfnamefont {T.}~\bibnamefont {Taniguchi}}, \emph
  {et~al.},\ }\bibfield  {title} {\bibinfo {title} {Ballistic {Majorana}
  nanowire devices},\ }\href {https://doi.org/10.1038/s41565-017-0032-8}
  {\bibfield  {journal} {\bibinfo  {journal} {Nat. nanotech.}\ }\textbf
  {\bibinfo {volume} {13}},\ \bibinfo {pages} {192} (\bibinfo {year}
  {2018})}\BibitemShut {NoStop}%
\bibitem [{\citenamefont {Hart}\ \emph {et~al.}(2014)\citenamefont {Hart},
  \citenamefont {Ren}, \citenamefont {Wagner}, \citenamefont {Leubner},
  \citenamefont {M{\"u}hlbauer}, \citenamefont {Br{\"u}ne}, \citenamefont
  {Buhmann}, \citenamefont {Molenkamp},\ and\ \citenamefont
  {Yacoby}}]{yacoby2014}%
  \BibitemOpen
  \bibfield  {author} {\bibinfo {author} {\bibfnamefont {S.}~\bibnamefont
  {Hart}}, \bibinfo {author} {\bibfnamefont {H.}~\bibnamefont {Ren}}, \bibinfo
  {author} {\bibfnamefont {T.}~\bibnamefont {Wagner}}, \bibinfo {author}
  {\bibfnamefont {P.}~\bibnamefont {Leubner}}, \bibinfo {author} {\bibfnamefont
  {M.}~\bibnamefont {M{\"u}hlbauer}}, \bibinfo {author} {\bibfnamefont
  {C.}~\bibnamefont {Br{\"u}ne}}, \bibinfo {author} {\bibfnamefont
  {H.}~\bibnamefont {Buhmann}}, \bibinfo {author} {\bibfnamefont {L.~W.}\
  \bibnamefont {Molenkamp}},\ and\ \bibinfo {author} {\bibfnamefont
  {A.}~\bibnamefont {Yacoby}},\ }\bibfield  {title} {\bibinfo {title} {Induced
  superconductivity in the quantum spin {Hall} edge},\ }\href
  {https://doi.org/10.1038/nphys3036} {\bibfield  {journal} {\bibinfo
  {journal} {Nat. Phys.}\ }\textbf {\bibinfo {volume} {10}},\ \bibinfo {pages}
  {638} (\bibinfo {year} {2014})}\BibitemShut {NoStop}%
\bibitem [{\citenamefont {Yu}\ \emph {et~al.}(2021)\citenamefont {Yu},
  \citenamefont {Chen}, \citenamefont {Gomanko}, \citenamefont {Badawy},
  \citenamefont {Bakkers}, \citenamefont {Zuo}, \citenamefont {Mourik},\ and\
  \citenamefont {Frolov}}]{yu2021non}%
  \BibitemOpen
  \bibfield  {author} {\bibinfo {author} {\bibfnamefont {P.}~\bibnamefont
  {Yu}}, \bibinfo {author} {\bibfnamefont {J.}~\bibnamefont {Chen}}, \bibinfo
  {author} {\bibfnamefont {M.}~\bibnamefont {Gomanko}}, \bibinfo {author}
  {\bibfnamefont {G.}~\bibnamefont {Badawy}}, \bibinfo {author} {\bibfnamefont
  {E.}~\bibnamefont {Bakkers}}, \bibinfo {author} {\bibfnamefont
  {K.}~\bibnamefont {Zuo}}, \bibinfo {author} {\bibfnamefont {V.}~\bibnamefont
  {Mourik}},\ and\ \bibinfo {author} {\bibfnamefont {S.}~\bibnamefont
  {Frolov}},\ }\bibfield  {title} {\bibinfo {title} {Non-{Majorana} states
  yield nearly quantized conductance in proximatized nanowires},\ }\href
  {https://doi.org/10.1038/s41567-020-01107-w} {\bibfield  {journal} {\bibinfo
  {journal} {Nat. Phys.}\ }\textbf {\bibinfo {volume} {17}},\ \bibinfo {pages}
  {482} (\bibinfo {year} {2021})}\BibitemShut {NoStop}%
\bibitem [{\citenamefont {Nadj-Perge}\ \emph {et~al.}(2014)\citenamefont
  {Nadj-Perge}, \citenamefont {Drozdov}, \citenamefont {Li}, \citenamefont
  {Chen}, \citenamefont {Jeon}, \citenamefont {Seo}, \citenamefont {MacDonald},
  \citenamefont {Bernevig},\ and\ \citenamefont {Yazdani}}]{yazdani2014}%
  \BibitemOpen
  \bibfield  {author} {\bibinfo {author} {\bibfnamefont {S.}~\bibnamefont
  {Nadj-Perge}}, \bibinfo {author} {\bibfnamefont {I.~K.}\ \bibnamefont
  {Drozdov}}, \bibinfo {author} {\bibfnamefont {J.}~\bibnamefont {Li}},
  \bibinfo {author} {\bibfnamefont {H.}~\bibnamefont {Chen}}, \bibinfo {author}
  {\bibfnamefont {S.}~\bibnamefont {Jeon}}, \bibinfo {author} {\bibfnamefont
  {J.}~\bibnamefont {Seo}}, \bibinfo {author} {\bibfnamefont {A.~H.}\
  \bibnamefont {MacDonald}}, \bibinfo {author} {\bibfnamefont {B.~A.}\
  \bibnamefont {Bernevig}},\ and\ \bibinfo {author} {\bibfnamefont
  {A.}~\bibnamefont {Yazdani}},\ }\bibfield  {title} {\bibinfo {title}
  {Observation of {Majorana} fermions in ferromagnetic atomic chains on a
  superconductor},\ }\href {https://doi.org/10.1126/science.1259327} {\bibfield
   {journal} {\bibinfo  {journal} {Sci.}\ }\textbf {\bibinfo {volume} {346}},\
  \bibinfo {pages} {602} (\bibinfo {year} {2014})}\BibitemShut {NoStop}%
\bibitem [{\citenamefont {Chen}\ \emph {et~al.}(2016)\citenamefont {Chen},
  \citenamefont {Fang}, \citenamefont {Chen}, \citenamefont {Li},\ and\
  \citenamefont {Tang}}]{tang2016}%
  \BibitemOpen
  \bibfield  {author} {\bibinfo {author} {\bibfnamefont {H.-J.}\ \bibnamefont
  {Chen}}, \bibinfo {author} {\bibfnamefont {X.-W.}\ \bibnamefont {Fang}},
  \bibinfo {author} {\bibfnamefont {C.-Z.}\ \bibnamefont {Chen}}, \bibinfo
  {author} {\bibfnamefont {Y.}~\bibnamefont {Li}},\ and\ \bibinfo {author}
  {\bibfnamefont {X.-D.}\ \bibnamefont {Tang}},\ }\bibfield  {title} {\bibinfo
  {title} {Robust signatures detection of {Majorana} fermions in
  superconducting iron chains},\ }\href {https://doi.org/10.1038/srep36600}
  {\bibfield  {journal} {\bibinfo  {journal} {Sci. Rep.}\ }\textbf {\bibinfo
  {volume} {6}},\ \bibinfo {pages} {36600} (\bibinfo {year}
  {2016})}\BibitemShut {NoStop}%
\bibitem [{\citenamefont {Pawlak}\ \emph {et~al.}(2016)\citenamefont {Pawlak},
  \citenamefont {Kisiel}, \citenamefont {Klinovaja}, \citenamefont {Meier},
  \citenamefont {Kawai}, \citenamefont {Glatzel}, \citenamefont {Loss},\ and\
  \citenamefont {Meyer}}]{loss-meyer_2016}%
  \BibitemOpen
  \bibfield  {author} {\bibinfo {author} {\bibfnamefont {R.}~\bibnamefont
  {Pawlak}}, \bibinfo {author} {\bibfnamefont {M.}~\bibnamefont {Kisiel}},
  \bibinfo {author} {\bibfnamefont {J.}~\bibnamefont {Klinovaja}}, \bibinfo
  {author} {\bibfnamefont {T.}~\bibnamefont {Meier}}, \bibinfo {author}
  {\bibfnamefont {S.}~\bibnamefont {Kawai}}, \bibinfo {author} {\bibfnamefont
  {T.}~\bibnamefont {Glatzel}}, \bibinfo {author} {\bibfnamefont
  {D.}~\bibnamefont {Loss}},\ and\ \bibinfo {author} {\bibfnamefont
  {E.}~\bibnamefont {Meyer}},\ }\bibfield  {title} {\bibinfo {title} {Probing
  atomic structure and majorana wavefunctions in mono-atomic fe chains on
  superconducting pb surface},\ }\href {https://doi.org/10.1038/npjqi.2016.35}
  {\bibfield  {journal} {\bibinfo  {journal} {npj Quant. Inf.}\ }\textbf
  {\bibinfo {volume} {2}},\ \bibinfo {pages} {1} (\bibinfo {year}
  {2016})}\BibitemShut {NoStop}%
\bibitem [{\citenamefont {Miao}\ \emph {et~al.}(2017)\citenamefont {Miao},
  \citenamefont {Jin}, \citenamefont {Zhang},\ and\ \citenamefont
  {Zhou}}]{zhou_PRL_2017}%
  \BibitemOpen
  \bibfield  {author} {\bibinfo {author} {\bibfnamefont {J.-J.}\ \bibnamefont
  {Miao}}, \bibinfo {author} {\bibfnamefont {H.-K.}\ \bibnamefont {Jin}},
  \bibinfo {author} {\bibfnamefont {F.-C.}\ \bibnamefont {Zhang}},\ and\
  \bibinfo {author} {\bibfnamefont {Y.}~\bibnamefont {Zhou}},\ }\bibfield
  {title} {\bibinfo {title} {Exact solution for the interacting {K}itaev chain
  at the symmetric point},\ }\href
  {https://doi.org/10.1103/PhysRevLett.118.267701} {\bibfield  {journal}
  {\bibinfo  {journal} {Phys. Rev. Lett.}\ }\textbf {\bibinfo {volume} {118}},\
  \bibinfo {pages} {267701} (\bibinfo {year} {2017})}\BibitemShut {NoStop}%
\bibitem [{\citenamefont {Wang}\ \emph {et~al.}(2017)\citenamefont {Wang},
  \citenamefont {Miao}, \citenamefont {Jin},\ and\ \citenamefont
  {Chen}}]{chen_PRB_2017}%
  \BibitemOpen
  \bibfield  {author} {\bibinfo {author} {\bibfnamefont {Y.}~\bibnamefont
  {Wang}}, \bibinfo {author} {\bibfnamefont {J.-J.}\ \bibnamefont {Miao}},
  \bibinfo {author} {\bibfnamefont {H.-K.}\ \bibnamefont {Jin}},\ and\ \bibinfo
  {author} {\bibfnamefont {S.}~\bibnamefont {Chen}},\ }\bibfield  {title}
  {\bibinfo {title} {Characterization of topological phases of dimerized
  {K}itaev chain via edge correlation functions},\ }\href
  {https://doi.org/10.1103/PhysRevB.96.205428} {\bibfield  {journal} {\bibinfo
  {journal} {Phys. Rev. B}\ }\textbf {\bibinfo {volume} {96}},\ \bibinfo
  {pages} {205428} (\bibinfo {year} {2017})}\BibitemShut {NoStop}%
\bibitem [{\citenamefont {Miao}\ \emph {et~al.}(2018)\citenamefont {Miao},
  \citenamefont {Jin}, \citenamefont {Zhang},\ and\ \citenamefont
  {Zhou}}]{zhou_scirep_2018}%
  \BibitemOpen
  \bibfield  {author} {\bibinfo {author} {\bibfnamefont {J.-J.}\ \bibnamefont
  {Miao}}, \bibinfo {author} {\bibfnamefont {H.-K.}\ \bibnamefont {Jin}},
  \bibinfo {author} {\bibfnamefont {F.-C.}\ \bibnamefont {Zhang}},\ and\
  \bibinfo {author} {\bibfnamefont {Y.}~\bibnamefont {Zhou}},\ }\bibfield
  {title} {\bibinfo {title} {Majorana zero modes and long range edge
  correlation in interacting {K}itaev chains: analytic solutions and
  density-matrix-renormalization-group study},\ }\href
  {https://doi.org/10.1038/s41598-017-17699-y} {\bibfield  {journal} {\bibinfo
  {journal} {Sc. Rep.}\ }\textbf {\bibinfo {volume} {8}},\ \bibinfo {pages}
  {488} (\bibinfo {year} {2018})}\BibitemShut {NoStop}%
\bibitem [{\citenamefont {Koga}\ \emph {et~al.}(2021)\citenamefont {Koga},
  \citenamefont {Murakami},\ and\ \citenamefont {Nasu}}]{murakami_PRB_2021}%
  \BibitemOpen
  \bibfield  {author} {\bibinfo {author} {\bibfnamefont {A.}~\bibnamefont
  {Koga}}, \bibinfo {author} {\bibfnamefont {Y.}~\bibnamefont {Murakami}},\
  and\ \bibinfo {author} {\bibfnamefont {J.}~\bibnamefont {Nasu}},\ }\bibfield
  {title} {\bibinfo {title} {Majorana correlations in the {K}itaev model with
  ordered-flux structures},\ }\href
  {https://doi.org/10.1103/PhysRevB.103.214421} {\bibfield  {journal} {\bibinfo
   {journal} {Phys. Rev. B}\ }\textbf {\bibinfo {volume} {103}},\ \bibinfo
  {pages} {214421} (\bibinfo {year} {2021})}\BibitemShut {NoStop}%
\bibitem [{\citenamefont {Ma}\ \emph {et~al.}()\citenamefont {Ma},
  \citenamefont {Zhang},\ and\ \citenamefont {Song}}]{ma-song-ArXiv_2023}%
  \BibitemOpen
  \bibfield  {author} {\bibinfo {author} {\bibfnamefont {E.}~\bibnamefont
  {Ma}}, \bibinfo {author} {\bibfnamefont {K.}~\bibnamefont {Zhang}},\ and\
  \bibinfo {author} {\bibfnamefont {Z.}~\bibnamefont {Song}},\ }\bibfield
  {title} {\bibinfo {title} {Topological bulk and edge correlations in a
  {K}itaev model on a square lattice},\ }\href
  {https://arxiv.org/abs/2309.16341} {\bibinfo  {journal} {cond-mat
  ArXiv:2309.16341}\ }\BibitemShut {NoStop}%
\bibitem [{\citenamefont {Vodola}\ \emph {et~al.}(2014)\citenamefont {Vodola},
  \citenamefont {Lepori}, \citenamefont {Ercolessi}, \citenamefont {Gorshkov},\
  and\ \citenamefont {Pupillo}}]{ercolessi_PRL_2014}%
  \BibitemOpen
\bibfield  {journal} {  }\bibfield  {author} {\bibinfo {author} {\bibfnamefont
  {D.}~\bibnamefont {Vodola}}, \bibinfo {author} {\bibfnamefont
  {L.}~\bibnamefont {Lepori}}, \bibinfo {author} {\bibfnamefont
  {E.}~\bibnamefont {Ercolessi}}, \bibinfo {author} {\bibfnamefont {A.~V.}\
  \bibnamefont {Gorshkov}},\ and\ \bibinfo {author} {\bibfnamefont
  {G.}~\bibnamefont {Pupillo}},\ }\bibfield  {title} {\bibinfo {title} {Kitaev
  chains with long-range pairing},\ }\href
  {https://doi.org/10.1103/PhysRevLett.113.156402} {\bibfield  {journal}
  {\bibinfo  {journal} {Phys. Rev. Lett.}\ }\textbf {\bibinfo {volume} {113}},\
  \bibinfo {pages} {156402} (\bibinfo {year} {2014})}\BibitemShut {NoStop}%
\bibitem [{\citenamefont {Vodola}\ \emph {et~al.}(2016)\citenamefont {Vodola},
  \citenamefont {Lepori}, \citenamefont {Ercolessi},\ and\ \citenamefont
  {Pupillo}}]{ercolessi_NJP_2016}%
  \BibitemOpen
  \bibfield  {author} {\bibinfo {author} {\bibfnamefont {D.}~\bibnamefont
  {Vodola}}, \bibinfo {author} {\bibfnamefont {L.}~\bibnamefont {Lepori}},
  \bibinfo {author} {\bibfnamefont {E.}~\bibnamefont {Ercolessi}},\ and\
  \bibinfo {author} {\bibfnamefont {G.}~\bibnamefont {Pupillo}},\ }\bibfield
  {title} {\bibinfo {title} {Long-range {I}sing and {K}itaev models: phases,
  correlations and edge modes},\ }\href
  {https://doi.org/10.1088/1367-2630/18/1/015001} {\bibfield  {journal}
  {\bibinfo  {journal} {New Journal of Physics}\ }\textbf {\bibinfo {volume}
  {18}},\ \bibinfo {pages} {015001} (\bibinfo {year} {2016})}\BibitemShut
  {NoStop}%
\bibitem [{\citenamefont {J\"ager}\ \emph {et~al.}(2020)\citenamefont
  {J\"ager}, \citenamefont {Dell'Anna},\ and\ \citenamefont
  {Morigi}}]{dellanna_PRB_2020}%
  \BibitemOpen
  \bibfield  {author} {\bibinfo {author} {\bibfnamefont {S.~B.}\ \bibnamefont
  {J\"ager}}, \bibinfo {author} {\bibfnamefont {L.}~\bibnamefont {Dell'Anna}},\
  and\ \bibinfo {author} {\bibfnamefont {G.}~\bibnamefont {Morigi}},\
  }\bibfield  {title} {\bibinfo {title} {Edge states of the long-range {K}itaev
  chain: An analytical study},\ }\href
  {https://doi.org/10.1103/PhysRevB.102.035152} {\bibfield  {journal} {\bibinfo
   {journal} {Phys. Rev. B}\ }\textbf {\bibinfo {volume} {102}},\ \bibinfo
  {pages} {035152} (\bibinfo {year} {2020})}\BibitemShut {NoStop}%
\bibitem [{\citenamefont {Francica}\ and\ \citenamefont
  {Dell'Anna}(2022)}]{dellanna_PRB_2022}%
  \BibitemOpen
  \bibfield  {author} {\bibinfo {author} {\bibfnamefont {G.}~\bibnamefont
  {Francica}}\ and\ \bibinfo {author} {\bibfnamefont {L.}~\bibnamefont
  {Dell'Anna}},\ }\bibfield  {title} {\bibinfo {title} {Correlations,
  long-range entanglement, and dynamics in long-range {K}itaev chains},\ }\href
  {https://doi.org/10.1103/PhysRevB.106.155126} {\bibfield  {journal} {\bibinfo
   {journal} {Phys. Rev. B}\ }\textbf {\bibinfo {volume} {106}},\ \bibinfo
  {pages} {155126} (\bibinfo {year} {2022})}\BibitemShut {NoStop}%
\bibitem [{\citenamefont {Takasan}\ \emph {et~al.}(2022)\citenamefont
  {Takasan}, \citenamefont {Sumita},\ and\ \citenamefont
  {Yanase}}]{takasan2022}%
  \BibitemOpen
  \bibfield  {author} {\bibinfo {author} {\bibfnamefont {K.}~\bibnamefont
  {Takasan}}, \bibinfo {author} {\bibfnamefont {S.}~\bibnamefont {Sumita}},\
  and\ \bibinfo {author} {\bibfnamefont {Y.}~\bibnamefont {Yanase}},\
  }\bibfield  {title} {\bibinfo {title} {Supercurrent-induced topological phase
  transitions},\ }\href {https://doi.org/10.1103/PhysRevB.106.014508}
  {\bibfield  {journal} {\bibinfo  {journal} {Phys. Rev. B}\ }\textbf {\bibinfo
  {volume} {106}},\ \bibinfo {pages} {014508} (\bibinfo {year}
  {2022})}\BibitemShut {NoStop}%
\bibitem [{\citenamefont {Kotetes}(2022)}]{kotetes2022}%
  \BibitemOpen
  \bibfield  {author} {\bibinfo {author} {\bibfnamefont {P.}~\bibnamefont
  {Kotetes}},\ }\bibfield  {title} {\bibinfo {title} {Diagnosing topological
  phase transitions in 1d superconductors using {B}erry singularity markers},\
  }\href {https://doi.org/10.1088/1361-648X/ac4f1e} {\bibfield  {journal}
  {\bibinfo  {journal} {J. Phys.: Cond. Mat.}\ }\textbf {\bibinfo {volume}
  {34}},\ \bibinfo {pages} {174003} (\bibinfo {year} {2022})}\BibitemShut
  {NoStop}%
\bibitem [{\citenamefont {Maiellaro}\ \emph {et~al.}(2023)\citenamefont
  {Maiellaro}, \citenamefont {Romeo}, \citenamefont {Illuminati},\ and\
  \citenamefont {Citro}}]{maiellaro2023}%
  \BibitemOpen
  \bibfield  {author} {\bibinfo {author} {\bibfnamefont {A.}~\bibnamefont
  {Maiellaro}}, \bibinfo {author} {\bibfnamefont {F.}~\bibnamefont {Romeo}},
  \bibinfo {author} {\bibfnamefont {F.}~\bibnamefont {Illuminati}},\ and\
  \bibinfo {author} {\bibfnamefont {R.}~\bibnamefont {Citro}},\ }\bibfield
  {title} {\bibinfo {title} {Resilience of topological superconductivity under
  particle current},\ }\href {https://doi.org/10.1103/PhysRevB.107.064505}
  {\bibfield  {journal} {\bibinfo  {journal} {Phys. Rev. B}\ }\textbf {\bibinfo
  {volume} {107}},\ \bibinfo {pages} {064505} (\bibinfo {year}
  {2023})}\BibitemShut {NoStop}%
\bibitem [{\citenamefont {Ma}\ and\ \citenamefont {Song}(2023)}]{ma2023}%
  \BibitemOpen
  \bibfield  {author} {\bibinfo {author} {\bibfnamefont {E.~S.}\ \bibnamefont
  {Ma}}\ and\ \bibinfo {author} {\bibfnamefont {Z.}~\bibnamefont {Song}},\
  }\bibfield  {title} {\bibinfo {title} {Off-diagonal long-range order in the
  ground state of the {K}itaev chain},\ }\href
  {https://doi.org/10.1103/PhysRevB.107.205117} {\bibfield  {journal} {\bibinfo
   {journal} {Phys. Rev. B}\ }\textbf {\bibinfo {volume} {107}},\ \bibinfo
  {pages} {205117} (\bibinfo {year} {2023})}\BibitemShut {NoStop}%
\bibitem [{\citenamefont {Medina~Cuy}\ \emph {et~al.}(2024)\citenamefont
  {Medina~Cuy}, \citenamefont {Buccheri},\ and\ \citenamefont
  {Dolcini}}]{FFF2024}%
  \BibitemOpen
  \bibfield  {author} {\bibinfo {author} {\bibfnamefont {F.~G.}\ \bibnamefont
  {Medina~Cuy}}, \bibinfo {author} {\bibfnamefont {F.}~\bibnamefont
  {Buccheri}},\ and\ \bibinfo {author} {\bibfnamefont {F.}~\bibnamefont
  {Dolcini}},\ }\bibfield  {title} {\bibinfo {title} {Lifshitz transitions and
  {W}eyl semimetals from a topological superconductor with supercurrent flow},\
  }\href {https://doi.org/10.1103/PhysRevResearch.6.033060} {\bibfield
  {journal} {\bibinfo  {journal} {Phys. Rev. Res.}\ }\textbf {\bibinfo {volume}
  {6}},\ \bibinfo {pages} {033060} (\bibinfo {year} {2024})}\BibitemShut
  {NoStop}%
\bibitem [{\citenamefont {Volovik}(2017)}]{volovik2017}%
  \BibitemOpen
  \bibfield  {author} {\bibinfo {author} {\bibfnamefont {G.}~\bibnamefont
  {Volovik}},\ }\bibfield  {title} {\bibinfo {title} {Topological lifshitz
  transitions},\ }\href {https://doi.org/10.1063/1.4974185} {\bibfield
  {journal} {\bibinfo  {journal} {Low Temp. Phys.}\ }\textbf {\bibinfo {volume}
  {43}},\ \bibinfo {pages} {47} (\bibinfo {year} {2017})}\BibitemShut {NoStop}%
\bibitem [{\citenamefont {Volovik}(2018)}]{Volovik2018}%
  \BibitemOpen
  \bibfield  {author} {\bibinfo {author} {\bibfnamefont {G.~E.}\ \bibnamefont
  {Volovik}},\ }\bibfield  {title} {\bibinfo {title} {Exotic {Lifshitz}
  transitions in topological materials},\ }\href
  {https://doi.org/10.3367/UFNe.2017.01.038218} {\bibfield  {journal} {\bibinfo
   {journal} {Phys. Usp.}\ }\textbf {\bibinfo {volume} {61}},\ \bibinfo {pages}
  {89} (\bibinfo {year} {2018})}\BibitemShut {NoStop}%
\bibitem [{\citenamefont {Prem}\ \emph {et~al.}(2017)\citenamefont {Prem},
  \citenamefont {Moroz}, \citenamefont {Gurarie},\ and\ \citenamefont
  {Radzihovsky}}]{prem2017}%
  \BibitemOpen
  \bibfield  {author} {\bibinfo {author} {\bibfnamefont {A.}~\bibnamefont
  {Prem}}, \bibinfo {author} {\bibfnamefont {S.}~\bibnamefont {Moroz}},
  \bibinfo {author} {\bibfnamefont {V.}~\bibnamefont {Gurarie}},\ and\ \bibinfo
  {author} {\bibfnamefont {L.}~\bibnamefont {Radzihovsky}},\ }\bibfield
  {title} {\bibinfo {title} {Multiply quantized vortices in fermionic
  superfluids: Angular momentum, unpaired fermions, and spectral asymmetry},\
  }\href {https://doi.org/10.1103/PhysRevLett.119.067003} {\bibfield  {journal}
  {\bibinfo  {journal} {Phys. Rev. Lett.}\ }\textbf {\bibinfo {volume} {119}},\
  \bibinfo {pages} {067003} (\bibinfo {year} {2017})}\BibitemShut {NoStop}%
\bibitem [{\citenamefont {Tada}(2018)}]{tada2018}%
  \BibitemOpen
  \bibfield  {author} {\bibinfo {author} {\bibfnamefont {Y.}~\bibnamefont
  {Tada}},\ }\bibfield  {title} {\bibinfo {title} {Nonthermodynamic nature of
  the orbital angular momentum in neutral fermionic superfluids},\ }\href
  {https://doi.org/10.1103/PhysRevB.97.214523} {\bibfield  {journal} {\bibinfo
  {journal} {Phys. Rev. B}\ }\textbf {\bibinfo {volume} {97}},\ \bibinfo
  {pages} {214523} (\bibinfo {year} {2018})}\BibitemShut {NoStop}%
\bibitem [{\citenamefont {Liu}\ and\ \citenamefont {Wilczek}(2003)}]{liu2003}%
  \BibitemOpen
  \bibfield  {author} {\bibinfo {author} {\bibfnamefont {W.~V.}\ \bibnamefont
  {Liu}}\ and\ \bibinfo {author} {\bibfnamefont {F.}~\bibnamefont {Wilczek}},\
  }\bibfield  {title} {\bibinfo {title} {Interior gap superfluidity},\ }\href
  {https://doi.org/10.1103/PhysRevLett.90.047002} {\bibfield  {journal}
  {\bibinfo  {journal} {Phys. Rev. Lett.}\ }\textbf {\bibinfo {volume} {90}},\
  \bibinfo {pages} {047002} (\bibinfo {year} {2003})}\BibitemShut {NoStop}%
\bibitem [{\citenamefont {Barouch}\ \emph {et~al.}(1970)\citenamefont
  {Barouch}, \citenamefont {McCoy},\ and\ \citenamefont
  {Dresden}}]{barouch1970}%
  \BibitemOpen
  \bibfield  {author} {\bibinfo {author} {\bibfnamefont {E.}~\bibnamefont
  {Barouch}}, \bibinfo {author} {\bibfnamefont {B.~M.}\ \bibnamefont {McCoy}},\
  and\ \bibinfo {author} {\bibfnamefont {M.}~\bibnamefont {Dresden}},\
  }\bibfield  {title} {\bibinfo {title} {Statistical mechanics of the
  $\mathrm{XY}$ model. i},\ }\href {https://doi.org/10.1103/PhysRevA.2.1075}
  {\bibfield  {journal} {\bibinfo  {journal} {Phys. Rev. A}\ }\textbf {\bibinfo
  {volume} {2}},\ \bibinfo {pages} {1075} (\bibinfo {year} {1970})}\BibitemShut
  {NoStop}%
\bibitem [{\citenamefont {Jordan}\ and\ \citenamefont
  {Wigner}(1928)}]{jw_1928}%
  \BibitemOpen
  \bibfield  {author} {\bibinfo {author} {\bibfnamefont {P.}~\bibnamefont
  {Jordan}}\ and\ \bibinfo {author} {\bibfnamefont {E.}~\bibnamefont
  {Wigner}},\ }\bibfield  {title} {\bibinfo {title} {{\"U}ber das paulische
  {\"a}quivalenzverbot},\ }\href {https://doi.org/10.1007/BF01331938}
  {\bibfield  {journal} {\bibinfo  {journal} {Z. Phys.}\ }\textbf {\bibinfo
  {volume} {47}},\ \bibinfo {pages} {631–651} (\bibinfo {year}
  {1928})}\BibitemShut {NoStop}%
\bibitem [{\citenamefont {Greiter}\ \emph {et~al.}(2014)\citenamefont
  {Greiter}, \citenamefont {Schnells},\ and\ \citenamefont
  {Thomale}}]{greiter2014}%
  \BibitemOpen
  \bibfield  {author} {\bibinfo {author} {\bibfnamefont {M.}~\bibnamefont
  {Greiter}}, \bibinfo {author} {\bibfnamefont {V.}~\bibnamefont {Schnells}},\
  and\ \bibinfo {author} {\bibfnamefont {R.}~\bibnamefont {Thomale}},\
  }\bibfield  {title} {\bibinfo {title} {The 1d {I}sing model and the
  topological phase of the {K}itaev chain},\ }\href
  {https://doi.org/https://doi.org/10.1016/j.aop.2014.08.013} {\bibfield
  {journal} {\bibinfo  {journal} {Ann. Phys.}\ }\textbf {\bibinfo {volume}
  {351}},\ \bibinfo {pages} {1026} (\bibinfo {year} {2014})}\BibitemShut
  {NoStop}%
\bibitem [{\citenamefont {Pan}\ and\ \citenamefont
  {Das~Sarma}(2023)}]{pan2023}%
  \BibitemOpen
  \bibfield  {author} {\bibinfo {author} {\bibfnamefont {H.}~\bibnamefont
  {Pan}}\ and\ \bibinfo {author} {\bibfnamefont {S.}~\bibnamefont
  {Das~Sarma}},\ }\bibfield  {title} {\bibinfo {title} {Majorana nanowires,
  {K}itaev chains, and spin models},\ }\href
  {https://doi.org/10.1103/PhysRevB.107.035440} {\bibfield  {journal} {\bibinfo
   {journal} {Phys. Rev. B}\ }\textbf {\bibinfo {volume} {107}},\ \bibinfo
  {pages} {035440} (\bibinfo {year} {2023})}\BibitemShut {NoStop}%
\bibitem [{\citenamefont {Dzyaloshinsky}(1958)}]{dzyaloshinsky_1958}%
  \BibitemOpen
  \bibfield  {author} {\bibinfo {author} {\bibfnamefont {I.}~\bibnamefont
  {Dzyaloshinsky}},\ }\bibfield  {title} {\bibinfo {title} {A thermodynamic
  theory of weak ferromagnetism of antiferromagnetics},\ }\href
  {https://doi.org/https://doi.org/10.1016/0022-3697(58)90076-3} {\bibfield
  {journal} {\bibinfo  {journal} {J. Phys. Chem. Sol.}\ }\textbf {\bibinfo
  {volume} {4}},\ \bibinfo {pages} {241} (\bibinfo {year} {1958})}\BibitemShut
  {NoStop}%
\bibitem [{\citenamefont {Moriya}(1960)}]{moriya_1960}%
  \BibitemOpen
  \bibfield  {author} {\bibinfo {author} {\bibfnamefont {T.}~\bibnamefont
  {Moriya}},\ }\bibfield  {title} {\bibinfo {title} {New mechanism of
  anisotropic superexchange interaction},\ }\href
  {https://doi.org/10.1103/PhysRevLett.4.228} {\bibfield  {journal} {\bibinfo
  {journal} {Phys. Rev. Lett.}\ }\textbf {\bibinfo {volume} {4}},\ \bibinfo
  {pages} {228} (\bibinfo {year} {1960})}\BibitemShut {NoStop}%
\bibitem [{\citenamefont {Bogdanov}\ and\ \citenamefont
  {Panagopoulos}(2020)}]{bogdanov2020}%
  \BibitemOpen
  \bibfield  {author} {\bibinfo {author} {\bibfnamefont {A.~N.}\ \bibnamefont
  {Bogdanov}}\ and\ \bibinfo {author} {\bibfnamefont {C.}~\bibnamefont
  {Panagopoulos}},\ }\bibfield  {title} {\bibinfo {title} {Physical foundations
  and basic properties of magnetic skyrmions},\ }\href
  {https://doi.org/https://doi.org/10.1038/s42254-020-0203-7} {\bibfield
  {journal} {\bibinfo  {journal} {Nat. Rev. Phys.}\ }\textbf {\bibinfo {volume}
  {2}},\ \bibinfo {pages} {492} (\bibinfo {year} {2020})}\BibitemShut {NoStop}%
\bibitem [{\citenamefont {Fert}\ \emph {et~al.}(2017)\citenamefont {Fert},
  \citenamefont {Reyren},\ and\ \citenamefont {Cros}}]{fert2017}%
  \BibitemOpen
  \bibfield  {author} {\bibinfo {author} {\bibfnamefont {A.}~\bibnamefont
  {Fert}}, \bibinfo {author} {\bibfnamefont {N.}~\bibnamefont {Reyren}},\ and\
  \bibinfo {author} {\bibfnamefont {V.}~\bibnamefont {Cros}},\ }\bibfield
  {title} {\bibinfo {title} {Magnetic skyrmions: advances in physics and
  potential applications},\ }\href
  {https://doi.org/https://doi.org/10.1038/natrevmats.2017.31} {\bibfield
  {journal} {\bibinfo  {journal} {Nat. Rev. Mat.}\ }\textbf {\bibinfo {volume}
  {2}},\ \bibinfo {pages} {1} (\bibinfo {year} {2017})}\BibitemShut {NoStop}%
\bibitem [{\citenamefont {Wiesendanger}(2016)}]{wiesendanger2016}%
  \BibitemOpen
  \bibfield  {author} {\bibinfo {author} {\bibfnamefont {R.}~\bibnamefont
  {Wiesendanger}},\ }\bibfield  {title} {\bibinfo {title} {Nanoscale magnetic
  skyrmions in metallic films and multilayers: a new twist for spintronics},\
  }\href {https://doi.org/https://doi.org/10.1038/natrevmats.2016.44}
  {\bibfield  {journal} {\bibinfo  {journal} {Nat. Rev. Mat.}\ }\textbf
  {\bibinfo {volume} {1}},\ \bibinfo {pages} {1} (\bibinfo {year}
  {2016})}\BibitemShut {NoStop}%
\bibitem [{\citenamefont {Mahdavifar}\ \emph {et~al.}(2024)\citenamefont
  {Mahdavifar}, \citenamefont {Salehpour}, \citenamefont {Cheraghi},\ and\
  \citenamefont {Afrousheh}}]{mahdavifar2024}%
  \BibitemOpen
  \bibfield  {author} {\bibinfo {author} {\bibfnamefont {S.}~\bibnamefont
  {Mahdavifar}}, \bibinfo {author} {\bibfnamefont {M.}~\bibnamefont
  {Salehpour}}, \bibinfo {author} {\bibfnamefont {H.}~\bibnamefont
  {Cheraghi}},\ and\ \bibinfo {author} {\bibfnamefont {K.}~\bibnamefont
  {Afrousheh}},\ }\bibfield  {title} {\bibinfo {title} {Resilience of quantum
  spin fluctuations against {D}zyaloshinskii--{M}oriya interaction},\ }\href
  {https://doi.org/https://doi.org/10.1038/s41598-024-60502-y} {\bibfield
  {journal} {\bibinfo  {journal} {Sc. Rep.}\ }\textbf {\bibinfo {volume}
  {14}},\ \bibinfo {pages} {10034} (\bibinfo {year} {2024})}\BibitemShut
  {NoStop}%
\bibitem [{\citenamefont {Hikihara}\ \emph {et~al.}(2001)\citenamefont
  {Hikihara}, \citenamefont {Kaburagi},\ and\ \citenamefont
  {Kawamura}}]{hikihara2001}%
  \BibitemOpen
  \bibfield  {author} {\bibinfo {author} {\bibfnamefont {T.}~\bibnamefont
  {Hikihara}}, \bibinfo {author} {\bibfnamefont {M.}~\bibnamefont {Kaburagi}},\
  and\ \bibinfo {author} {\bibfnamefont {H.}~\bibnamefont {Kawamura}},\
  }\bibfield  {title} {\bibinfo {title} {Ground-state phase diagrams of
  frustrated spin-{S} {XXZ} chains: Chiral ordered phases},\ }\href
  {https://doi.org/10.1103/PhysRevB.63.174430} {\bibfield  {journal} {\bibinfo
  {journal} {Phys. Rev. B}\ }\textbf {\bibinfo {volume} {63}},\ \bibinfo
  {pages} {174430} (\bibinfo {year} {2001})}\BibitemShut {NoStop}%
\bibitem [{\citenamefont {Roy}\ \emph {et~al.}(2019)\citenamefont {Roy},
  \citenamefont {Chanda}, \citenamefont {Das}, \citenamefont {Sadhukhan},
  \citenamefont {Sen(De)},\ and\ \citenamefont {Sen}}]{roy2019}%
  \BibitemOpen
  \bibfield  {author} {\bibinfo {author} {\bibfnamefont {S.}~\bibnamefont
  {Roy}}, \bibinfo {author} {\bibfnamefont {T.}~\bibnamefont {Chanda}},
  \bibinfo {author} {\bibfnamefont {T.}~\bibnamefont {Das}}, \bibinfo {author}
  {\bibfnamefont {D.}~\bibnamefont {Sadhukhan}}, \bibinfo {author}
  {\bibfnamefont {A.}~\bibnamefont {Sen(De)}},\ and\ \bibinfo {author}
  {\bibfnamefont {U.}~\bibnamefont {Sen}},\ }\bibfield  {title} {\bibinfo
  {title} {Phase boundaries in an alternating-field quantum {XY} model with
  {D}zyaloshinskii-{M}oriya interaction: Sustainable entanglement in
  dynamics},\ }\href {https://doi.org/10.1103/PhysRevB.99.064422} {\bibfield
  {journal} {\bibinfo  {journal} {Phys. Rev. B}\ }\textbf {\bibinfo {volume}
  {99}},\ \bibinfo {pages} {064422} (\bibinfo {year} {2019})}\BibitemShut
  {NoStop}%
\bibitem [{\citenamefont {Wu}\ \emph {et~al.}(2008)\citenamefont {Wu},
  \citenamefont {Ji}, \citenamefont {Chou}, \citenamefont {Li},\ and\
  \citenamefont {Chi}}]{wu_APL_2008}%
  \BibitemOpen
  \bibfield  {author} {\bibinfo {author} {\bibfnamefont {S.}~\bibnamefont
  {Wu}}, \bibinfo {author} {\bibfnamefont {J.-Y.}\ \bibnamefont {Ji}}, \bibinfo
  {author} {\bibfnamefont {M.}~\bibnamefont {Chou}}, \bibinfo {author}
  {\bibfnamefont {W.-H.}\ \bibnamefont {Li}},\ and\ \bibinfo {author}
  {\bibfnamefont {G.}~\bibnamefont {Chi}},\ }\bibfield  {title} {\bibinfo
  {title} {Low-temperature phase separation in {GaN} nanowires: An in situ
  x-ray investigation},\ }\bibfield  {journal} {\bibinfo  {journal} {App. Phys.
  Lett.}\ }\textbf {\bibinfo {volume} {92}},\ \href
  {https://doi.org/10.1063/1.2913207} {10.1063/1.2913207} (\bibinfo {year}
  {2008})\BibitemShut {NoStop}%
\bibitem [{\citenamefont {Tran}\ \emph {et~al.}(2021)\citenamefont {Tran},
  \citenamefont {Weng}, \citenamefont {Hennes}, \citenamefont {Demaille},
  \citenamefont {Coati}, \citenamefont {Vlad}, \citenamefont {Garreau},
  \citenamefont {Sauvage-Simkin}, \citenamefont {Sacchi}, \citenamefont
  {Vidal},\ and\ \citenamefont {Zheng}}]{zheng_2021}%
  \BibitemOpen
  \bibfield  {author} {\bibinfo {author} {\bibfnamefont {T.}~\bibnamefont
  {Tran}}, \bibinfo {author} {\bibfnamefont {X.}~\bibnamefont {Weng}}, \bibinfo
  {author} {\bibfnamefont {M.}~\bibnamefont {Hennes}}, \bibinfo {author}
  {\bibfnamefont {D.}~\bibnamefont {Demaille}}, \bibinfo {author}
  {\bibfnamefont {A.}~\bibnamefont {Coati}}, \bibinfo {author} {\bibfnamefont
  {A.}~\bibnamefont {Vlad}}, \bibinfo {author} {\bibfnamefont {Y.}~\bibnamefont
  {Garreau}}, \bibinfo {author} {\bibfnamefont {M.}~\bibnamefont
  {Sauvage-Simkin}}, \bibinfo {author} {\bibfnamefont {M.}~\bibnamefont
  {Sacchi}}, \bibinfo {author} {\bibfnamefont {F.}~\bibnamefont {Vidal}},\ and\
  \bibinfo {author} {\bibfnamefont {Y.}~\bibnamefont {Zheng}},\ }\bibfield
  {title} {\bibinfo {title} {Spatial correlation of embedded nanowires probed
  by x-ray off-bragg scattering of the host matrix},\ }\href
  {https://doi.org/10.1107/S1600576721006579} {\bibfield  {journal} {\bibinfo
  {journal} {J. App. Crys.}\ }\textbf {\bibinfo {volume} {54}},\ \bibinfo
  {pages} {1173 – 1178} (\bibinfo {year} {2021})}\BibitemShut {NoStop}%
\bibitem [{\citenamefont {Miyoshi}\ \emph {et~al.}(2005)\citenamefont
  {Miyoshi}, \citenamefont {Bugoslavsky},\ and\ \citenamefont
  {Cohen}}]{miyoshi_2005}%
  \BibitemOpen
  \bibfield  {author} {\bibinfo {author} {\bibfnamefont {Y.}~\bibnamefont
  {Miyoshi}}, \bibinfo {author} {\bibfnamefont {Y.}~\bibnamefont
  {Bugoslavsky}},\ and\ \bibinfo {author} {\bibfnamefont {L.~F.}\ \bibnamefont
  {Cohen}},\ }\bibfield  {title} {\bibinfo {title} {Andreev reflection
  spectroscopy of niobium point contacts in a magnetic field},\ }\href
  {https://doi.org/10.1103/PhysRevB.72.012502} {\bibfield  {journal} {\bibinfo
  {journal} {Phys. Rev. B}\ }\textbf {\bibinfo {volume} {72}},\ \bibinfo
  {pages} {012502} (\bibinfo {year} {2005})}\BibitemShut {NoStop}%
\bibitem [{\citenamefont {Daghero}\ \emph {et~al.}(2011)\citenamefont
  {Daghero}, \citenamefont {Tortello}, \citenamefont {Ummarino},\ and\
  \citenamefont {Gonnelli}}]{gonnelli_2011}%
  \BibitemOpen
  \bibfield  {author} {\bibinfo {author} {\bibfnamefont {D.}~\bibnamefont
  {Daghero}}, \bibinfo {author} {\bibfnamefont {M.}~\bibnamefont {Tortello}},
  \bibinfo {author} {\bibfnamefont {G.}~\bibnamefont {Ummarino}},\ and\
  \bibinfo {author} {\bibfnamefont {R.}~\bibnamefont {Gonnelli}},\ }\bibfield
  {title} {\bibinfo {title} {Directional point-contact {A}ndreev-reflection
  spectroscopy of fe-based superconductors: Fermi surface topology, gap
  symmetry, and electron-boson interaction},\ }\bibfield  {journal} {\bibinfo
  {journal} {Reports on Progress in Physics}\ }\textbf {\bibinfo {volume}
  {74}},\ \href {https://doi.org/10.1088/0034-4885/74/12/124509}
  {10.1088/0034-4885/74/12/124509} (\bibinfo {year} {2011})\BibitemShut
  {NoStop}%
\bibitem [{\citenamefont {Beckmann}\ \emph {et~al.}(2004)\citenamefont
  {Beckmann}, \citenamefont {Weber},\ and\ \citenamefont
  {V.~L\"ohneysen}}]{beckmann_2004}%
  \BibitemOpen
  \bibfield  {author} {\bibinfo {author} {\bibfnamefont {D.}~\bibnamefont
  {Beckmann}}, \bibinfo {author} {\bibfnamefont {H.~B.}\ \bibnamefont
  {Weber}},\ and\ \bibinfo {author} {\bibfnamefont {H.}~\bibnamefont
  {V.~L\"ohneysen}},\ }\bibfield  {title} {\bibinfo {title} {Evidence for
  crossed {A}ndreev reflection in superconductor-ferromagnet hybrid
  structures},\ }\href {https://doi.org/10.1103/PhysRevLett.93.197003}
  {\bibfield  {journal} {\bibinfo  {journal} {Phys. Rev. Lett.}\ }\textbf
  {\bibinfo {volume} {93}},\ \bibinfo {pages} {197003} (\bibinfo {year}
  {2004})}\BibitemShut {NoStop}%
\bibitem [{\citenamefont {Das}\ \emph {et~al.}(2012{\natexlab{b}})\citenamefont
  {Das}, \citenamefont {Ronen}, \citenamefont {Heiblum}, \citenamefont
  {Mahalu}, \citenamefont {Kretinin},\ and\ \citenamefont
  {Shtrikman}}]{das2012}%
  \BibitemOpen
  \bibfield  {author} {\bibinfo {author} {\bibfnamefont {A.}~\bibnamefont
  {Das}}, \bibinfo {author} {\bibfnamefont {Y.}~\bibnamefont {Ronen}}, \bibinfo
  {author} {\bibfnamefont {M.}~\bibnamefont {Heiblum}}, \bibinfo {author}
  {\bibfnamefont {D.}~\bibnamefont {Mahalu}}, \bibinfo {author} {\bibfnamefont
  {A.~V.}\ \bibnamefont {Kretinin}},\ and\ \bibinfo {author} {\bibfnamefont
  {H.}~\bibnamefont {Shtrikman}},\ }\bibfield  {title} {\bibinfo {title}
  {High-efficiency cooper pair splitting demonstrated by two-particle
  conductance resonance and positive noise cross-correlation},\ }\href
  {https://doi.org/https://doi.org/10.1038/ncomms2169} {\bibfield  {journal}
  {\bibinfo  {journal} {Nat. Comm.}\ }\textbf {\bibinfo {volume} {3}},\
  \bibinfo {pages} {1165} (\bibinfo {year} {2012}{\natexlab{b}})}\BibitemShut
  {NoStop}%
\bibitem [{\citenamefont {He}\ \emph {et~al.}(2014)\citenamefont {He},
  \citenamefont {Wu}, \citenamefont {Choy}, \citenamefont {Liu}, \citenamefont
  {Tanaka},\ and\ \citenamefont {Law}}]{he_2014}%
  \BibitemOpen
  \bibfield  {author} {\bibinfo {author} {\bibfnamefont {J.~J.}\ \bibnamefont
  {He}}, \bibinfo {author} {\bibfnamefont {J.}~\bibnamefont {Wu}}, \bibinfo
  {author} {\bibfnamefont {T.-P.}\ \bibnamefont {Choy}}, \bibinfo {author}
  {\bibfnamefont {X.-J.}\ \bibnamefont {Liu}}, \bibinfo {author} {\bibfnamefont
  {Y.}~\bibnamefont {Tanaka}},\ and\ \bibinfo {author} {\bibfnamefont
  {K.}~\bibnamefont {Law}},\ }\bibfield  {title} {\bibinfo {title} {Correlated
  spin currents generated by resonant-crossed {A}ndreev reflections in
  topological superconductors},\ }\bibfield  {journal} {\bibinfo  {journal}
  {Nat. Comm.}\ }\textbf {\bibinfo {volume} {5}},\ \href
  {https://doi.org/10.1038/ncomms4232} {10.1038/ncomms4232} (\bibinfo {year}
  {2014})\BibitemShut {NoStop}%
\bibitem [{\citenamefont {P{\"o}schl}\ \emph {et~al.}(2022)\citenamefont
  {P{\"o}schl}, \citenamefont {Danilenko}, \citenamefont {Sabonis},
  \citenamefont {Kristjuhan}, \citenamefont {Lindemann}, \citenamefont
  {Thomas}, \citenamefont {Manfra},\ and\ \citenamefont {Marcus}}]{poschl2022}%
  \BibitemOpen
  \bibfield  {author} {\bibinfo {author} {\bibfnamefont {A.}~\bibnamefont
  {P{\"o}schl}}, \bibinfo {author} {\bibfnamefont {A.}~\bibnamefont
  {Danilenko}}, \bibinfo {author} {\bibfnamefont {D.}~\bibnamefont {Sabonis}},
  \bibinfo {author} {\bibfnamefont {K.}~\bibnamefont {Kristjuhan}}, \bibinfo
  {author} {\bibfnamefont {T.}~\bibnamefont {Lindemann}}, \bibinfo {author}
  {\bibfnamefont {C.}~\bibnamefont {Thomas}}, \bibinfo {author} {\bibfnamefont
  {M.~J.}\ \bibnamefont {Manfra}},\ and\ \bibinfo {author} {\bibfnamefont
  {C.~M.}\ \bibnamefont {Marcus}},\ }\bibfield  {title} {\bibinfo {title}
  {Nonlocal conductance spectroscopy of {A}ndreev bound states in gate-defined
  {InAs/Al} nanowires},\ }\href {https://doi.org/10.1103/PhysRevB.106.L241301}
  {\bibfield  {journal} {\bibinfo  {journal} {Phys. Rev. B}\ }\textbf {\bibinfo
  {volume} {106}},\ \bibinfo {pages} {L241301} (\bibinfo {year}
  {2022})}\BibitemShut {NoStop}%
\bibitem [{\citenamefont {Wei}\ \emph {et~al.}(2018)\citenamefont {Wei},
  \citenamefont {Ramanathan},\ and\ \citenamefont
  {Cappellaro}}]{cappellaro_PRL_108}%
  \BibitemOpen
  \bibfield  {author} {\bibinfo {author} {\bibfnamefont {K.~X.}\ \bibnamefont
  {Wei}}, \bibinfo {author} {\bibfnamefont {C.}~\bibnamefont {Ramanathan}},\
  and\ \bibinfo {author} {\bibfnamefont {P.}~\bibnamefont {Cappellaro}},\
  }\bibfield  {title} {\bibinfo {title} {Exploring localization in nuclear spin
  chains},\ }\href {https://doi.org/10.1103/PhysRevLett.120.070501} {\bibfield
  {journal} {\bibinfo  {journal} {Phys. Rev. Lett.}\ }\textbf {\bibinfo
  {volume} {120}},\ \bibinfo {pages} {070501} (\bibinfo {year}
  {2018})}\BibitemShut {NoStop}%
\bibitem [{\citenamefont {Richerme}\ \emph {et~al.}(2014)\citenamefont
  {Richerme}, \citenamefont {Gong}, \citenamefont {Lee}, \citenamefont {Senko},
  \citenamefont {Smith}, \citenamefont {Foss-Feig}, \citenamefont {Michalakis},
  \citenamefont {Gorshkov},\ and\ \citenamefont {Monroe}}]{richerme2014}%
  \BibitemOpen
  \bibfield  {author} {\bibinfo {author} {\bibfnamefont {P.}~\bibnamefont
  {Richerme}}, \bibinfo {author} {\bibfnamefont {Z.-X.}\ \bibnamefont {Gong}},
  \bibinfo {author} {\bibfnamefont {A.}~\bibnamefont {Lee}}, \bibinfo {author}
  {\bibfnamefont {C.}~\bibnamefont {Senko}}, \bibinfo {author} {\bibfnamefont
  {J.}~\bibnamefont {Smith}}, \bibinfo {author} {\bibfnamefont
  {M.}~\bibnamefont {Foss-Feig}}, \bibinfo {author} {\bibfnamefont
  {S.}~\bibnamefont {Michalakis}}, \bibinfo {author} {\bibfnamefont {A.~V.}\
  \bibnamefont {Gorshkov}},\ and\ \bibinfo {author} {\bibfnamefont
  {C.}~\bibnamefont {Monroe}},\ }\bibfield  {title} {\bibinfo {title}
  {Non-local propagation of correlations in quantum systems with long-range
  interactions},\ }\href {https://doi.org/https://doi.org/10.1038/nature13450}
  {\bibfield  {journal} {\bibinfo  {journal} {Nat.}\ }\textbf {\bibinfo
  {volume} {511}},\ \bibinfo {pages} {198} (\bibinfo {year}
  {2014})}\BibitemShut {NoStop}%
\bibitem [{\citenamefont {Jurcevic}\ \emph {et~al.}(2014)\citenamefont
  {Jurcevic}, \citenamefont {Lanyon}, \citenamefont {Hauke}, \citenamefont
  {Hempel}, \citenamefont {Zoller}, \citenamefont {Blatt},\ and\ \citenamefont
  {Roos}}]{jurcevic2014}%
  \BibitemOpen
  \bibfield  {author} {\bibinfo {author} {\bibfnamefont {P.}~\bibnamefont
  {Jurcevic}}, \bibinfo {author} {\bibfnamefont {B.~P.}\ \bibnamefont
  {Lanyon}}, \bibinfo {author} {\bibfnamefont {P.}~\bibnamefont {Hauke}},
  \bibinfo {author} {\bibfnamefont {C.}~\bibnamefont {Hempel}}, \bibinfo
  {author} {\bibfnamefont {P.}~\bibnamefont {Zoller}}, \bibinfo {author}
  {\bibfnamefont {R.}~\bibnamefont {Blatt}},\ and\ \bibinfo {author}
  {\bibfnamefont {C.~F.}\ \bibnamefont {Roos}},\ }\bibfield  {title} {\bibinfo
  {title} {Quasiparticle engineering and entanglement propagation in a quantum
  many-body system},\ }\href
  {https://doi.org/https://doi.org/10.1038/nature13461} {\bibfield  {journal}
  {\bibinfo  {journal} {Nat.}\ }\textbf {\bibinfo {volume} {511}},\ \bibinfo
  {pages} {202} (\bibinfo {year} {2014})}\BibitemShut {NoStop}%
\end{thebibliography}%

\end{document}